\documentclass{jfm}

\usepackage{graphicx}
\usepackage[caption=false]{subfig}
\usepackage{natbib}
\usepackage{amsmath,amssymb}

\ifCUPmtlplainloaded \else
  \checkfont{eurm10}
  \iffontfound
    \IfFileExists{upmath.sty}
      {\typeout{^^JFound AMS Euler Roman fonts on the system,
                   using the 'upmath' package.^^J}%
       \usepackage{upmath}}
      {\typeout{^^JFound AMS Euler Roman fonts on the system, but you
                   dont seem to have the}%
       \typeout{'upmath' package installed. JFM.cls can take advantage
                 of these fonts,^^Jif you use 'upmath' package.^^J}%
       \Prananovidecommand\upi{\upi}%
      }
  \else
    \providecommand\upi{\upi}%
  \fi
\fi

\ifCUPmtlplainloaded \else
  \checkfont{msam10}
  \iffontfound
    \IfFileExists{amssymb.sty}
      {\typeout{^^JFound AMS Symbol fonts on the system, using the
                'amssymb' package.^^J}%
       \usepackage{amssymb}%
       \let\le=\leqslant  \let\leq=\leqslant
       \let\ge=\geqslant  
      }{}
  \fi
\fi

\ifCUPmtlplainloaded \else
  \IfFileExists{amsbsy.sty}
    {\typeout{^^JFound the 'amsbsy' package on the system, using it.^^J}%
     \usepackage{amsbsy}}
    {\providecommand\boldsymbol[1]{\mbox{\boldmath $##1$}}}
\fi

\providecommand\bnabla{\boldsymbol{\nabla}}
\providecommand\bcdot{\boldsymbol{\cdot}}

\newcommand\Pran{\mbox{\textit{Pr}}} 
\newcommand\Pe{\mbox{\textit{Pe}}}  
\newcommand\Ra{\mbox{\textit{Ra}}}  
\newcommand\Nu{\mbox{\textit{Nu}}}  

\newcommand{\zhat}{\hat{\boldsymbol{z}}}

\newcommand{\m}{\mathrm{max}}
\newcommand{\M}{\mathrm{MAX}}
\newcommand {\vel}{\boldsymbol{v}}

\newsavebox{\astrutbox}
\sbox{\astrutbox}{\rule[-5pt]{0pt}{20pt}}

\title[Wall to Wall Optimal Transport]{Wall to Wall Optimal Transport}

\author[P. Hassanzadeh, G. P. Chini and C. R. Doering]%
{Pedram Hassanzadeh$^1$\thanks{Current address: Center for the Environment and Department of Earth and Planetary Sciences, Harvard University, Cambridge, MA 02139, hassanzadeh@fas.harvard.edu},\ns
Gregory P. Chini$^2$,\ns
\& Charles R. Doering$^3$\thanks{Email address for correspondence: doering@umich.edu}}

\affiliation{$^1$Department of Mechanical Engineering, University of California, Berkeley, CA 94720, USA\\[\affilskip]
$^2$Department of Mechanical Engineering, Program in Integrated Applied Mathematics, and Center for Fluid Physics, University of New Hampshire, Durham, NH 03824, USA\\[\affilskip] 
$^3$Department of Mathematics, Department of Physics, and Center for the Study of Complex Systems, University of Michigan, Ann Arbor, Michigan 48109, USA
}

\pubyear{2010}
\volume{650}
\pagerange{119--126}
\date{?; revised ?; accepted ?. - To be entered by editorial office}
\begin{document}

\maketitle

\begin{abstract}
The calculus of variations is employed to find steady divergence-free velocity fields that maximize transport of a tracer between two parallel walls held
at fixed concentration for one of two constraints on flow strength: a fixed value of the kinetic energy (mean square velocity) or a fixed value of the enstrophy (mean square vorticity).
The optimizing flows consist of an array of (convection) cells of a particular aspect ratio $\varGamma$.
We solve the nonlinear Euler--Lagrange equations analytically for weak flows and numerically---and via matched asymptotic analysis in the fixed energy case---for strong flows.
We report the results in terms of the Nusselt number $\Nu$, a dimensionless measure of the tracer transport, as a function of the P\'eclet number $\Pe$, a dimensionless measure of the strength of the flow.
For both constraints the maximum transport $\Nu_\mathrm{MAX}(\Pe)$ is realized in cells of decreasing aspect ratio $\varGamma_\mathrm{opt}(\Pe)$ as $\Pe$ increases.
For the fixed energy problem, $\Nu_\mathrm{MAX} \sim \Pe$ and $\varGamma_\mathrm{opt} \sim \Pe^{-1/2}$, while for the fixed enstrophy scenario, $\Nu_\mathrm{MAX} \sim \Pe^{10/17}$ and $\varGamma_\mathrm{opt} \sim \Pe^{-0.36}$.
We interpret our results in the context of buoyancy-driven Rayleigh--B\'enard convection problems that satisfy the flow intensity constraints, enabling us to investigate how the transport scalings compare with upper bounds on $\Nu$ expressed as a function of the Rayleigh number $\Ra$.
For steady convection in porous media, corresponding to the fixed energy problem, we find $\Nu_\mathrm{MAX} \sim \Ra$ and $\varGamma_\mathrm{opt} \sim \Ra^{-1/2}$, while for steady convection in a pure fluid layer between stress-free isothermal walls, corresponding to fixed enstrophy transport, $\Nu_\mathrm{MAX} \sim \Ra^{5/12}$ and $\varGamma_\mathrm{opt} \sim \Ra^{-1/4}$.
\end{abstract}

\begin{keywords}
\end{keywords}

\section{Introduction} \label{sec:intro}

Transport and mixing by incompressible flows are ubiquitous phenomena in science and engineering.
In some applications, e.g., cooling or heating, the aim may be to maximize transport of an advected and diffused quantity.
In other problems, such as pollutant spills, the goal may be to move a substance from one location to another as quickly as possible while minimizing mixing. 
In some systems, naturally occurring or engineered, the transported quantity is passive, moved by the fluid but not forcing the flow, while in other situations, including thermal and compositional convection, it is active and drives its own displacement.

In every case once the relevant equations of motions are agreed upon key theoretical challenges include (1) qualitatively understanding the physical mechanisms of transport and mixing in the mathematical model, (2) quantitatively estimating the magnitude and/or efficiency of these processes as functions of initial and boundary conditions, source and sink distributions, and/or applied forces, (3) determining fundamental limits on the transport and/or mixing effectiveness of relevant classes of fluid flows in terms of rigorous bounds, and (4) understanding how such limits might be approached, or even achieved.

The topic of maximal transport has played a significant role in theoretical and mathematical fluid mechanics at least since \cite{Malkus} raised the question of limits on heat transfer in thermal convection $60$ years ago.
Indeed, Malkus' speculation that convective turbulence might optimize heat transport inspired \cite{Howard63} to derive upper bounds on heat transport in Rayleigh-B\'enard convection from the Boussinesq approximation to the Navier-Stokes equations.
His approach combined the connection between bulk averaged convective heat flux and turbulent energy dissipation, a mild statistical hypothesis, and a clever variational analysis to produce the first effective rigorous limits on turbulent transport.
Howard's variational problem was soon subjected to sophisticated asymptotic analysis by \cite{Busse69}, and was generalized to derive limits on momentum transport in shear flows \cite[]{Nickerson69,Busse70}, thermal convection in a fluid saturated porous layer \cite[]{BJ1972,GJ1973}, and, later, aspects of passive tracer transport in shear-driven turbulence \cite[]{KS1987}.

An alternative variational formulation for extremes of heat and momentum transport in boundary-driven flows was introduced over two decades ago by \cite{DC1992,DC1994,DC1996}.  
In some scenarios this so-called `background method' reproduces the mathematical estimates following from Howard's approach \cite[]{Kerswell98}, but it has also produced a steady stream of original results for systems including turbulent precession \cite[]{Kerswell96}, shear and stress driven flows \cite[]{DSW2000,TCY2004,HD2010,HD2013}, turbulent mixing \cite[]{CP2001,TCK2009}, and a variety of buoyancy driven flows \cite[]{SKB2004,DOR2006,Whitehead11} including Rayleigh-B\'enard convection in a fluid saturated porous layer \cite[]{Doering98,WCDD2013}.

A distinct mathematical strategy applicable to body force--rather than boundary--driven flows was proposed a dozen years ago by \cite{DF2002} and subsequently used to produce turbulent energy dissipation rate bounds for a variety of Navier-Stokes flows \cite[]{DES2003,PLD2005,CDP2007,RDD2011}.
That approach was also generalized to deduce limits on passive tracer mixing as measured by the suppression of tracer concentration variance (and other norms) in the presence of scalar sources and sinks; see, for example, \cite{TDG2004}, \cite{PY2006}, \cite{Stirring07}, and \cite{Stirring08}.

In active transport problems, the maximally transporting `optimal' flows produced by bounding analyses are normally not solutions of the underlying
equations of motion for the fluid.  The bounding methods mentioned above involve extremizing transport or dissipation over a larger set of flows, a superset of the solutions sharing certain features including incompressibility, boundary conditions, and a selection of suitably averaged energy, momentum, and/or other bulk balances.
Of course the precision of a bound may be questioned if the optimizing flow is not realized as an exact solution of the equations of motion.\footnote{The high Reynolds number scaling of a dissipation rate bound is known to be sharp, realized by an exact solution of the Navier-Stokes equations, in at least one case  \cite[]{DSW2000}.}
Nevertheless, such optimal flows provide insight into the structure of maximally effective modes of transport whether or not they are naturally realizable.

Several researchers recently formulated some maximal transient mixing models as optimal control problems.
The idea here is to determine, from within a specified class of admissible flows, those stirring protocols that extremize an appropriate measure of mixing at a specific instant of time, or its rate of change as a function of time; see, for example, \cite{d1999control}, \cite{Optimal07}, \cite{Optimal08}, \cite{Optimal11}, and \cite{gubanov2010towards,Cost12}.
Constraints on the admissible class of flows include incompressibility and boundary conditions, a limit (of some sort) on flow intensity, and possibly other structural restrictions.
Although the flows are not required {\it a priori} to solve any particular physical equations of motion, any such divergence-free velocity field can formally be considered a solution of the incompressible Stokes, Navier-Stokes, or even Euler equations satisfying their boundary conditions subjected to suitable body forces.
The realizability issue is then an `engineering' challenge to implement the body forces that generate the desired flow.
Again, whether or not it is practical or even possible to design such a system, the optimal flows provide physical insight into the mechanics of mixing and the rigorous bounds provide a concrete target to strive for.

These three score years of study serve as the context and provide the inspiration and motivation for the research reported in this paper.
As a natural next step along this line of investigation, we address two general and generic questions:
\begin{itemize}
\item What is the maximum rate at which a tracer (a scalar concentration that we will refer to as temperature) can be transported by divergence-free velocity fields satisfying some given boundary conditions and flow intensity constraints?
\item What does the optimizing velocity field look like?
\end{itemize}

The precise problem we pose is a combination of the continuing challenge to derive sharp bounds on heat transport in Rayleigh-B\'enard convection and a precursor to an optimal control approach to transport problems in Navier-Stokes flows.
The conditions and constraints considered here are (i) that the flow is confined between two impermeable parallel boundaries and is endowed with a specified root mean square speed or vorticity, and (ii) that the concentration of the tracer is constant on each boundary, i.e., the boundaries are isothermal.
For simplicity we restrict attention to steady flows in two spatial dimensions. We do not require the incompressible velocity field to satisfy any familiar momentum equation (e.g., for motion driven by the buoyancy force) but only a suitable bulk amplitude constraint which, after the fact, may be interpreted in terms of well-known dissipation-transport balances (e.g., for classical Rayleigh-B\'enard convection).  
Unlike prior approaches to deriving rigorous bounds, however, in this analysis the full advection--diffusion equation for the transported scalar field is enforced as a constraint.

In the following section we present the full formulation of the problem.
Subsequently in sections~\ref{sec:problem1} and \ref{sec:problem2} we analyze the problem with, respectively, the fixed energy (mean square speed) and fixed enstrophy (mean square vorticity) constraints on stirring strength.
We then employ the calculus of variations to maximize the relevant transport, a functional of the flow, subject to the relevant constraints.
The nonlinear Euler--Lagrange equations are linearized for weak flows and solved analytically, and for stronger flows we solve the Euler--Lagrange equations using numerical continuation.
Exploiting observed symmetries, we solve the Euler--Lagrange equations via matched asymptotic analysis in the fixed energy case.
The maximal transport obtained thereby serves as an upper bound for problems that share the boundary conditions and flow intensity constraints.
We compare these results for relevant Rayleigh--B\'enard problems: convection in a fluid saturated porous layer (fixed energy) and convection in a Boussinesq fluid with stress-free boundaries (fixed enstrophy).
The concluding section~\ref{sec:conc} includes a discussion of future work utilizing this and related approaches to study extreme behavior in fluid dynamical systems.   

\section{Mathematical Formulation} \label{sec:math}

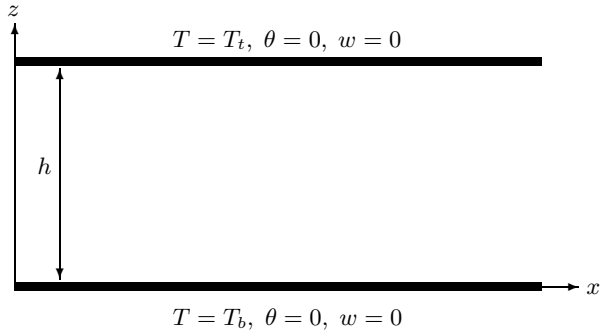
\begin{figure}
\setlength{\unitlength}{1cm}
\begin{picture}(8,5)
  \linethickness{1mm}
  \put(3.5 ,3.5){\line(1,0){7}}  
  \linethickness{1mm}
  \put(3.5 ,0.5){\line(1,0){7}}
  \linethickness{0.1mm}
  \put(4.1, 0.6){\vector(0, 1){2.8}}
  \put(4.1, 3.4){\vector(0, -1){2.8}}
\put(3.8, 2){$h$} 
\put(5.6, 3.7){$T=T_t, \; \theta=0, \; w=0$} 
\put(5.6, 0){$T=T_b,  \; \theta=0,  \; w=0$} 
  \put(3.5, .5){\vector(0, 1){3.5}}
  \put(3.5, .5){\vector(1, 0){7.5}}
\put(3.4, 4.1){$z$} 
\put(11.1, .4){$x$}
\end{picture}
\caption{Schematic of the configuration. The top and bottom boundaries are impermeable and held at fixed temperatures.}
\label{schem}
\end{figure}

We focus on heat transport in two dimensions in terms of a temperature field $T(x,z,t)$ satisfying the advection-diffusion equation
\begin{eqnarray}
\dot{T} + \vel \bcdot \bnabla T = \kappa \Delta T, \label{AdDiff}
\end{eqnarray}
where $\dot{T} = \partial T/\partial t$, $\Delta=\partial^2/\partial x^2+\partial^2/\partial z^2$, $\kappa$ is the (constant) thermal diffusivity of the fluid, and  $\vel(x,z,t)= u \hat{\boldsymbol{x}} + w \zhat$ is an incompressible flow field, i.e.,
\begin{equation}
\bnabla \bcdot \vel = 0 \label{Div}.
\end{equation}
The geometry of the problem is shown in figure~\ref{schem}.
The domain $D = [0,L] \times [0,h]$ is bounded by two parallel impermeable walls held at fixed temperatures.
Respecting tradition we align the walls horizontally in the $x$ direction separated by distance $h$ in the vertical $z$ direction.
All system variables are periodic in the $x$ direction with horizontal period $L$.

We consider velocity fields that have either fixed mean squared speed
\begin{eqnarray}
U^2 = \frac{1}{hL}\int_D | \vel |^2 \,  \mathrm{d}x \,\mathrm{d}z , \label{C1}
\end{eqnarray}    
or fixed enstrophy density
\begin{eqnarray}
\Omega^2 = \frac{1}{hL}\int_D | \boldsymbol{\omega}|^2 \,  \mathrm{d}x \, \mathrm{d}z =
\frac{1}{hL}\int_D | \bnabla \vel |^2 \,  \mathrm{d}x \, \mathrm{d}z,  \label{C2} 
\end{eqnarray} 
where $\boldsymbol{\omega} = \bnabla \times \vel$ is the vorticity and $| \bnabla \vel |^2 = \bnabla \vel \, \boldsymbol{\colon} \bnabla \vel$.
The second equality in (\ref{C2}) follows for many boundary conditions on impermeable horizontal boundaries including no-slip or stress-free conditions.\footnote{By 'stress-free' we mean homogeneous Neumann boundary conditions on the tangential velocity component ($\partial u/\partial z = 0$) corresponding to stress-free boundaries for Newtonian fluids.}
The physical significance of the second expression is that, when multiplied by viscosity, it yields the viscous energy dissipation rate for Newtonian fluids in these domains.
Hence the enstrophy constraint is natural to consider in situations where a power budget limits the strength of the flow in a viscous fluid.

We non-dimensionalize the system using the spacing between the walls $h$ and the diffusion time scale $h^2/\kappa$.
The dimensionless temperature is naturally $(T-T_t)/(T_b-T_t)$ taking values $1$ and $0$ on the bottom and top boundaries.
The (dimensionless) Pe\'clet number $\Pe$, measuring the relative strength of the flow, is the ratio of the diffusive time scale to the advective time scale (i.e., a measure of the strength of advection relative to diffusion).
For the problem with fixed energy we define
\begin{eqnarray}
\Pe &\equiv& \frac{Uh}{\kappa},  \label{PeU}
\end{eqnarray} 
or for the problem with fixed enstrophy  
\begin{eqnarray}
\Pe &\equiv& \frac{\Omega h^2}{\kappa}.  \label{PeO}
\end{eqnarray} 
We also define the aspect ratio
\begin{eqnarray}
\varGamma \equiv \frac{L}{2h},
\end{eqnarray}
the normalized width of a \emph{single} convection cell associated with the optimal flow (see, e.g., figure~\ref{fig:rolls}).
In the absence of advection (when $\vel=0$) the transport is purely conductive with steady dimensionless temperature field $1-z$ so we define the dimensionless temperature deviation variable $\theta \equiv (T-T_t)/(T_b-T_t) - 1 + z$ which vanishes on the boundaries.
Hereafter all variables (i.e., $\vel$, $T$, $\theta$, $x$, $z$, $t$) are dimensionless.

Non-dimensionalizing equations~(\ref{AdDiff})--(\ref{Div}) and the boundary conditions yields  
\begin{eqnarray}
\dot{\theta} + \vel \bcdot \bnabla \theta &=& \Delta \theta + w, \label{ad} \\
\bnabla \bcdot \vel &=& 0, \label{div} \\
\theta(x,0,t) &=& \theta(x,1,t) = 0, \label{bct} \\
w(x,0,t) &=& w(x,1,t) = 0. \label{bcv}
\end{eqnarray}   
Define the long time-space average by angle brackets $\left< \cdot \right>$:  
\begin{eqnarray}
\left< \boldsymbol{a}(x,z,t) \right> \equiv  \underset{t \rightarrow \infty}{\mathrm{lim}} \frac{1}{t} \int_0^{t} \left\{ \frac{1}{2\varGamma} \int_D \boldsymbol{a}(x,z,s) \;  \mathrm{d}x \, \mathrm{d}z \right\} \mathrm{d}s. 
\end{eqnarray}
The fixed energy constraint (\ref{C1}) is then
\begin{eqnarray}
\Pe^2 = \left< |\vel|^2 \right>, \label{ke}
\end{eqnarray} 
while the fixed enstrophy constraint (\ref{C2}) is
\begin{eqnarray}
\Pe^2 = \left< |\bnabla \vel|^2 \right>. \label{en}
\end{eqnarray} 

The Nusselt number $\Nu$ measures the scalar transport by advection and is defined as the ratio of the total flux in the presence of advection to the heat flux by pure conduction.
We are interested in the vertical transport between horizontal walls so
\begin{eqnarray}
\Nu = 1+ \left<  w T \right> = 1 + \left<  w \theta \right>. 
\label{Nu}
\end{eqnarray}
Note that $\left< w \, f(\cdot) \right>=0$ for any $f(z)$ as a result of incompressibility and the boundary conditions. 

\subsection{Objective}\label{sec:goal}

With the strength of advection gauged by $\Pe$, the geometry of the flow parameterized by $\varGamma$, and strength of transport measured by $\Nu$ as defined above, we are now in a position to fully formulate the task at hand:
\smallskip
\begin{description}
\item{1)} Search over all steady divergence-free velocity fields $\vel$ with given $(\Pe,\varGamma)$ that satisfy (\ref{bcv}) to find the maximum possible value of $\Nu$ in (\ref{Nu}), noting that $\vel$, (\ref{ad}) and (\ref{bct}) uniquely determine a steady-state $\theta(x,z)$.
This maximum is denoted
\begin{eqnarray}
\Nu_{\mathrm{max}}(\Pe,\varGamma) \equiv \underset{\vel}{\mathrm{sup}} \{\Nu(\vel)\}. \label{sup1}
\end{eqnarray} 
\item{2)} For the same $\Pe$, step 1 may be repeated for various values of $\varGamma$.
For each value of $\Pe$ the largest value of $\Nu_{\mathrm{max}}(\Pe,\varGamma)$ will be called
\begin{eqnarray}
\Nu_{\mathrm{MAX}}(\Pe) \equiv \underset{\varGamma}{\mathrm{sup}} \{\Nu_{\mathrm{max}}(\Pe,\varGamma)\}. \label{sup2}
\end{eqnarray}   
\item{3)} And for each value of $\Pe$, the $\varGamma$ from step 2 that realizes $\Nu_{\mathrm{MAX}}$ is dubbed the optimal aspect ratio and denoted $\varGamma_{\mathrm{opt}}(\Pe)$.
That is, $\Nu_{\mathrm{MAX}}(\Pe) = \Nu_{\mathrm{max}}(\Pe,\varGamma_{\mathrm{opt}}(\Pe)).$
\end{description}
\smallskip

The time (in)dependence of the flow merits further discussion.
The effect of unsteadiness on optimal transport is not fully understood and whether a time-dependent flow can transport more than a steady flow (with the same amount of energy or enstrophy) remains an open question.
Of course the question can be answered by performing the optimization in step 1 over both space and time, i.e., by considering $\vel = \vel(x,z,t)$.
Such an analysis is a problem of optimal control theory and is left for the future.
Here we focus on steady flows (i.e., $\vel = \vel(x,z)$) and use calculus of variations to carry out step 1.

\section[Fixed Energy]{Optimal Transport with Fixed Energy} \label{sec:problem1}

The steady-state fixed energy system, equations~(\ref{ad})--(\ref{bcv}) and (\ref{ke}), is
\begin{eqnarray}
\vel \bcdot \bnabla \theta &=& \Delta \theta + w, \label{adS1} \\
\bnabla \bcdot \vel &=& 0, \label{divS1} \\
\Pe^2 &=& \left< |\vel|^2\right>, \label{keS1} \\
\theta(x,0) &=& \theta(x,1) = 0, \label{bctS1} \\
w(x,0) &=& w(x,1) = 0. \label{bcvS1}
\end{eqnarray}   
Before solving this system we note that a simple analysis yields a rigorous {\it a priori} upper bound on $\Nu$.
Starting from (\ref{Nu}), recalling the Cauchy--Schwarz inequality and remembering that the maximum principle assures $0 \le T \le 1$ so $|T-1/2| \le 1/2$,
\begin{eqnarray}
\Nu &=& 1+\left< wT \right> = 1+\left< w(T-1/2) \right> \nonumber \\
&\leq& 1 + \left<  w^2 \right>^{1/2} \left< (T-1/2)^2 \right>^{1/2} \nonumber \\
&\leq& 1 + \frac{\left< |\vel|^2 \right>^{1/2}}{2} = 1 + \frac{\Pe}{2}. \label{NuPe1}
\end{eqnarray}

\subsection{Variational Formulation for Steady Flows} \label{sec:var} 

The variational problem to maximize $\Nu=1+\left<\theta w\right>$ given constraints (\ref{adS1})--(\ref{keS1}) and boundary conditions (\ref{bctS1})--(\ref{bcvS1}) is to extremize the functional
\begin{eqnarray}
\mathcal{F} = \left< w \theta  - \phi(x,z) \left(\vel \bcdot \bnabla \theta - \Delta \theta - w \right) + p(x,z) \left(\bnabla \bcdot \vel \right) - \frac{\mu}{2} \left(|\vel|^2-\Pe^2 \right) \right> 
\label{var}
\end{eqnarray} 
where $\phi(x,z)$, $p(x,z)$, and $\mu$ are Lagrange multipliers; $\phi$ and $p$ are functions of $x$ and $z$ to enforce the differential constraints (\ref{adS1}) and (\ref{divS1}) point wise in space while $\mu$ is a real number.
The Euler-Lagrange equations are
\begin{eqnarray}
0 = \frac{\delta \mathcal{F}}{\delta \vel} &=& \left( \theta + \phi \right) \zhat + \theta \, \bnabla \phi - \bnabla p - \mu \vel,   \label{du} \\
0 = \frac{\delta \mathcal{F}}{\delta \theta} &=& \vel \bcdot \bnabla \phi  + \Delta \phi + w,   \label{dt} \\
0 = \frac{\delta \mathcal{F}}{\delta \phi} &=& -\vel \bcdot \bnabla \theta + \Delta \theta + w,  \label{dphi}\\
0 = \frac{\delta \mathcal{F}}{\delta p} &=& \bnabla \bcdot \vel, \label{div2} \\
0 = \frac{\partial \mathcal{F}}{\partial \mu} &=&\frac{1}{2}\left(\Pe ^2 - \left<|\vel|^2 \right>\right). 
\label{dmu}
\end{eqnarray}
In deriving (\ref{dt}) the natural boundary conditions that $\phi$ vanishes at $z=0$ and $z=1$ were employed to eliminate a surface term arising from the integration by parts of $\phi \Delta \theta$.
Therefore, the boundary conditions accompanying these partial differential equations are
\begin{eqnarray}
w(x,0)=w(x,1)&=& 0, \label{bc1} \\
\theta(x,0)=\theta(x,1)&=& 0, \label{bc2} \\
\phi(x,0)=\phi(x,1)&=& 0. \label{bc3}
\end{eqnarray}
Note that the $+ \theta \, \bnabla \phi$ term is equivalent to $- \phi \, \bnabla \theta$ in (\ref{du}), the difference being a perfect gradient that can be absorbed into $\bnabla p$.

Equations~(\ref{du}) and (\ref{dphi})--(\ref{div2}) exhibit certain similarities with those governing steady Rayleigh--B\'enard convection in a fluid saturated porous layer at infinite Prandtl--Darcy number \citep[see, e.g.,][]{Doering98}.
But here an extra scalar, the Lagrange multiplier or ``adjoint'' field $\phi$, enters the problem.
This resemblance, which will also be observed in the linear analysis in the next section, will be discussed further in \S\ref{sec:pm}.        

\subsection{The Limit of Small $\Pe$: Asymptotic Solution} \label{sec:smallPe}
In the limit of small $\Pe$ the flow field $|\vel| \ll 1$ (at least in the $L^2$ sense) which, along with (\ref{dt})--(\ref{dphi}) and the maximum principle, implies that $|\theta| \ll 1$ and $|\phi| \ll 1$.
Therefore, in this limit we can linearize equations (\ref{du})--(\ref{dphi}) to obtain the system
\begin{eqnarray}
\mu \, \vel + \bnabla p &=& (\theta +\phi) \zhat, \label{l1} \\
\Delta \phi + w &=& 0, \label{l2} \\
\Delta \theta + w &=& 0, \label{l3}\\
\bnabla \bcdot \vel &=&  0. \label{l4}
\end{eqnarray}
Subtracting (\ref{l3}) from (\ref{l2}) and using (\ref{bc2})--(\ref{bc3}) then yields $\theta=\phi$.
Taking the $z$-component of the double-curl of (\ref{l1}) and using (\ref{l4}) we deduce
\begin{equation}
\mu \Delta w = 2 \, \theta_{xx} \label{aux}
\end{equation}     
where, as usual, the $x$ subscript indicates the partial derivative.

Equations~(\ref{l3})--(\ref{aux}) and boundary conditions~(\ref{bc1})--(\ref{bc2}) can be solved analytically in this small--$\Pe$ limit.
Indeed, Fourier transforming in the $x$ direction, \eqref{l3} and \eqref{aux} become
\begin{eqnarray}
(\mathrm{D}_z^2-k^2) \, \hat{\theta}_k(z) + \hat{w}_k(z) &=& 0, \label{eq1} \\
\mu (\mathrm{D}_z^2-k^2) \, \hat{w}_k(z) + 2 k^2 \, \hat{\theta}_k(z) &=& 0, \label{eq2} 
\end{eqnarray}
where $\mathrm{D}_z=d/d z$ and $\hat{w}_k(z)$ and $\hat{\theta}_k(z)$ are the Fourier coefficients of $w$ and $\theta$ with horizontal (dimensionless) wavenumber $k \equiv \upi / \varGamma$.
The solution is
\begin{eqnarray}
\hat{w}_k(z) &=& A_k \, \sin{(m \upi z)} \label{wk}, \\
\hat{\theta}_k(z) &=& B_k \, \sin{(m \upi z)} \label{tk},
\end{eqnarray}     
where $m$ is the vertical (dimensionless) wavenumber, and $A_k$ and $B_k$ are still undetermined. Substituting into (\ref{eq1}) and (\ref{eq2}) gives
\begin{eqnarray}
\mu &=& (2 \, k^2)/(m^2\upi^2+k^2)^2, \label{mu} \\
A_k &=& (m^2 \upi^2+k^2) \, B_k. \label{AB}
\end{eqnarray}
Using equation~(\ref{l4}), 
\begin{eqnarray}
\hat{u}_k(z) &=& \mathrm{i} \, \frac{m \upi}{k} \, A_k \, \cos{(m \upi z)} \label{uk}.
\end{eqnarray} 
Substituting (\ref{uk}) and (\ref{wk}) into (\ref{dmu}) yields
\begin{equation}
\left< |\vel|^2 \right> = \left(A^2_k+ \frac{m^2\upi^2}{k^2} \, A^2_k \right) = \Pe^2 \Rightarrow A_k = \frac{\, k}{(m^2\upi^2+k^2)^{1/2}} \, \Pe \label{Ak}
\end{equation} 
which, along with (\ref{AB}), gives
\begin{eqnarray}
B_k = \frac{\, k}{(m^2\upi^2+k^2)^{3/2}} \, \Pe.
\label{Bk}
\end{eqnarray}
Knowing $A_k$ and $B_k$, $\Nu$ is obtained from (\ref{Nu}):
\begin{eqnarray}
\Nu = 1+ A_k B_k = 1 + \frac{k^2}{(m^2\upi^2+k^2)^{2}} \, \Pe^2. \label{smallPeB}
\end{eqnarray}
Then for a given parameter set $(\Pe, \varGamma=\upi/k)$, $\Nu$ is maximized at $m=1$.
As a result, using the notation defined in \S\ref{sec:goal}, 
\begin{eqnarray}
\Nu_\m(\Pe,\varGamma) = 1 + \frac{\varGamma^2}{\upi^2(\varGamma^2+1)^{2}} \, \Pe^2. \label{smallPe}
\end{eqnarray}
The largest value of $\Nu_\m(\Pe,\varGamma)$, i.e., $\Nu_\M$, is achieved at $\varGamma_{\mathrm{opt}}=1$: 
\begin{eqnarray}
\Nu_\M(\Pe)=1+\frac{\Pe^2}{4\upi^2}. \label{smallPemax}
\end{eqnarray} 
Note that $(k,m)=(\upi,1)$ corresponds to the maximum value of $\mu=1/(2\upi)^2$ (see (\ref{mu})).

\begin{figure}
\centering
  \subfloat[]{\label{fig:rolls1}\includegraphics[width=.75\textwidth]{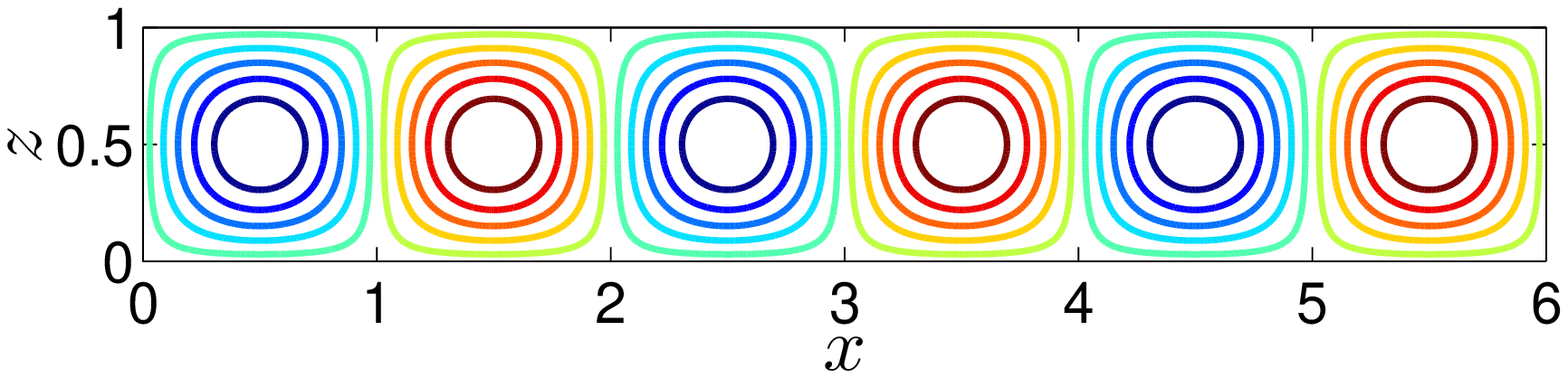}}  
\\           
  \subfloat[]{\label{fig:rolls2}\includegraphics[width=.75\textwidth]{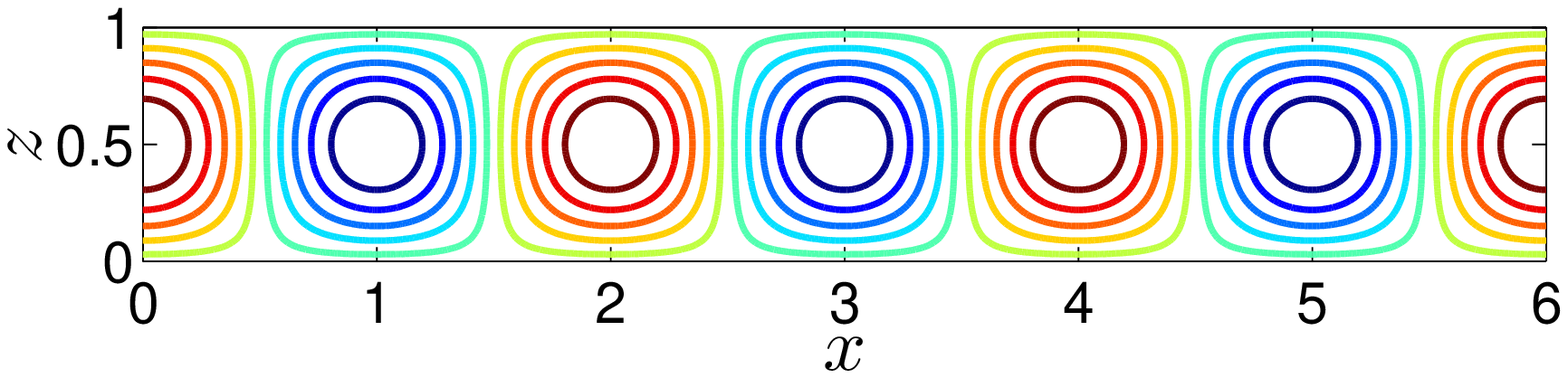}}               
  \caption{Optimal flow field in the small--$\Pe$ limit for the fixed energy problem: (a)~streamlines $\psi$, (b)~temperature $\theta$.}
  \label{fig:rolls}
\end{figure}

Therefore, in the limit of small $\Pe$ (i.e., large $\mu$) the maximum transport is realized by an array of square \textit{convection} cells (rolls) with optimal aspect ratio $\varGamma_{\mathrm{opt}}=1$.
Figure~\ref{fig:rolls} shows this flow field where the square convection cells are clearly seen. This solution is identical to the first unstable mode at onset for Rayleigh--B\'enard convection in a fluid saturated porous media \citep{Doering98}.
Additionally, we note (and discuss further in \S\ref{sec:pm}) that the factor $4\upi^2$ arising in (\ref{smallPemax}) is the critical $\Ra$ for the instability.

\subsection{Small to Large $\Pe$: Numerical Solution}\label{sec:numres}
For $Pe>\mathit{O}(1)$, the full nonlinear system~(\ref{du})--(\ref{div2}) must be solved. 
Defining the stream function $\psi$ by $u=\partial \psi/\partial z$ and $w=-\partial \psi/\partial x$, equations (\ref{du})--(\ref{div2}) reduce to
\begin{eqnarray}
\mathrm{J}(\theta,\phi) + \mu \, \Delta \psi + (\theta+\phi)_x &=& 0, \label{N1} \\
-\mathrm{J}(\psi,\phi) + \Delta \phi - \psi_x &=& 0, \label{N2} \\
-\mathrm{J}(\psi,\theta) - \Delta \theta + \psi_x &=& 0, \label{N3} 
\end{eqnarray}  
where $\mathrm{J}(a,b)=\frac{\partial a}{\partial x} \frac{\partial b}{\partial z} - \frac{\partial a}{\partial z} \frac{\partial b}{\partial x}$.
The boundary conditions (\ref{bc1})--(\ref{bc3}) become
\begin{eqnarray}
\psi(x,0) = \psi(x,1) &=& 0, \label{R1}\\ 
\theta(x,0) = \theta(x,1) &=& 0, \label{R2}\\
\phi(x,0) = \phi(x,1) &=& 0. \label{R3}
\end{eqnarray}
These equations and boundary conditions imply an interesting symmetry between $\theta$ and $\phi$ that will be exposed in the numerical results and exploited later to obtain
asymptotic solutions for large $\Pe$. 

\subsubsection{Numerical Continuation} \label{sec:cont}
Numerical continuation is a strategy to systematically trace a branch of solutions starting from an initial iterate \citep[][Appx. D]{Boyd}.
In the problem considered here the solutions are known analytically in the limit of small $\Pe$ (i.e., large $\mu$) 
for given values of $\varGamma$ (\S\ref{sec:smallPe}). These solutions provide suitable initial guesses for solutions at larger values of $\Pe$ (i.e., smaller $\mu$),
which are computed numerically according to the following continuation algorithm:
\medskip
\begin{description}
\item{1)} Start from the analytical solution for large $\mu$ for a given value of $\varGamma$.
\item{2)} Incrementally reduce $\mu$ such that, at iteration $N+1$, $\mu^{N+1}$ is set to be $0.1\%$--$5\%$ smaller than $\mu^{N}$. Use the solution at iteration $N$ (with $\mu^N$) as the initial guess to iteratively compute the new solution at iteration $N+1$ (with $\mu^{N+1}$).
\item{3)} Using the converged solution from step~2, calculate $\Pe(\mu^{N+1},\varGamma)$ and $\Nu_\m(\mu^{N+1},\varGamma)$ from
the relations
\begin{eqnarray}
\Pe^2 &=& \left< \psi^2_x\ + \psi^2_z \right>, \label{PepsiS1} \\ 
\Nu_\m &=& 1 - \left< \psi_x \theta \right>. \label{NupsiS1} 
\end{eqnarray}
\item{4)} Repeat steps~2 and 3 reducing $\mu$ (and increasing $\Pe$) by several orders of magnitude.  
\item{5)} Repeat steps 1--4 for a variety of values of $\varGamma$.
\end{description}  
\medskip
This procedure produces $\Nu_\m(\Pe,\varGamma)$ for a wide range of $\Pe$ and $\varGamma$.
Note that in step~1 the vertical wavenumber $m$ must be chosen for the linear solution.
Empirically, we find that $\Nu_\M$ always emerges from solutions continued from linear solutions with $m=1$.
Therefore, for most cases we set $m=1$, although cases with $m=2$ and linearly superposed solutions with different values of $m$ were also studied (see \S\ref{sec:numresults}).
The percentage reduction of $\mu$ in step~2 depends on the degree of nonlinearity of the problem: not unexpectedly, $\mu$ must be varied more slowly as $\Pe$ increases.

\subsubsection{Numerical Method}\label{sec:nummethod}
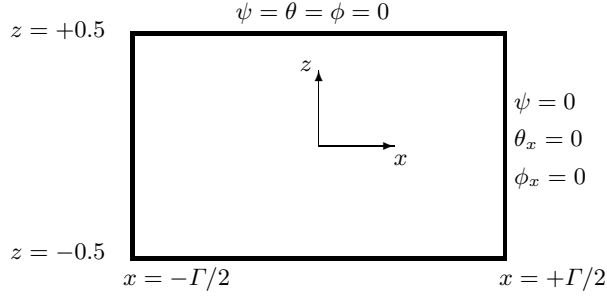
\begin{figure}
\setlength{\unitlength}{1cm}
\begin{picture}(8,5)
  \linethickness{0.5mm}
  \put(5      ,3.5){\line(1,0){5}}  
  \put(5      ,0.5){\line(1,0){5}}
  \put(5.025  ,0.5){\line(0,1){3}}  
  \put(9.975  ,0.5){\line(0,1){3}}  
  \put(3.4, 0.4975){$z=-0.5$} 
  \put(3.4, 3.45){$z=+0.5$} 
  \put(4.9, 0.15){$x=-\varGamma/2$} 
  \put(9.9, 0.15){$x=+\varGamma/2$} 
  \put(6.4,  3.7){$\psi=\theta=\phi=0$} 
  \put(10.1,    2.5){$\psi = 0$} 
  \put(10.1,    2.0){$\theta_x =0$}  
  \put(10.1,    1.5){$\phi_x =0$} 
  \linethickness{0.05mm} 
  \put(7.5, 2){\vector(0, 1){1}}
  \put(7.5, 2){\vector(1, 0){1}}
  \put(7.25, 3){$z$} 
  \put(8.5, 1.75){$x$}
\end{picture}
\caption{The fixed energy problem: the geometry and boundary conditions of the computational domain corresponding to a single two-dimensional (convection) cell. Boundary conditions on the bottom (left) boundary are the same as the top (right) boundary.}
\label{fig:schem2}
\end{figure}

In this section we present details of the iterative numerical method used in step~2 of the continuation algorithm.  Specifically,     
we use a Newton--Kantorovich iteration scheme \citep[][Appx. C]{Boyd} with a pseudospectral Chebyshev collocation method \citep{Trefethen,Boyd} to solve (\ref{N1})--(\ref{N3}).
Instead of solving these equations in a large horizontally periodic domain including multiple cells (e.g. as shown in figure~\ref{fig:rolls1}), 
we choose the computational domain to include only a single cell (figure~\ref{fig:schem2}).
This unicellular approach has been successfully used to study other, related problems, including Rayleigh--B\'enard convection in pure fluids \citep{Greg} and in porous media \citep{Lind}.
In this configuration symmetry boundary conditions are imposed on the vertical sides of the domain at $x=\pm \varGamma/2$.
Note that in the computational domain the horizontal walls are located at $z=\pm 0.5$ (rather than at $z=0,1$, as before) for convenience since Chebyshev polynomials are employed.       

The Newton--Kantorovich method used in step~2 of the continuation algorithm (\S\ref{sec:cont}) is implemented as follows.
The known solution at the $N$th iterate $(\psi^N,\theta^N,\phi^N)$ is used as a first guess to iteratively approximate the true solution at the ($N+1$)th iteration $(\psi^{N+1},\theta^{N+1},\phi^{N+1})$. Taylor expanding (\ref{N1})--(\ref{N3}) about $(\psi^{N},\theta^{N},\phi^{N})$, defining ``$\udelta$'' of any quantity as the difference between its value at iterations $N+1$ and $N$, and neglecting higher-order terms,  we obtain the system of three linear differential equations (\ref{e1})--(\ref{e3}) given in Appendix~\ref{app:NK}. Applying a pseudospectral Chebyshev collocation method in both the $x$ and $z$ directions yields the following linear matrix equation:
\begin{eqnarray}\nonumber
 \begin{bmatrix}
  \mu \mathsfbi{\Delta}  & (\mathsfbi{I}+\boldsymbol{\phi}^N_z)\mathsfbi{D}_x - \boldsymbol{\phi}^N_x \mathsfbi{D}_z & (\mathsfbi{I}-\boldsymbol{\theta}^N_z)\mathsfbi{D}_x + \boldsymbol{\theta}^N_x \mathsfbi{D}_z \\
 -(\mathsfbi{I}-\boldsymbol{\theta}^N_z)\mathsfbi{D}_x - \boldsymbol{\theta}^N_x \mathsfbi{D}_z & \mathsfbi{\Delta} - \boldsymbol{\psi}^N_z \mathsfbi{D}_x + \boldsymbol{\psi}^N_x \mathsfbi{D}_z & \mathsfbi{O} \\
 -(\mathsfbi{I}+\boldsymbol{\phi}^N_z)\mathsfbi{D}_x + \boldsymbol{\phi}^N_x \mathsfbi{D}_z & \mathsfbi{O} & \mathsfbi{\Delta} + \boldsymbol{\psi}^N_z \mathsfbi{D}_x - \boldsymbol{\psi}^N_x \mathsfbi{D}_z \\
 \end{bmatrix}
 \begin{bmatrix}
   \udelta \boldsymbol{\psi} \\
  \udelta \boldsymbol{\theta} \\
\udelta \boldsymbol{\phi}
 \end{bmatrix}
&& \\ =
 \begin{bmatrix}
   - \mu \mathsfbi{\Delta} \boldsymbol{\psi}^N -(\mathsfbi{I}+\boldsymbol{\phi}^N_z)\boldsymbol{\theta}^N_x -(\mathsfbi{I}-\boldsymbol{\theta}^N_z)\boldsymbol{\phi}^N_x \\
   -  \mathsfbi{\Delta} \boldsymbol{\theta}^N +(\mathsfbi{I}-\boldsymbol{\theta}^N_z)\boldsymbol{\psi}^N_x + \boldsymbol{\psi}^N_z \boldsymbol{\theta}^N_x\\
   -  \mathsfbi{\Delta} \boldsymbol{\phi}^N +(\mathsfbi{I}+\boldsymbol{\phi}^N_z)\boldsymbol{\psi}^N_x - \boldsymbol{\psi}^N_z \boldsymbol{\phi}^N_x\\
 \end{bmatrix}
&& \label{bigmat},
\end{eqnarray}
where $\mathsfbi{I}$ and $\mathsfbi{O}$ are $M^2  \times M^2$ identity and zero matrices, respectively, and $M$ is the number of collocation grid points.
$\mathsfbi{D}_x$ and $\mathsfbi{D}_z$ are the $x$ and $z$ differentiation matrices, and $\mathsfbi{\Delta} = \mathsfbi{D}_{xx}+\mathsfbi{D}_{zz}$.
These $M^2  \times M^2$ matrices are constructed using tensor products (a.k.a. Kronecker products) as described in detail in \citet{Trefethen}. $\boldsymbol{\psi}$, $\boldsymbol{\theta}$, and $\boldsymbol{\phi}$ are vectors of length $M^2$. The boundary conditions are implemented by modifying the rows corresponding to the boundary grid points in the coefficient matrix and the right-hand side matrix in (\ref{bigmat}).

A MATLAB code was developed to construct the elements of the matrices in (\ref{bigmat}) and to solve the linear algebraic system by direct matrix inversion. Once $\udelta \boldsymbol{\psi}$, $\udelta \boldsymbol{\theta}$, and $\udelta \boldsymbol{\phi}$ are calculated, the solution is updated by setting $(\cdot)^{N+1}=(\cdot)^{N}+\udelta (\cdot)$.
The iterations stop when $\udelta (\cdot) /\|(\cdot)\|_\infty \le 10^{-10}$ for all three variables $\boldsymbol{\psi}$, $\boldsymbol{\theta}$, and $\boldsymbol{\phi}$. Finally, 
Clenshaw--Curtis quadrature \citep{Trefethen} is used for all spatial integrations, for example to calculate $\Pe^2$ from (\ref{PepsiS1}) and $\Nu_\m$ from (\ref{NupsiS1}). 


\subsubsection{Numerical Results}\label{sec:numresults}      
\begin{figure}                   
\centering     \subfloat[]{\label{fig:res0psid}\includegraphics[width=.45\textwidth]{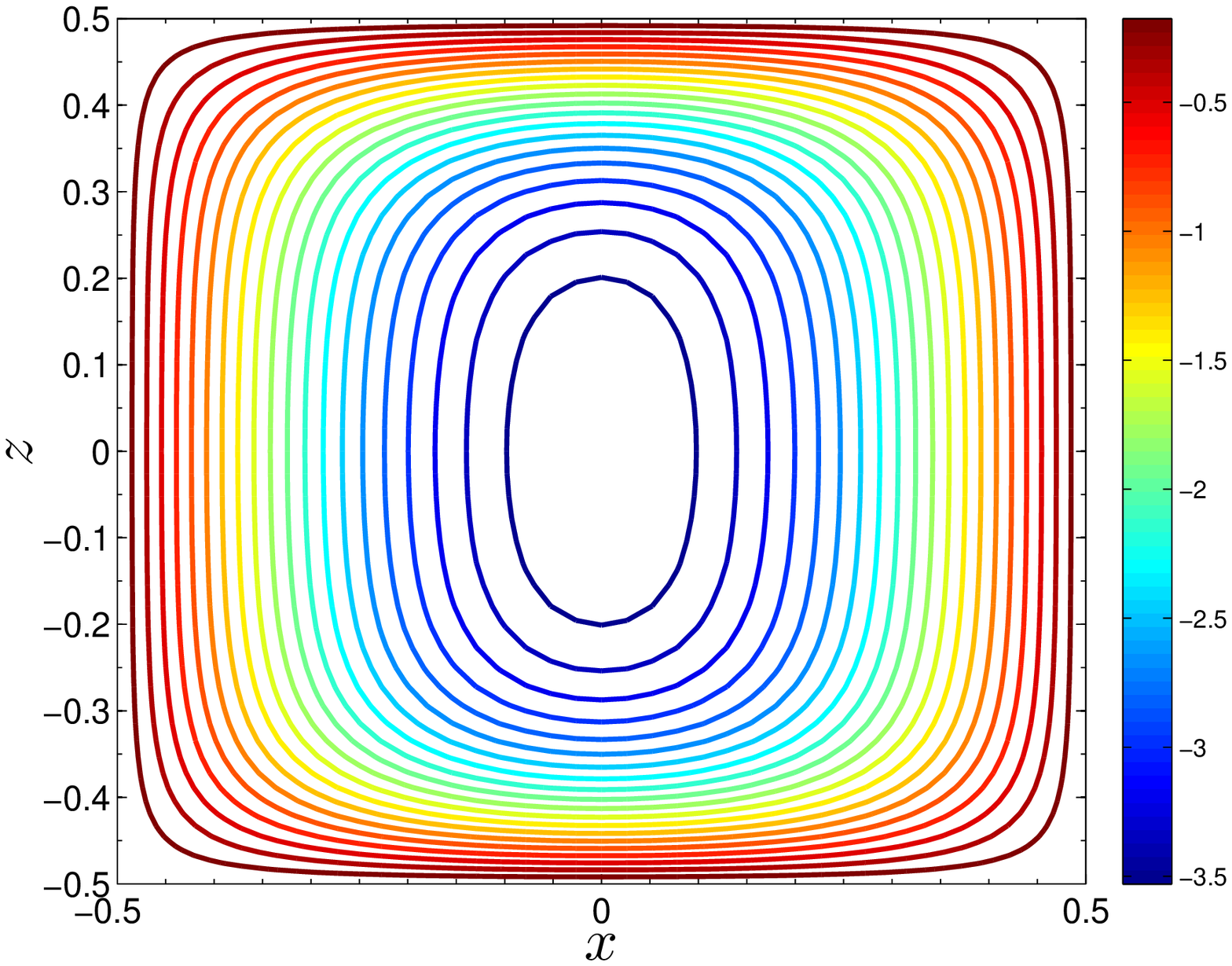}}  
  \subfloat[]{\label{fig:res0thtd}\includegraphics[width=.45\textwidth]{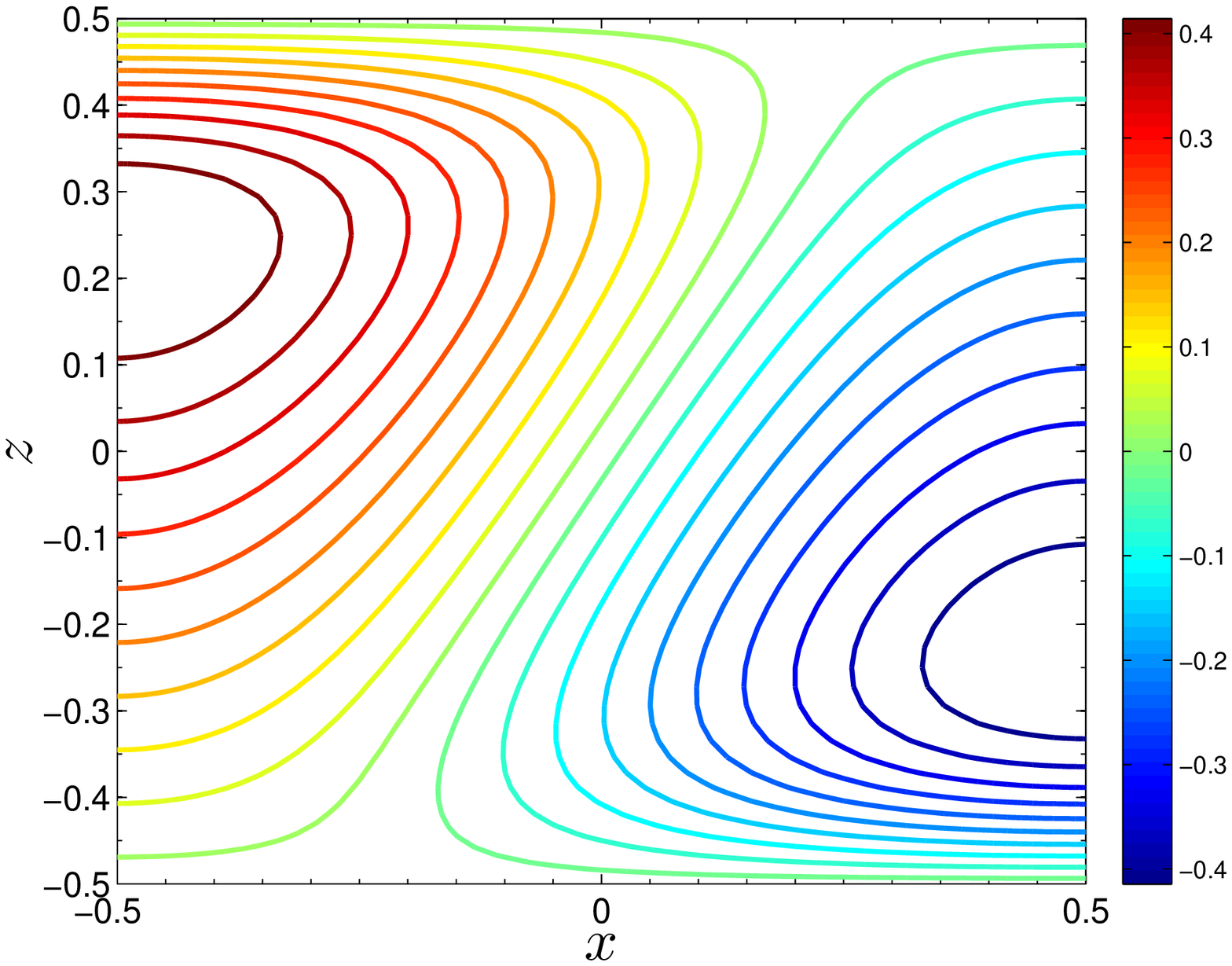}}  
\\           
  \subfloat[]{\label{fig:res0psie}\includegraphics[width=.45\textwidth]{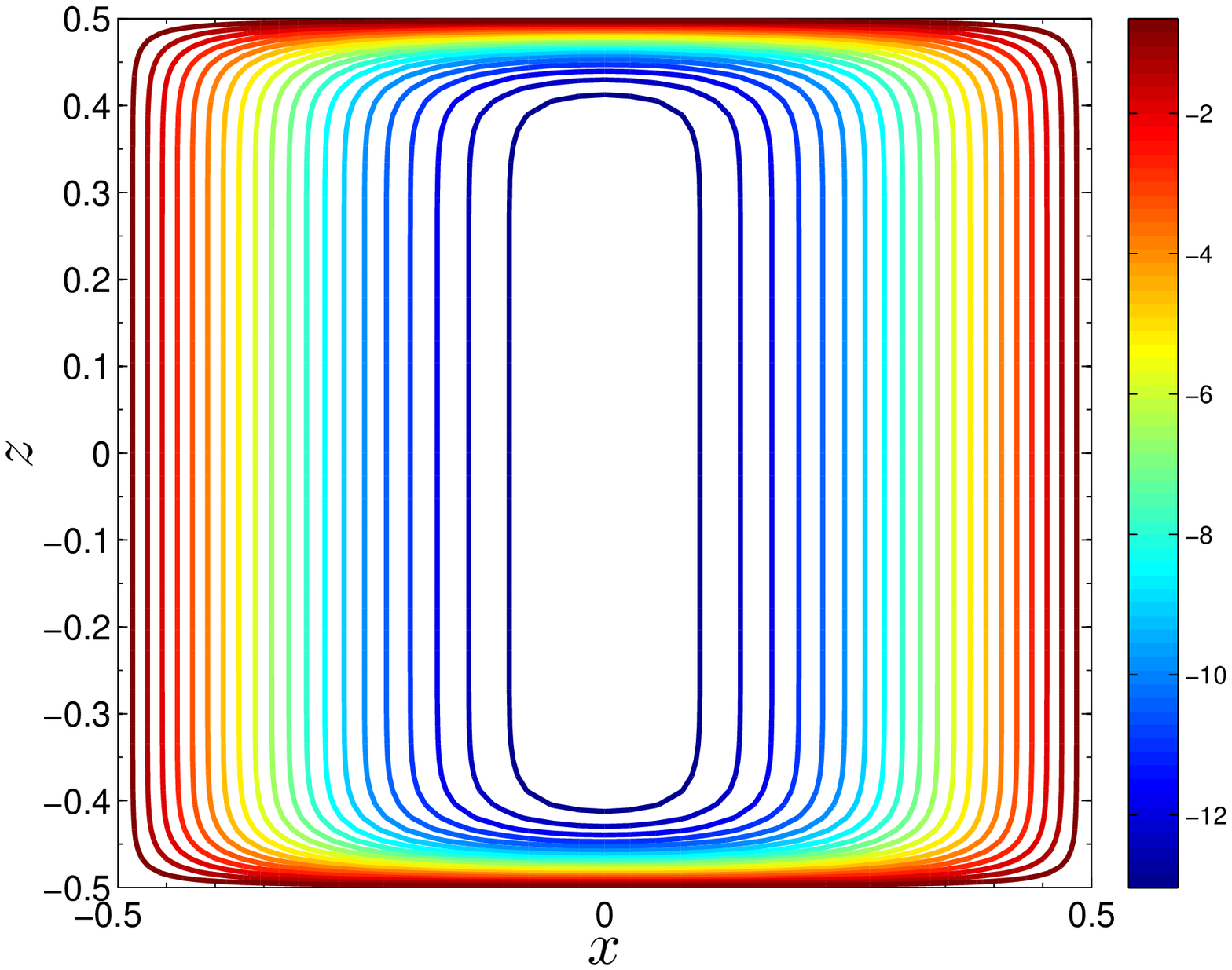}}  
  \subfloat[]{\label{fig:res0thte}\includegraphics[width=.45\textwidth]{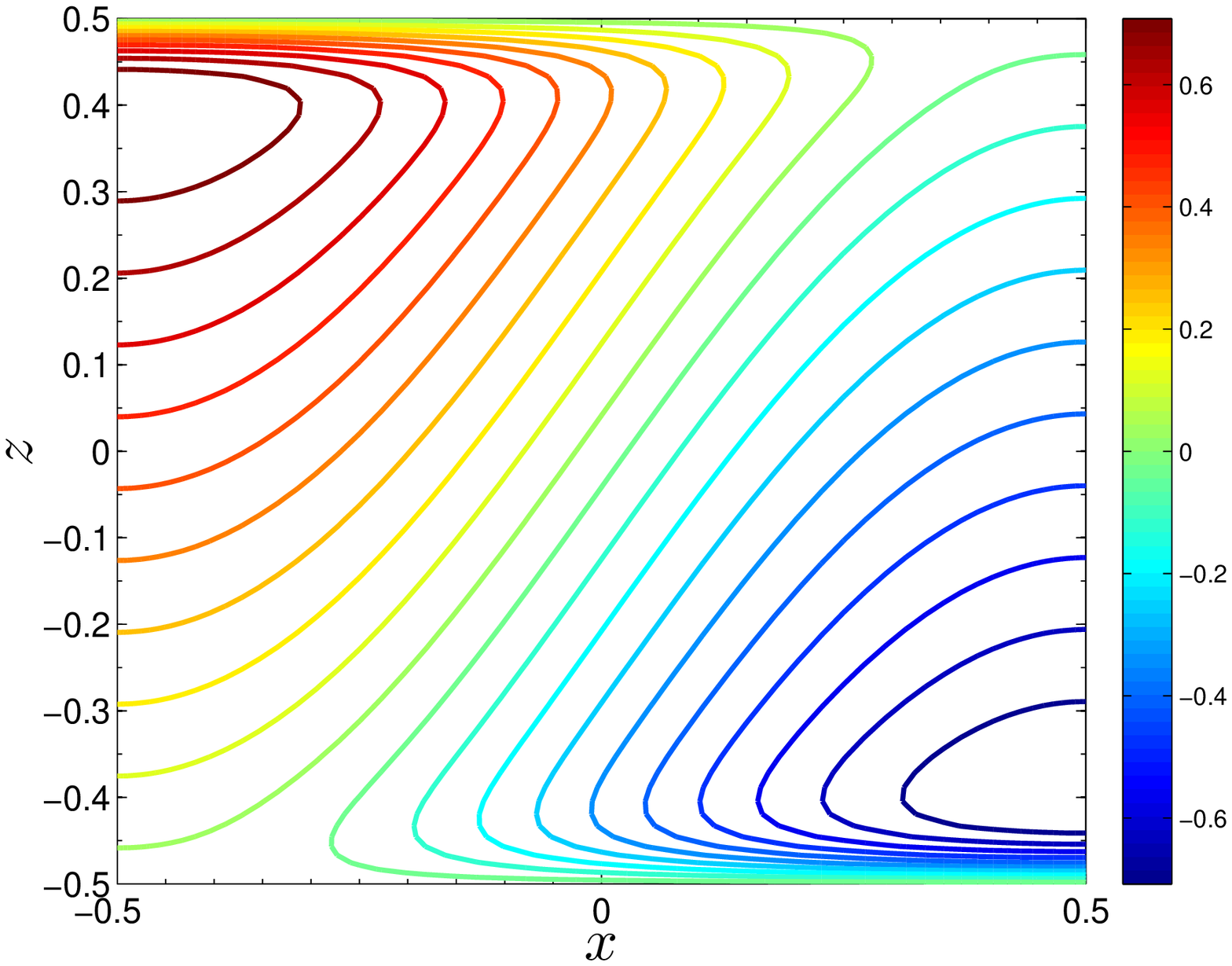}}  
\\           
  \subfloat[]{\label{fig:res0psib}\includegraphics[width=.45\textwidth]{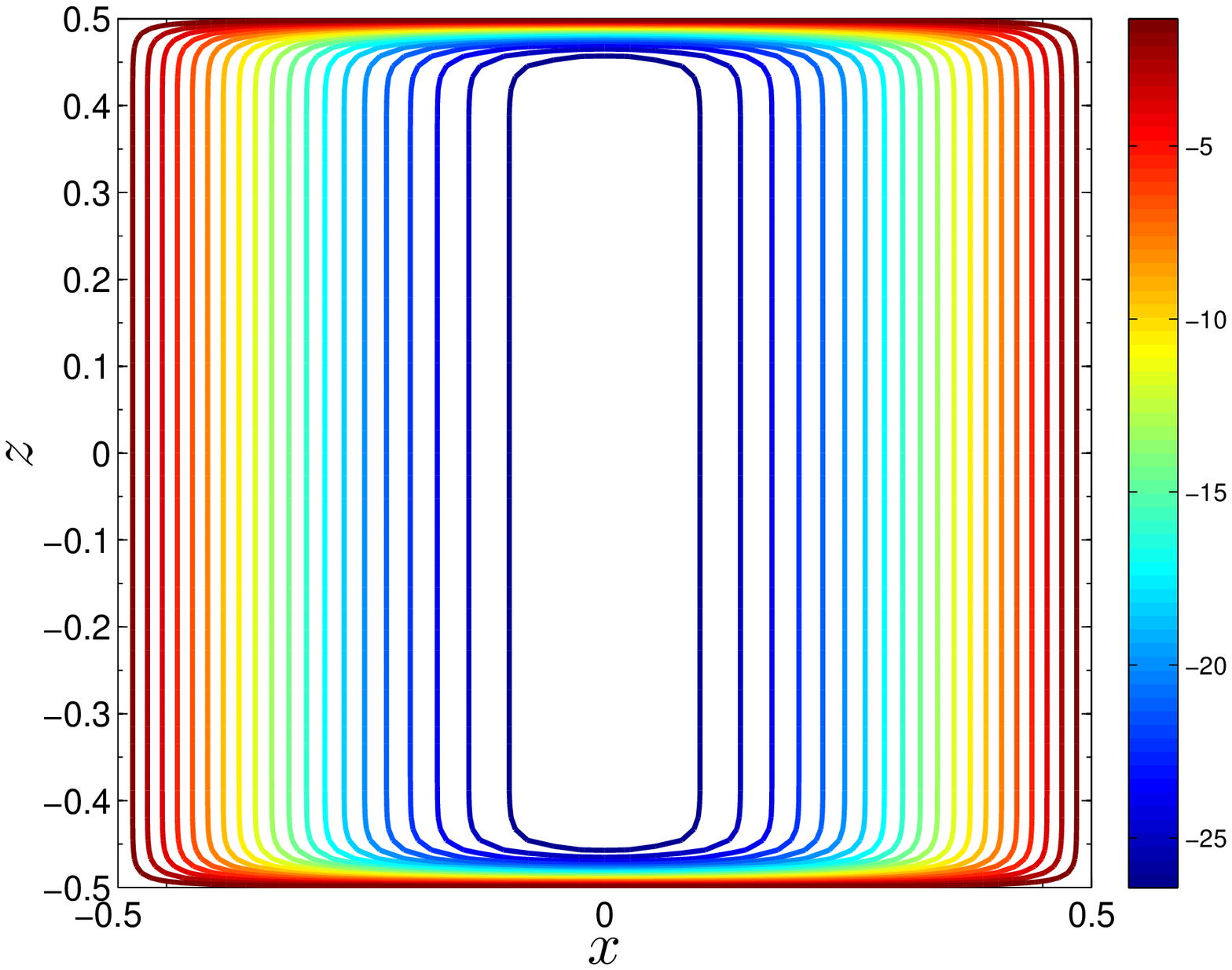}}  
  \subfloat[]{\label{fig:res0thtb}\includegraphics[width=.45\textwidth]{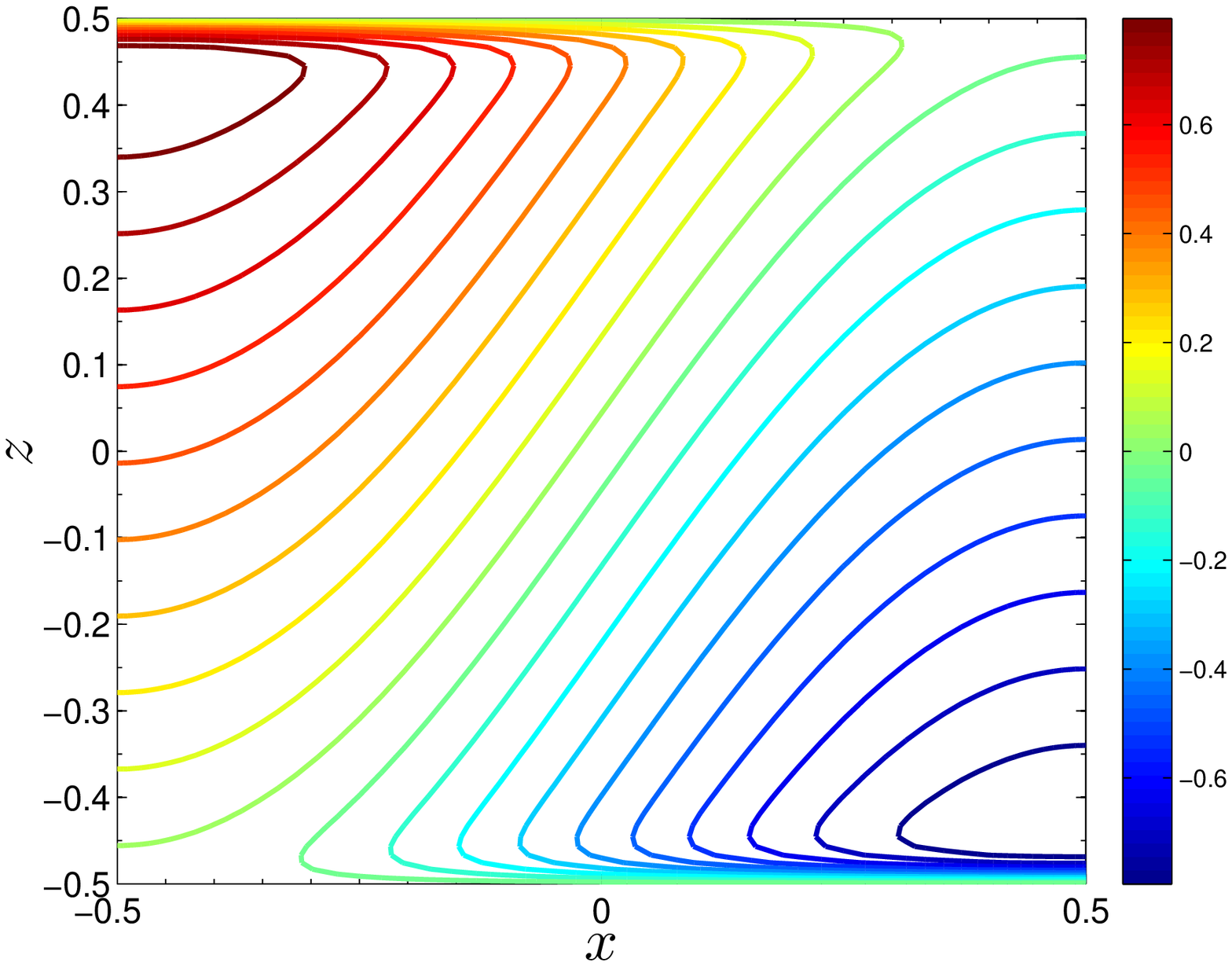}}              
  \caption{Evolution of the flow fields with $\Pe$ for the case with $\varGamma=1$. Panels on the left show $\psi$ and panels on the right show $\theta$. (a) and (b) $\Pe=10.0$, $\Nu_\m=2.4$; (c) and (d) $\Pe=59.4$, $\Nu_\m=9.7$; (e) and (f) $\Pe=161.3$, $\Nu_\m=20.7$. The resolution is $61^2$.}
  \label{fig:res0}
\end{figure}
The results presented here used $M=61$ or $91$. The iterative solution always converged in less than $6$ iterations. Figure \ref{fig:res0} shows $\psi$ and $\theta$ for $\varGamma=1$ and increasing values of $\Pe$.
The flow shown in figures~\ref{fig:res0psid} and \ref{fig:res0thtd} is still in the linear regime.
The bulk flow structure changes and boundary layers emerge in both $\psi$ and $\theta$  as $\Pe$ increases (figures~\ref{fig:res0psie}--\ref{fig:res0thtb}).

\begin{figure}
\centering
  \includegraphics[width=1\textwidth]{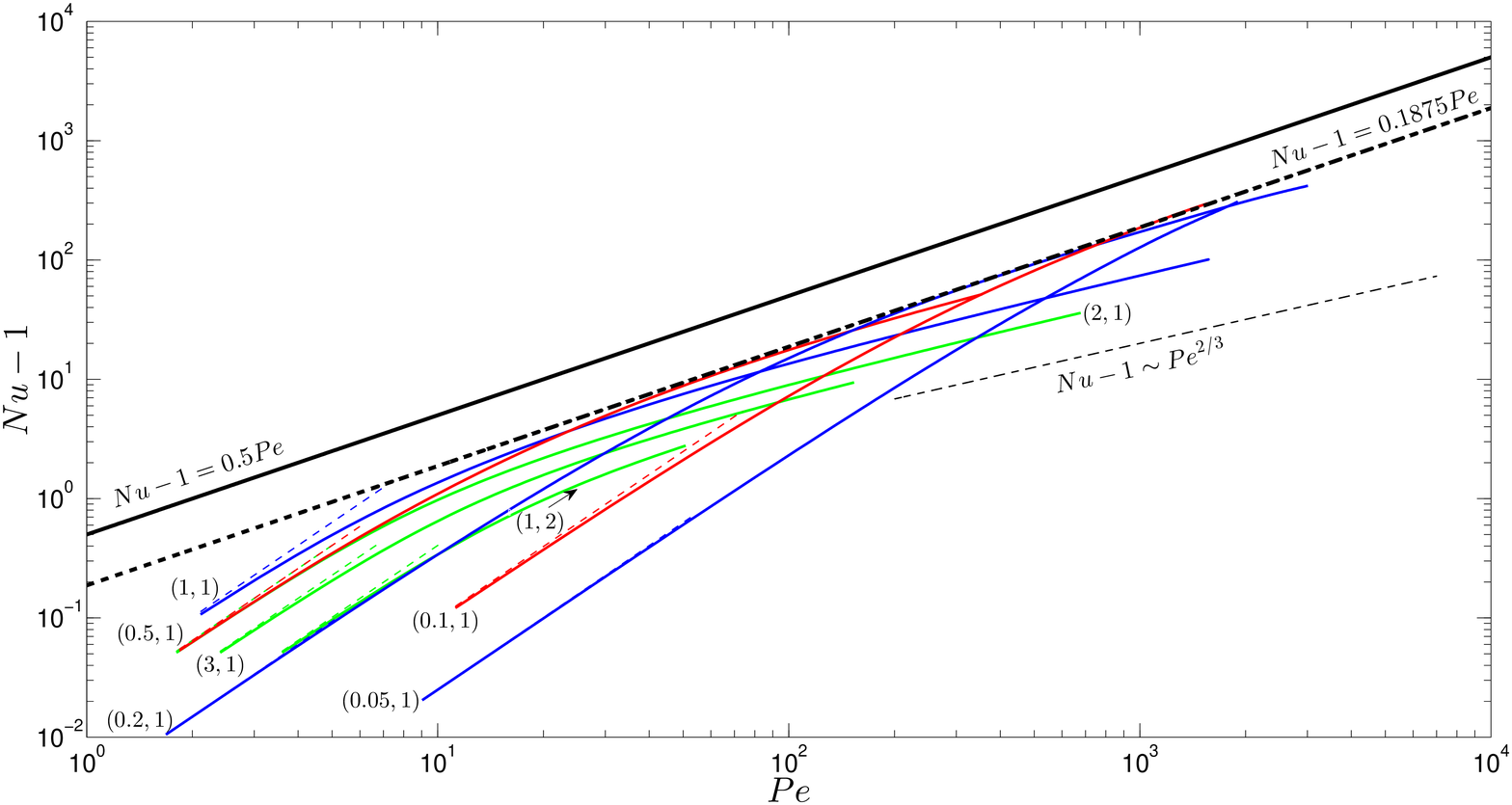}          
  \caption{The numerically obtained $\Nu_\m$ as a function of $\Pe$ for various values of $\varGamma$ (non-black lines). The labels show $(\varGamma,m)$. For each case, the (short) dashed line of the same color, visible for most cases, shows the analytical $\Nu_\m$ given in (\ref{smallPe}) in the small--$\Pe$ limit. The thick black solid line shows the absolute upper bound (\ref{NuPe1}), and the thick black dashed line shows the analytically obtained $\Nu_\M$ given in (\ref{NuPeUlt}), which is derived in the large--$\Pe$ limit. The thin black dashed line indicates the $\Pe^{2/3}$ slope. Cases with $\varGamma>1$ or $m>1$ (green lines) do not produce $\Nu_\M$. All results shown here have resolution $M=61$. Using a higher resolution $M=91$ results in negligible changes to the plot.}
  \label{fig:MapNum}
\end{figure}

Figure \ref{fig:MapNum} is a plot of $\Nu_\m(\Pe,\varGamma)$ for several values of $\varGamma$.
This figure exhibits a number of interesting features.
Firstly, $\Nu_\m$ agrees with (\ref{smallPe}) in the limit of small $\Pe$, which serves to some extent as a benchmark for the code.
Moreover, the absolute upper bound (\ref{NuPe1}) quantitatively overestimates the maximum possible high P\'eclet number heat transport only by an $O(1)$ factor: it captures the correct high P\'eclet number linear scaling.
It also shows that $\Nu_\M(\Pe)$ is obtained from solutions continued from linear solutions with $m=1$.  
This is not unexpected because flows with $m>1$ include horizontal transport in the bulk far from the boundaries, which evidently is not an efficient use of the available energy.\footnote{Indeed, we computed several cases with $m>1$ as well as a few cases continuing from superposed solutions of two $m$ (only one case is shown in this figure) and they all confirmed this conclusion.} 

Figure \ref{fig:MapNum} also suggests that $\Nu_\m$ scales as $K(\varGamma) \, \Pe^{2/3}$ as $\Pe$ increases, and that $\Nu_\M$ is obtained by flows with smaller $\varGamma$ as $\Pe$ increases.
The ensemble of $\Nu_\m$ plotted against $\Pe$ for different values of $\varGamma$ forms an envelope that determines $\Nu_\M$, and the numerical results suggest that 
$\Nu_\M \rightarrow C \, \Pe$ where $C$ is an absolute constant prefactor.   
The prefactors $K(\varGamma)$ and $C$ can, of course, be determined from the numerical results but this proves unnecessary because, guided by the numerics, we 
are able to obtain asymptotic solutions of (\ref{N1})--(\ref{N3}), and hence analytical expressions for $\Nu_\m$, $\Nu_\M$, and $\varGamma_\mathrm{opt}$,
in the limit of large $\Pe$ in \S\ref{sec:largePe}.  (The analytically derived $\Nu_\M$, given in \eqref{NuPeUlt}, is also plotted in figure~\ref{fig:MapNum} and is seen to agree very well with the envelope 
produced by the numerical results.)   

\subsection{The Large--$\Pe$ Asymptotic Solution}\label{sec:largePe}
\begin{figure}
\centering
  \subfloat[]{\label{fig:res1psi}\includegraphics[width=.45\textwidth]{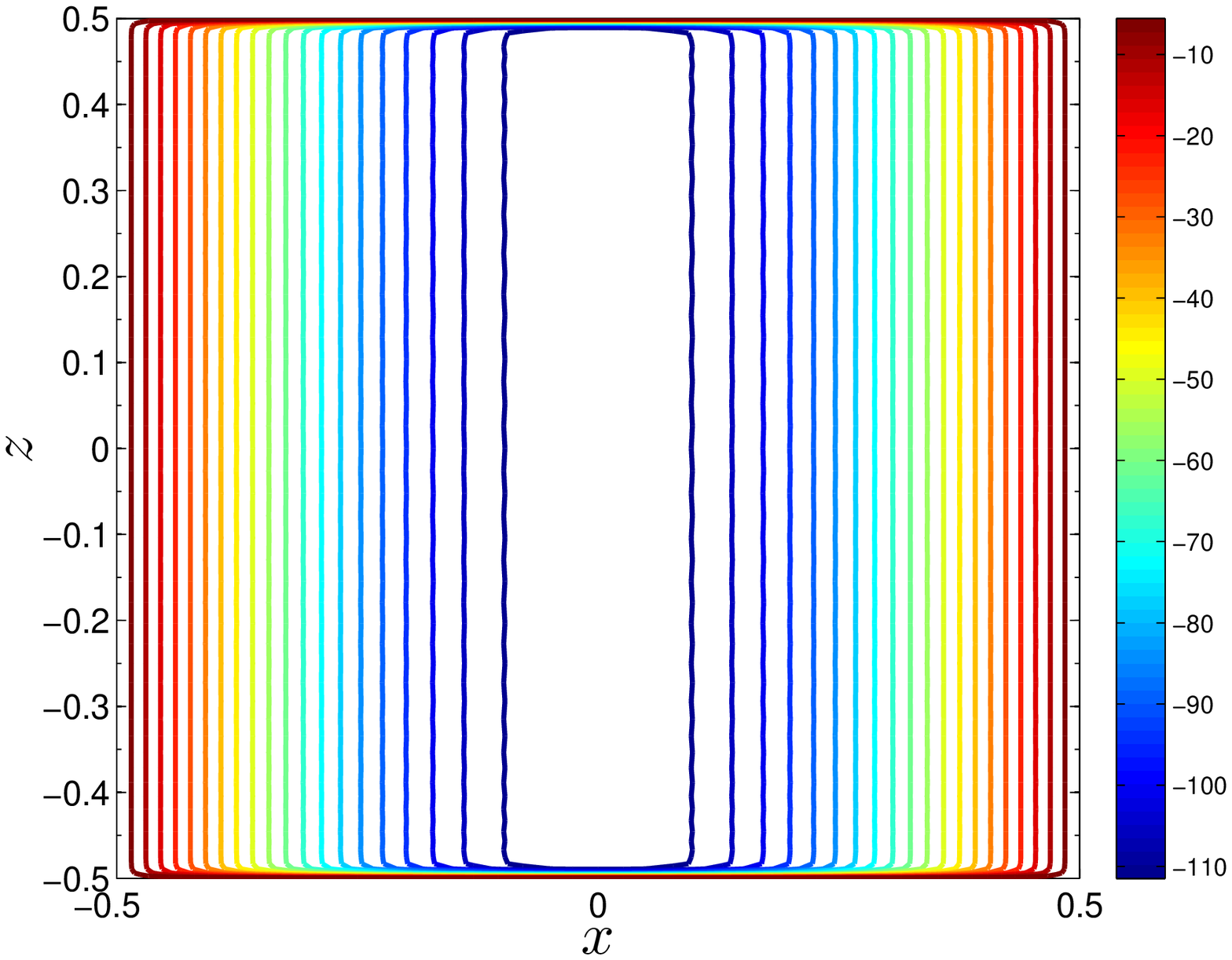}}  
  \subfloat[]{\label{fig:res1psi2}\includegraphics[width=.45\textwidth]{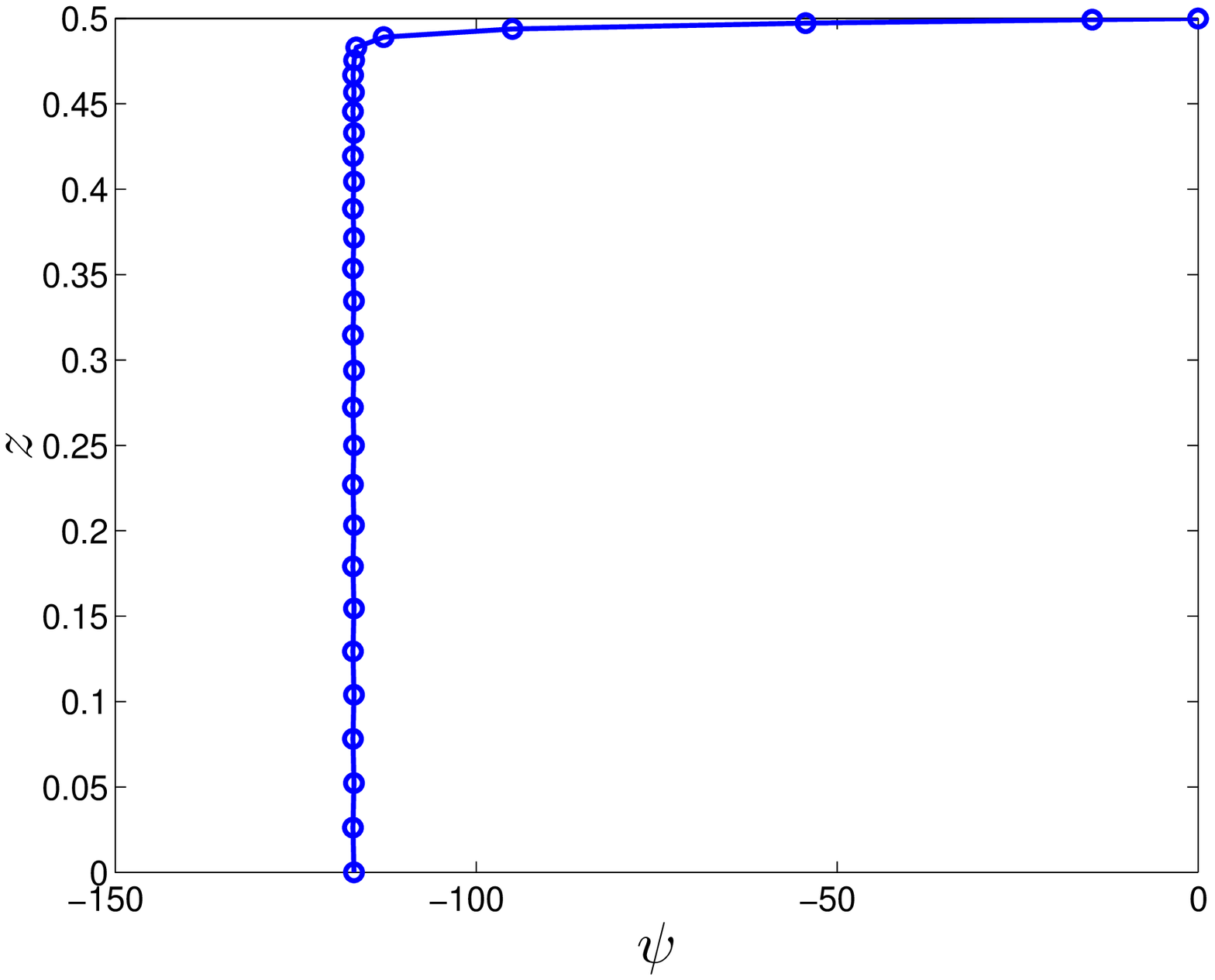}}  
\\           
  \subfloat[]{\label{fig:res1tht}\includegraphics[width=.45\textwidth]{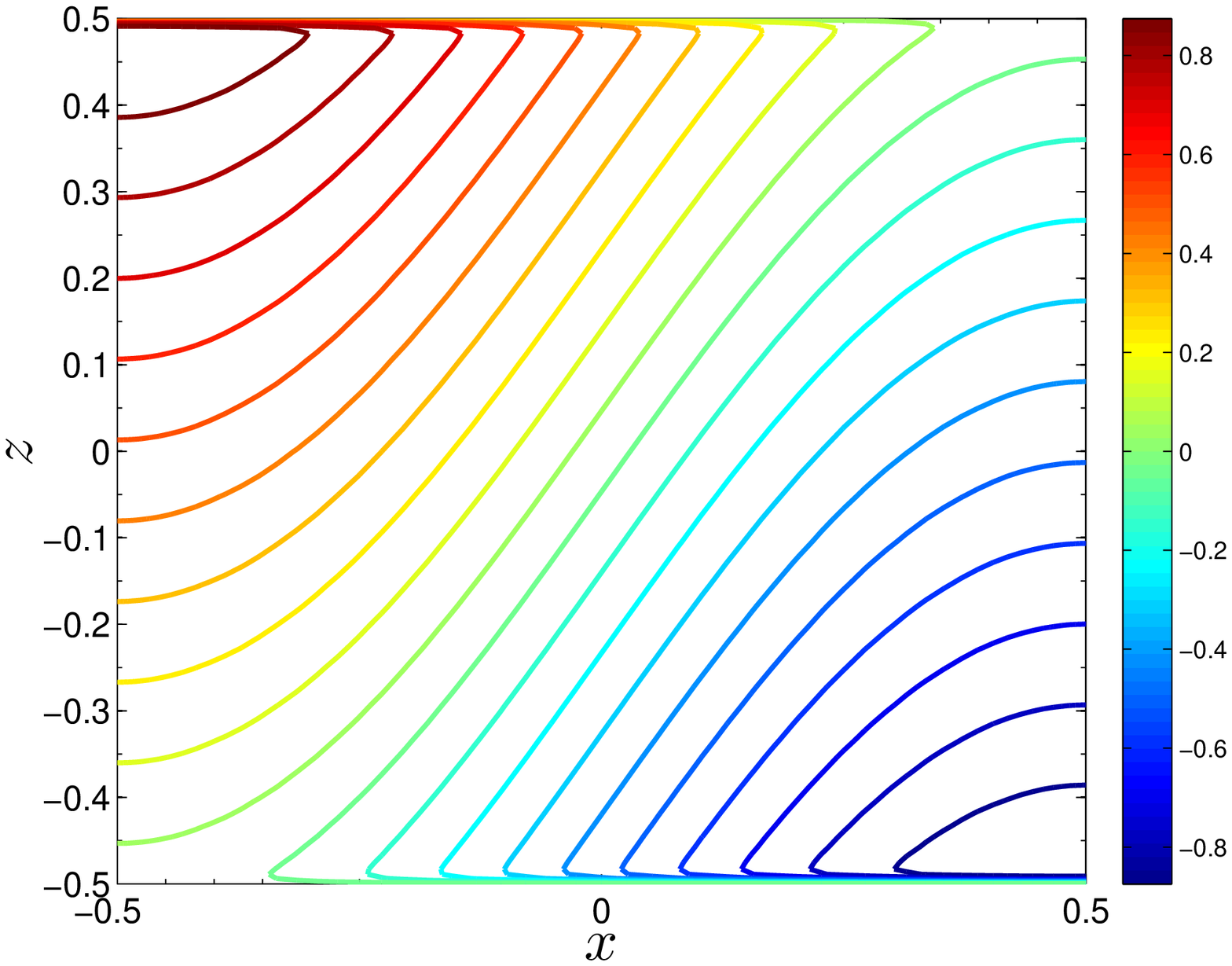}}   
  \subfloat[]{\label{fig:res1phi}\includegraphics[width=.45\textwidth]{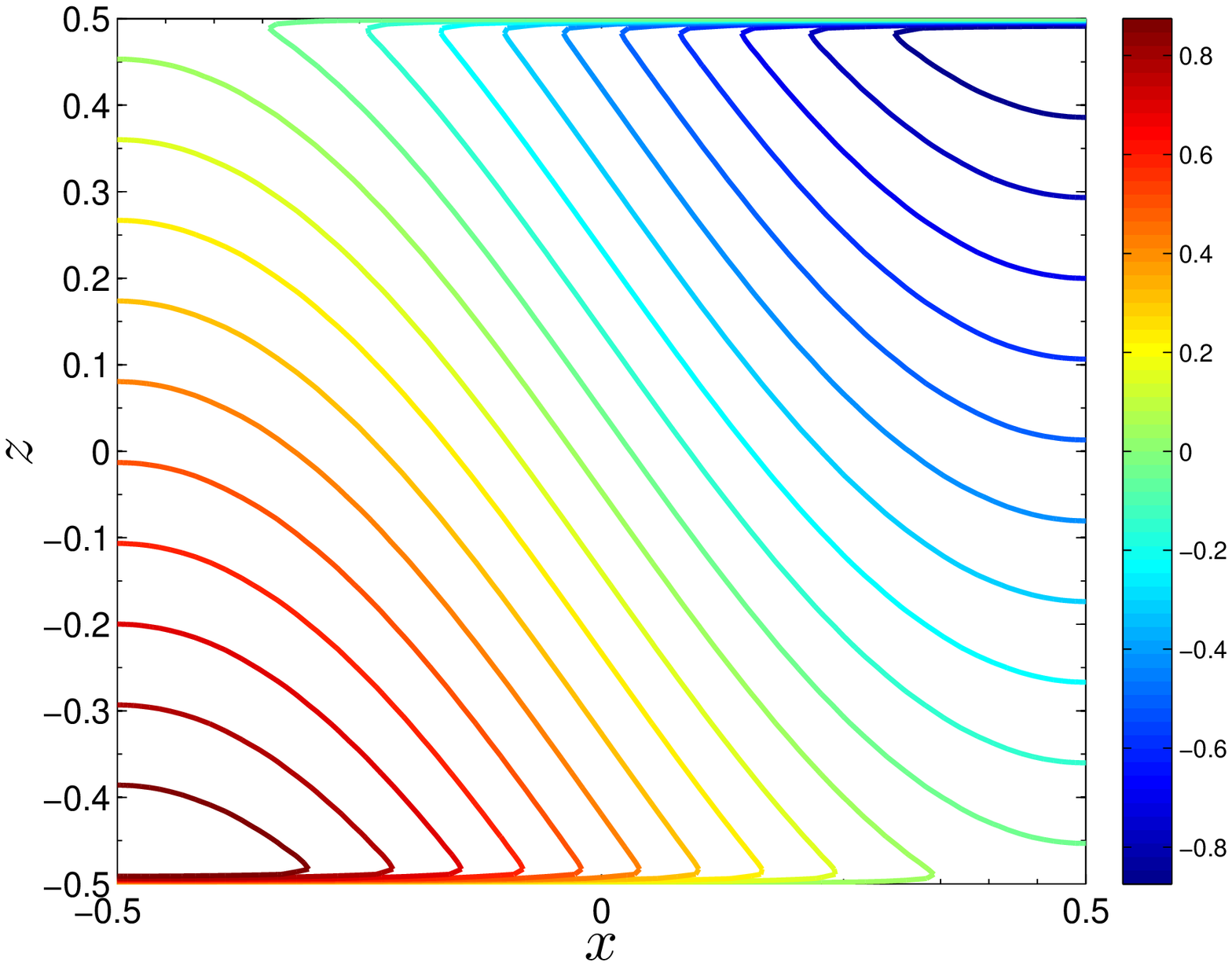}} 
\\           
  \subfloat[]{\label{fig:res1xi}\includegraphics[width=.45\textwidth]{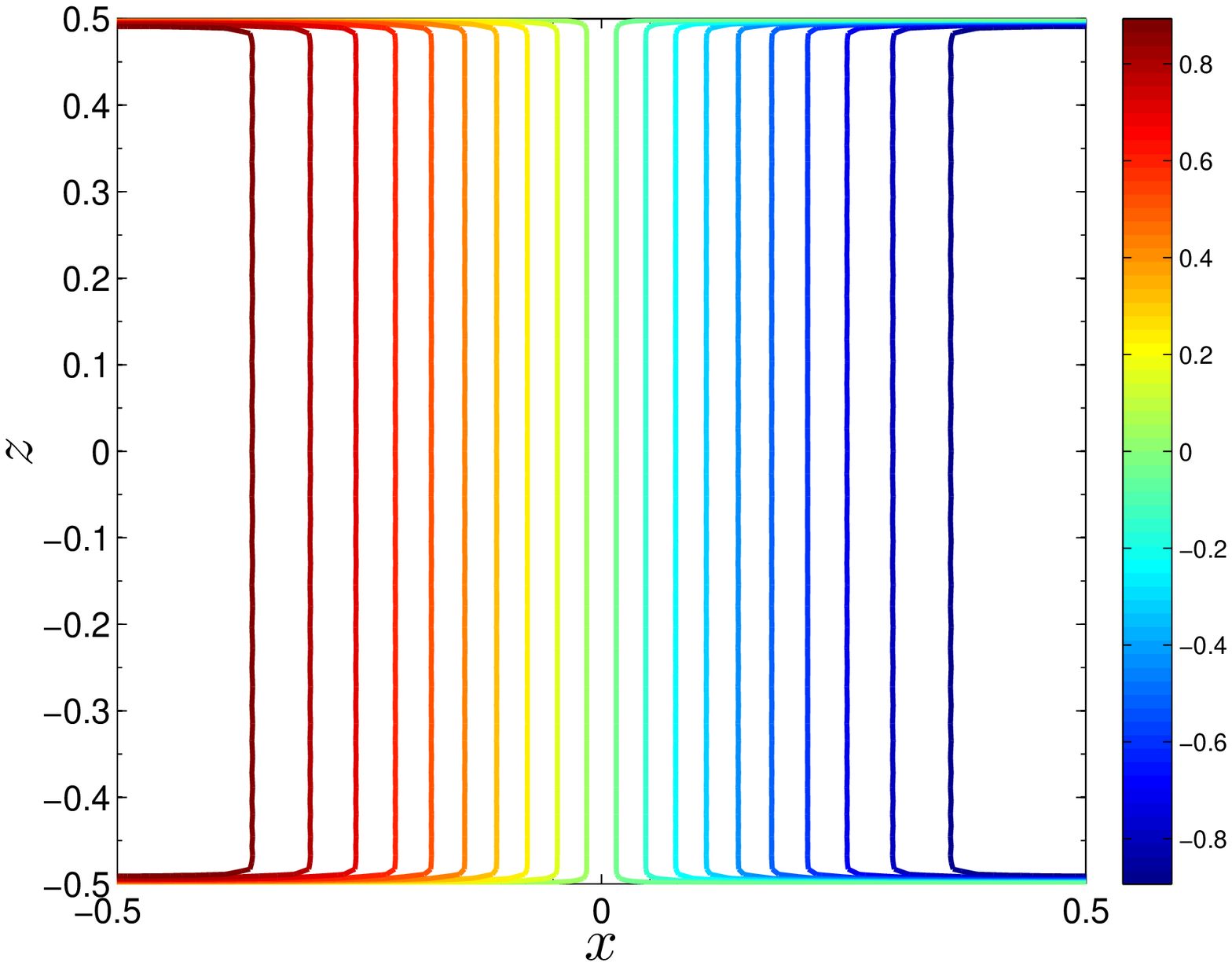}}   
  \subfloat[]{\label{fig:res1eta}\includegraphics[width=.45\textwidth]{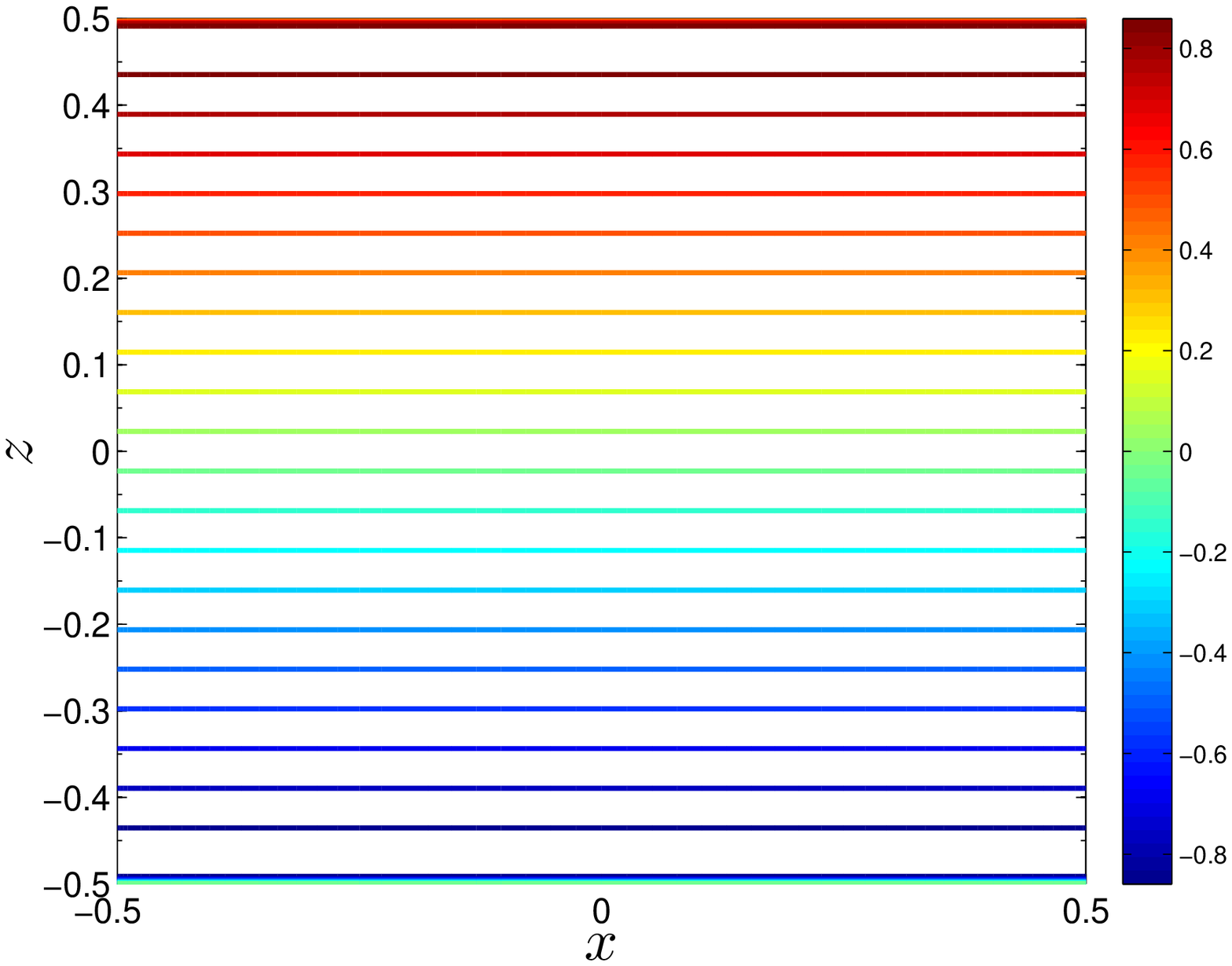}}             
  \caption{Flow field for $\varGamma=1$, $\mu=3.557 \times 10^{-5}$, $\Pe=1320.5$, and $\Nu_\m=90.7$. (a) $\psi$, (b)~$\psi$ along $x=0$, (c) $\theta$, (d) $\phi$, (e) $\xi \equiv \theta+\phi$, and (f) $\eta \equiv \theta - \phi$. The resolution is $61^2$.}
  \label{fig:res1}
\end{figure}

The high--$\Pe$ (small $\mu$) numerical solutions plotted in figure~\ref{fig:res1} \citep[and in figure~2.7 of][]{PedramMA} display some striking features: $\psi$ is nearly independent of $z$ in the bulk and depends on $x$ as $\cos{(\upi x/\varGamma)}$ in both the bulk and boundary layers.
And while $\theta$ and $\phi$ do not have such simple structure in the bulk or boundary layers, the variables
\begin{eqnarray}
\xi(x,z) &\equiv& \phi(x,z) + \theta(x,z) \label{xi}, \\
\eta(x,z) &\equiv& \theta(x,z)-\phi(x,z) \label{eta}
\end{eqnarray}
do: $\xi$ (like $\psi$) is nearly independent of $z$ except close to the top and bottom boundaries and $\eta$ is only a function of $z$ everywhere (see figure~\ref{fig:res1}).
This observation suggests formulating equations for $(\psi,\xi,\eta)$ and using matched asymptotic analysis to solve the resulting equations in the large--$\Pe$ limit.  

Equations (\ref{N1})--(\ref{N3}) imply that the equations for $\psi$, $\xi$, and $\eta$ are
\begin{eqnarray}
- \mathrm{J}(\xi,\eta) + 2 \mu \Delta \psi + 2 \xi_x &=& 0, \label{F1} \\
\mathrm{J}(\psi,\xi) + \Delta \eta &=& 0,  \label{F2} \\
\mathrm{J}(\psi,\eta) + \Delta \xi - 2 \psi_x &=& 0.  \label{F3} 
\end{eqnarray} 
The computational results suggest the ansatzen
\begin{eqnarray}
\psi &=& \bar{\psi}(x) \; A\left(\frac{1/2+z}{\delta}\right) A\left(\frac{1/2-z}{\delta} \right), \label{psibar}\\
\xi  &=& \bar{\xi}(x)  \; B\left(\frac{1/2+z}{\delta}\right) B\left(\frac{1/2-z}{\delta} \right), \label{xibar}\\
\eta &=& \bar{\eta}(z) \; C\left(\frac{1/2+z}{\delta}\right) C\left(\frac{1/2-z}{\delta} \right), \label{etabar}
\end{eqnarray}
where $\delta$ denotes the boundary layer thickness.\footnote{We introduce a single boundary layer scale here; we originally allowed for three independent boundary layer thicknesses but later deduced that they are, in fact, identical \cite[][pp.~19--20]{PedramMA}.} The overbar indicates interior solution, i.e., the \textit{outer} solutions far from the top and bottom boundary layers.  The boundary layer functions $A(Z)$, $B(Z)$, and $C(Z)$ are the \textit{inner} solutions that vanish at the boundaries $Z = 0$ and approach unity in the interior where $Z \gg 1$.

 
\subsubsection{Interior Solution}\label{sec:intS1}
Inserting \eqref{psibar}--\eqref{etabar} into~(\ref{F1})--(\ref{F3}) and setting $A=B=C=1$ yields
\begin{eqnarray}
2 \mu \, \bar{\psi}_{xx} + (2-\bar{\eta}_z) \, \bar{\xi}_x &=& 0, \label{A1} \\
\bar{\eta}_{zz} &=& 0, \label{A2} \\
\bar{\xi}_{xx} + (\bar{\eta}_{z}- 2) \bar{\psi}_{x} &=& 0. \label{A3} 
\end{eqnarray}
Equation~(\ref{A2}) implies that $\bar{\eta}$ is an affine function of $z$ which could also be inferred from (\ref{A1}) and (\ref{A3}) since $\bar{\psi}$ and $\bar{\xi}$ are only functions of $x$.
The symmetry of the solution with respect to reflection about $z=0$ implies
\begin{equation}
\bar{\eta}(z) = \bar{\eta}_o \, z, \label{etasol}
\end{equation}
where $\bar{\eta}_o$ is to be determined. 

Eliminating $\bar{\psi}$ between (\ref{A1}) and (\ref{A3}) gives
\begin{eqnarray}
\bar{\xi}_{xxx} + \left( \frac{\bar{\eta}_o-2}{\sqrt{2 \mu}} \right)^2 \, \bar{\xi}_{x} &=& 0. \label{xix} 
\end{eqnarray}  
Periodicity with period $2 \varGamma$ in $x$ together with $\xi_{x}(\pm \varGamma/2,z)=0$ implies
\begin{equation}
\bar{\xi} = \pm \bar{\xi}_o \, \sin{(\upi \, x/\varGamma)},  \label{xisol} \\
\end{equation}
and 
\begin{equation}
\bar{\eta}_o = 2-\frac{\upi}{\varGamma} \, \sqrt{2\mu},  \label{eta_o}
\end{equation}
where $\bar{\xi}_o>0$ is another yet-to-be determined constant. Another possible solution, $\bar{\eta}_o = 2+({\upi}/{\varGamma}) \, \sqrt{2\mu}>2$, is discarded in light of the numerical results. Moreover, in \S\ref{sec:compS1} it will be seen that $\bar{\eta}_o \le 2$ because of the maximum principle, confirming that (\ref{eta_o}) is the only admissible solution. 

Equation~(\ref{A3}) then yields
\begin{eqnarray}
\bar{\psi} &=& \frac{\pm\bar{\xi}_o}{\sqrt{2\mu}} \, \cos{(\upi \, x/\varGamma)}  \label{psisol}
\end{eqnarray}
so that the interior fields (i.e., the outer solutions) are known up to the one undetermined constant $\bar{\xi}_o$. (Notice that in (\ref{xisol}) and (\ref{psisol}) either $-\bar{\xi}_o$ or $+\bar{\xi}_o$ should be chosen for both $\bar{\psi}$ and $\bar{\xi}$.)

\subsubsection{Boundary Layer Solution}\label{sec:blsol}
The boundary conditions on the inner solutions are $A(0)=B(0)=C(0)=0$ and $A(+\infty)=B(+\infty)=C(+\infty)=1$. 
Focusing on the boundary layer adjacent to $z=+0.5$, (\ref{F1})--(\ref{F3}) give, upon substitution of (\ref{psibar})--(\ref{etabar}),
\begin{eqnarray}
0 &=& 2\mu \, (\bar{\psi}_{xx} \, A + \bar{\psi}  \, A'' / \delta^2) + (2-( \bar{\eta}_z \, C-\bar{\eta} \, C'/\delta)) \,  \bar{\xi}_x \, B, \label{S1} \\
0 &=& \bar{\eta}_{zz} \, C - 2 \bar{\eta}_z \, C'/\delta + \bar{\eta} \, C''/\delta^2 - \bar{\psi}_x \, A \, \bar{\xi} \, B'/\delta + \bar{\xi}_x \, B \, \bar{\psi} \, A'/\delta, \label{S2} \\
0 &=& \bar{\xi}_{xx} \, B + \bar{\xi} \, B''/\delta^2 - (2-(\bar{\eta}_z \, C-\bar{\eta} \, C'/\delta))  \bar{\psi}_x \, A, \label{S3} 
\end{eqnarray}
where a prime denotes $d/dZ$ (e.g., $A' \equiv dA/dZ$).
Using the interior solutions (\ref{etasol}) and (\ref{xisol})--(\ref{psisol}) and observing that $\bar{\eta} \rightarrow \bar{\eta}_o/2$ as $z \rightarrow 0.5$, the above equations become
\begin{eqnarray}
0 &=& \sqrt{2\mu} \, \left[-\left( \frac{\upi}{\varGamma} \right)^2 A + \, \frac{1}{\delta^2} A'' \right] + \frac{\upi}{\varGamma} \left[ 2-\left(2-\frac{\upi}{\varGamma}\sqrt{2\mu} \right) \left( C - \frac{1}{2\delta} C' \right) \right]  B, \label{SS1} \\
0 &=& \left(2-\frac{\upi}{\varGamma}\sqrt{2\mu} \right) \left[- \frac{2}{\delta} C' + \frac{1}{2\delta^2} C'' \right] + \frac{\upi}{2\varGamma} \frac{\bar{\xi}^2_o}{\sqrt{2\mu}}\left[ \frac{1}{\delta} A B'+ \frac{1}{\delta} B A' \right], \label{SS2} \\
0 &=& \sqrt{2\mu} \, \left[-\left( \frac{\upi}{\varGamma} \right)^2 B + \, \frac{1}{\delta^2} B'' \right] + \frac{\upi}{\varGamma} \left[ 2-\left(2-\frac{\upi}{\varGamma}\sqrt{2\mu} \right) \left( C - \frac{1}{2\delta} C' \right) \right]  A, \label{SS3} 
\end{eqnarray}
where (\ref{SS2}) has been averaged over $x$ to eliminate the $\sin^2{(\upi x/\varGamma)}$ and $\cos^2{(\upi x/\varGamma)}$ terms.

To balance the leading order terms, we need to determine the boundary layer thickness $\delta$ as a function of a small parameter $\epsilon$ defined based on $\mu$ and $\varGamma$.
We know that $\mu \ll 1$ to achieve the large $\Pe$ limit, although from the above equations it seems that $\sqrt{\mu} \ll 1$ may be a more appropriate parameter in this problem.
Here we restrict our analysis to $\varGamma \le 1$, because the numerical results of \S\ref{sec:numres} showed that $\varGamma > 1$ does not maximize the transport.

With $\mu \ll 1$ and $\varGamma \le 1$ we define the small parameter
\begin{equation}
\epsilon \equiv \frac{\varGamma \sqrt{2\mu}}{\upi} \label{epsil},
\end{equation}
(where the constants are included to simplify the algebra), and recognize that we must also consider the magnitude of
\begin{equation}
\sigma \equiv \varGamma/\sqrt{\mu}.\label{gammeroo}
\end{equation}
If $\varGamma = O(1)$, then $\sigma \gg 1$. The numerical results (figure~\ref{fig:MapNum}) suggest that $\varGamma_\mathrm{opt}$ decreases as $\Pe$ increases.
Therefore, we should also allow for the possibility that $\varGamma \ll 1$, implying $\sigma =  O(1)$ or even $\sigma \ll 1$.
Close examination of (\ref{SS1})--(\ref{SS3}) reveals that $\sigma \gg 1$ and $\sigma = O(1)$ conveniently give the same balance and result in the same scaling for boundary layer thickness.
Therefore, one solution covers both limits.
Additionally, the distinguished limit $\sigma = O(1)$ guarantees that the solution is uniformly valid in $\varGamma$.
Here we focus on these two limits.  As demonstrated subsequently, the justification for excluding the scenario in which $\sigma \ll 1$ from our analysis is that
$\Nu_\m$ for a fixed value of $\varGamma$ in the limit of large $\Pe$ is obtained with $\sigma \gg 1$, and that $\Nu_\M$ for large $\Pe$ is achieved when $\sigma = O(1)$;
see \S\S\ref{sec:NuPe}--\ref{sec:NuPeUlt}. 

Using definitions (\ref{epsil}) and (\ref{gammeroo}) in (\ref{SS1}) and balancing the leading order terms gives
\begin{equation}
A'' + \left(1-\frac{\upi}{ \sigma \sqrt{2}}  \right) \, B \, C' = 0, \label{Add}
\end{equation} 
upon identifying
\begin{equation}
\delta = \epsilon. \label{balance1}
\end{equation} 
Note that based on our restriction on $\sigma$, the term in the parentheses is $O(1)$.
The same procedure applied to (\ref{SS3}) produces
\begin{equation}
B'' + \left(1-\frac{\upi}{ \sigma \sqrt{2}}  \right) \, A \, C' = 0, \label{Bdd} \\ 
\end{equation} 
while equation~(\ref{SS2}) yields 
\begin{equation}
\left(1-\frac{\upi}{ \sigma \sqrt{2}}  \right) \, C'' +  \frac{\bar{\xi}^2_o}{2}  \, (A B' + B A') = 0. \label{Cdd} \\  
\end{equation} 
Integrating equation~(\ref{Cdd}) we see that
\begin{eqnarray}
\left(1-\frac{\upi}{\sigma \sqrt{2}}  \right) \, C'&=&- \frac{\bar{\xi}^2_o}{2} \, (AB-1), \label{Cprime}
\end{eqnarray} 
where the constant of integration is determined by $C' \rightarrow 0$ and $A$ and $B \rightarrow 1$ as $Z \rightarrow \infty$.

Substitution of (\ref{Cprime}) into (\ref{Add}) and (\ref{Bdd}) shows that $A$ and $B$ satisfy the same equation.  Since they also 
satisfy the same boundary conditions, we conclude $A=B$.  In fact, making the ansatz $A=B=\rho C$ renders
(\ref{Cdd}) identical to (\ref{Add}) and (\ref{Bdd}) upon setting the constant $\rho=(1-\pi/(\sigma\sqrt{2}))/\bar{\xi}_o$.  
Making this ansatz, again noting that $C$ satisfies the same boundary conditions as do $A$ and $B$ (i.e., $A(+\infty)=B(+\infty)=C(+\infty)=1$), we conclude that $A(Z)=B(Z)=C(Z)$ (i.e., $\rho=1$). Therefore,
\begin{eqnarray}
A'' + \bar{\xi}_o \, AA' = 0, \label{Afinal}
\end{eqnarray}
and 
\begin{eqnarray}
\bar{\xi}_o= 1-\frac{\pi}{\sigma\sqrt{2}} \equiv 1-\frac{\upi}{2 \varGamma} \sqrt{2 \mu} = \frac{\bar{\eta}_o}{2}.
\end{eqnarray}
Equation~(\ref{Afinal}) can be integrated, giving
\begin{eqnarray}
A' + \frac{\bar{\xi}_o}{2} A^2 = \frac{\bar{\xi}_o}{2} \label{Afinal2},
\end{eqnarray}
where again the constant of integration is determined from the boundary conditions $A' \rightarrow 0$ and $A  \rightarrow 1$ as $Z \rightarrow \infty$.
The exact solution is
\begin{eqnarray}
A(Z) = \frac{1-\exp(-\bar{\xi}_o Z)}{1+\exp(-\bar{\xi}_o Z)} = \tanh{\left[ \frac{\bar{\xi}_o}{2} Z \right]}.
\end{eqnarray}

\subsubsection{The Complete Solution: Matching}\label{sec:compS1}
Including the bottom boundary layer and matching the three regions (the bulk and the two boundary layers), the complete asymptotic approximation is
\begin{eqnarray}
\psi(x,z) &\sim&\frac{1}{\sqrt{2\mu}} \, \left( 1 - \frac{\upi}{2\varGamma}  \sqrt{2 \mu} \right) 
 \cos{\left(\frac{\upi}{\varGamma} x \right)} \, 
H(z), \label{X1}  \\
\xi(x,z) &\sim& \; \; \; \; \; \; \; \; \; \left( 1 - \frac{\upi}{2\varGamma}  \sqrt{2 \mu} \right) \sin{\left(\frac{\upi}{\varGamma} x \right)} \, 
 H(z),  \label{X2} \\
\eta(x,z) &\sim& \; \; \; \; \; \; 2\, \left(1-\frac{\upi}{2\varGamma} \sqrt{2\mu}  \right) \, z \, H(z),   \label{X3}  
\end{eqnarray}
where
\begin{eqnarray} \nonumber
H(z) &=& A\left(\frac{0.5-z}{\delta}\right) A\left(\frac{0.5+z}{\delta}\right) \\
&=& \tanh{\left[  \frac{\upi}{2} \left(1-\frac{\upi}{2 \varGamma} \sqrt{2 \mu} \right) \frac{(0.5-z)}{\varGamma \sqrt{2\mu}} \right]} \tanh{\left[  \frac{\upi}{2} \left(1-\frac{\upi}{2 \varGamma} \sqrt{2 \mu} \right) \frac{(0.5+z)}{\varGamma \sqrt{2\mu}} \right]}.
\end{eqnarray}
Therefore, to leading order assuming that $\varGamma \sqrt{\mu} \ll 1$ and $\varGamma / \sqrt{\mu}$ is finite or large,
\begin{eqnarray}
u(x,z) &\sim& \; \; \frac{1}{\sqrt{2\mu}} \, \left( 1 - \frac{\upi}{2\varGamma}  \sqrt{2 \mu} \right) 
\, \cos{\left(\frac{\upi}{\varGamma} x \right)} \, 
H'(z), \label{usol}  \\
w(x,z) &\sim& \frac{1}{\sqrt{2\mu}} \left(\frac{\upi}{\varGamma}\right) \, \left( 1 - \frac{\upi}{2\varGamma}  \sqrt{2 \mu} \right) 
\, \sin{\left(\frac{\upi}{\varGamma} x \right)} \, 
H(z),  \label{wsol}\\
\theta(x,z) &\sim& \frac{1}{2} \left( 1 - \frac{\upi}{2\varGamma}  \sqrt{2 \mu} \right)
\, \left(  \sin{\left(\frac{\upi}{\varGamma} x \right)} + 2z  \right) \, H(z). \label{tsol}  
\end{eqnarray}
Many features of these asymptotic solutions are compared with the numerical results in \citet[][figures~2.8--2.9]{PedramMA} revealing excellent agreement for $\Pe \ge O(1)$, uniformly for a wide range of $\varGamma$.


We note here that the maximum principle requires $|\theta| \le 1$.
Consequently, in (\ref{tsol}), the first term in the parentheses, i.e., $\bar{\xi}_o = \bar{\eta}_o/2 = 1 - \frac{\upi}{2\varGamma}  \sqrt{2 \mu}$, has to be smaller than $1$. This analysis justifies discarding the other solution for $\bar{\eta}_o$ in \S\ref{sec:intS1}.     

\begin{figure}
\centering
  \subfloat[]{\label{fig:compPe}\includegraphics[width=.5\textwidth]{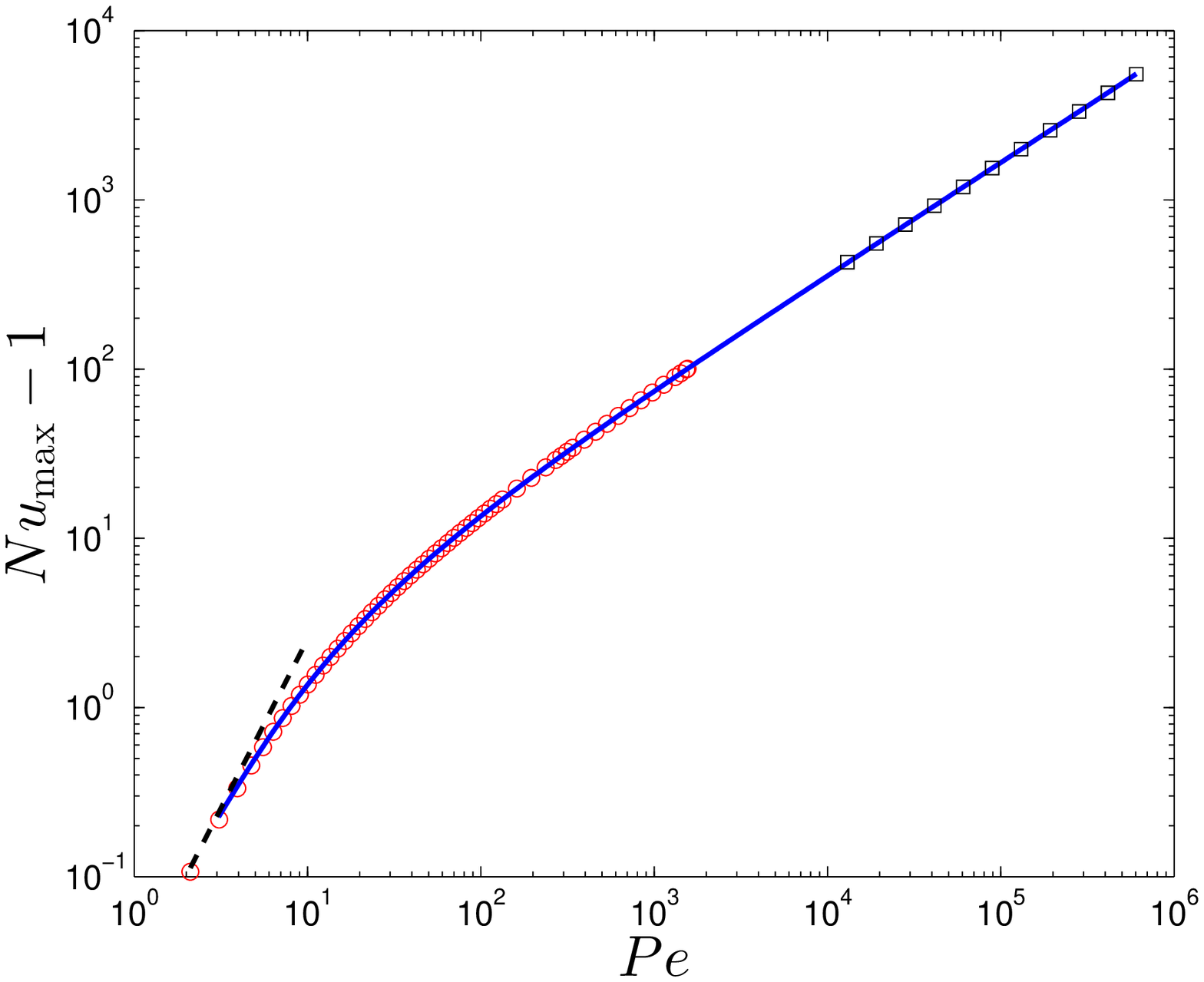}}  
  \subfloat[]{\label{fig:compNu}\includegraphics[width=.5\textwidth]{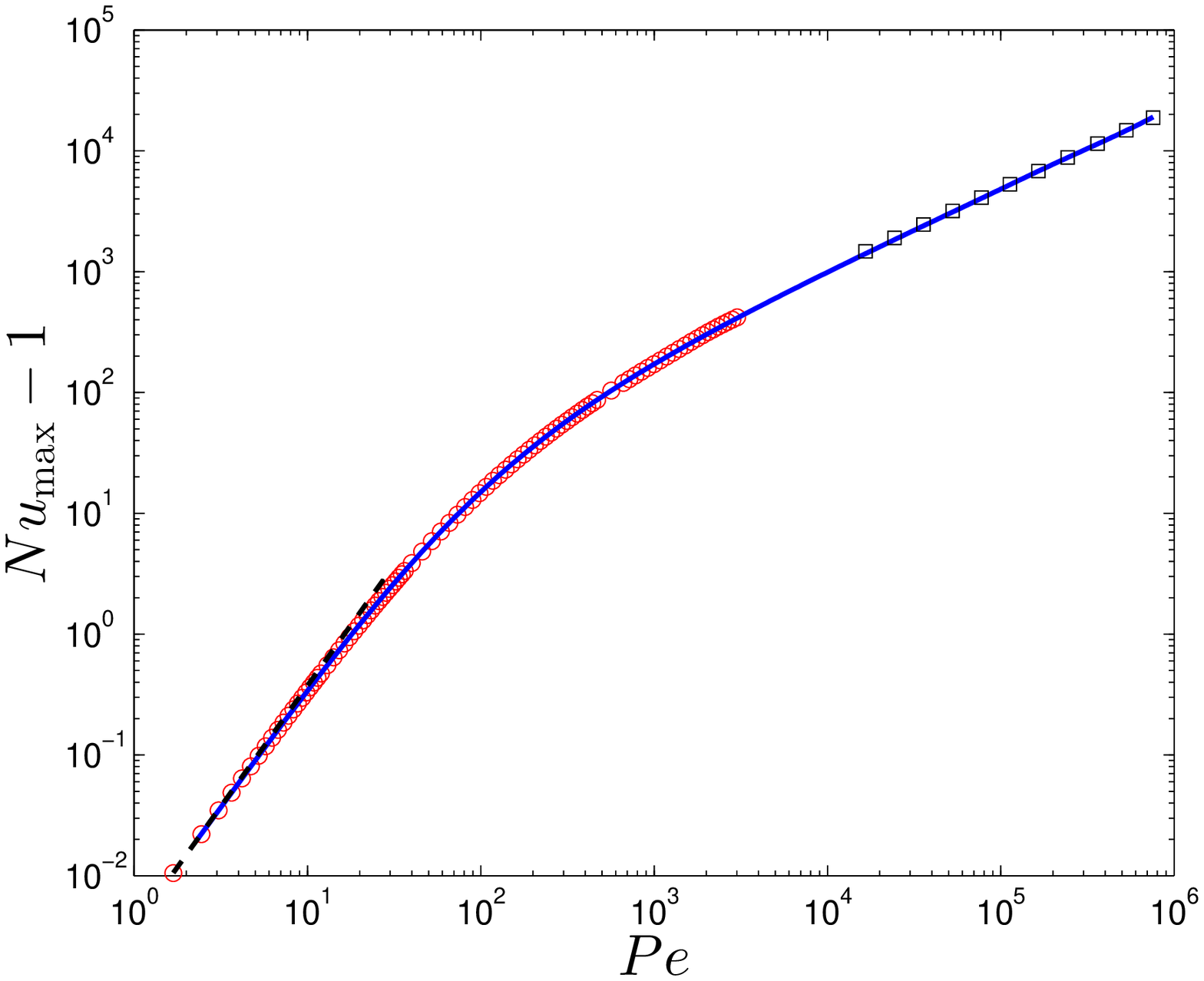}}            
  \caption{$\Nu_\m$ from the numerical solutions (red circles), the small--$\Pe$ analytical solution (\ref{smallPe}) (black dashed lines), and the large--$\Pe$ asymptotic solution (\ref{PeAnalytical})--(\ref{NuAnalytical}) (blue solid lines) for two cases: (a) $\varGamma=1$ and (b) $\varGamma=0.2$. The black squares show (\ref{NuFixedGamma}).}
  \label{fig:compNuPe}
\end{figure}

Asymptotic approximations for $\Pe$ and $\Nu_\m$ can be calculated analytically from (\ref{usol})--(\ref{tsol}):
\begin{eqnarray}
\Pe \equiv \sqrt{\left<  u^2+w^2 \right>} &=& \frac{1}{2\sqrt{\mu}} \left(1 - \frac{\upi}{2\varGamma}  \sqrt{2 \mu} \right) \, \sqrt{\int^{0.5}_{-0.5} \left\{(H')^2+(\upi/\varGamma)^2 H^2\right\} \mathrm{d}z}, \label{PeAnalytical}\\
\Nu_\m \equiv 1 + \left<  w\theta \right> &=& 1+ \frac{1}{4\sqrt{2\mu}} \left( \frac{\upi}{\varGamma}\right)\left(1 - \frac{\upi}{2\varGamma}  \sqrt{2 \mu} \right)^2  \, \int^{0.5}_{-0.5} H^2 \mathrm{d}z.  \label{NuAnalytical} 
\end{eqnarray}
Because $H(z)$ depends on $\mu$ and $\varGamma$, it is difficult to find an explicit expression for $\Nu_\m(\Pe,\varGamma)$.
However, (\ref{PeAnalytical}) and (\ref{NuAnalytical}) can be easily evaluated numerically for a given pair of $(\mu,\varGamma)$.
Figure~\ref{fig:compNuPe} compares the values of $\Nu_\m(\Pe,\varGamma)$ from the numerical solutions with the values given by (\ref{PeAnalytical}) and (\ref{NuAnalytical}) for $\varGamma=0.2$ and $1$.
The numerical and analytical results agree well, even for relatively small values of $\Pe$, suggesting that the higher-order terms may be transcendentally small in $\epsilon$.\footnote{The numerically inspired ansatzen $\psi = a(z) \, \cos{(\upi x/\varGamma)}$, $\xi = b(z) \, \sin{(\upi x/\varGamma)}$, and $\eta=c(z)$ separate equations~(\ref{F1})--(\ref{F3}) into ODEs for $a(z)$, $b(z)$, and $c(z)$ without approximation. The exact solutions of those ODEs involve elliptic functions which are well known to be exponentially accurately approximated by products of $\tanh$ functions.}     

\subsubsection{$\Nu_\m(\Pe,\varGamma)$: $\epsilon \ll 1$ and $\sigma \gg 1$ Limit}\label{sec:NuPe}
In the limit of relatively small $\epsilon \equiv \varGamma \sqrt{2\mu}/\upi$, the integrals in (\ref{PeAnalytical}) and (\ref{NuAnalytical}) are 
approximately
\begin{eqnarray}
\int^{0.5}_{-0.5} H^2 \mathrm{d}z &\approx& 1-\left[\frac{4\epsilon}{1-\frac{\upi}{\sigma \sqrt{2}}}\right], \label{Happrox} \\
\int^{0.5}_{-0.5} (H')^2 \mathrm{d}z &\approx& \frac{2}{3}\left[\frac{1-\frac{\upi}{\sigma \sqrt{2}}}{\epsilon}\right],\label{DHapprox}
\end{eqnarray} 
as has been confirmed numerically (again, recall that $\sigma \equiv \varGamma /\sqrt{\mu}$ is not small compared to 1). 
Consider (\ref{PeAnalytical}) and (\ref{NuAnalytical}) in the limit of vanishing $\mu$ and fixed $\varGamma$, i.e.,  $\epsilon \ll 1$ and $\sigma \gg 1$.
In this limit, (\ref{Happrox}) and (\ref{DHapprox}) further simplify to $1$ and $2/(3\epsilon)$, respectively.
Then (\ref{PeAnalytical}) and (\ref{NuAnalytical}) yield
\begin{eqnarray}
\Pe &=& \frac{1}{\sqrt{2}\epsilon}  \sqrt{\frac{\sqrt{2}}{3\upi}\sigma}, \label{PeUlt} \\
\Nu_\m &=& 1+ \frac{1}{4\epsilon}. \label{NuUlt}
\end{eqnarray}
Solving (\ref{PeUlt}) for $\mu$ and substituting into (\ref{NuUlt}) yields $\Nu_\m(\Pe,\varGamma)$ as
\begin{eqnarray}
\Nu_\m = 1 + \frac{1}{4} \left[   \frac{3 \upi^2}{\varGamma^2} \right]^{1/3}  \Pe^{2/3}. \label{NuFixedGamma}
\end{eqnarray}
The accuracy of this approximation is demonstrated in figure~\ref{fig:compNuPe} for $\varGamma=0.2$ and $1$.
 
Equation~(\ref{NuFixedGamma}) gives $\Nu_\m$ as a function of $\Pe$ for a fixed value of $\varGamma$, which is not the same as $\Nu_\M$, i.e., the maximum achievable $\Nu_\m$ at that $\Pe$.
Determining $\Nu_\M(\Pe)$ requires letting $\varGamma$ shrink as $\Pe$ increases, as analyzed next.

\subsubsection{$\Nu_\M(\Pe)$: $\epsilon \ll 1$ and $\sigma = O(1)$ Limit}\label{sec:NuPeUlt}
Here we examine the limit $\sqrt{\mu} \ll 1$ and $\varGamma \ll 1$ when their ratio is finite.
Physically this means that we allow the cells to narrow as $\Pe$ increases.
It is in this distinguished limit that $\Nu_\M$ for a given $\Pe$, i.e., the optimal transport for a given amount of energy, is achieved. 

In this limit, (\ref{Happrox}) again reduces to $1$, but (\ref{DHapprox}) cannot be further simplified. Using these approximations for the integrals in (\ref{PeAnalytical})--(\ref{NuAnalytical}) implies
\begin{eqnarray}
\Pe &=& \frac{1}{\sqrt{2}\epsilon} \left(1-\frac{\upi}{\sigma\sqrt{2}}\right) \sqrt{\frac{2}{3}+\frac{\sqrt{2}}{3\upi}\sigma}, \label{PeUlt3} \\
\Nu_\m-1 &=& \frac{1}{4\epsilon} \left(1-\frac{\upi}{\sigma\sqrt{2}}\right)^2. \label{NuUlt3}
\end{eqnarray}
Dividing (\ref{NuUlt3}) by (\ref{PeUlt3}) to eliminate $\epsilon$ yields
\begin{eqnarray}
\frac{\Nu_\m-1}{\Pe} = \frac{1}{2\sqrt{2}} \frac{\left(1-\frac{\upi}{\sigma\sqrt{2}}\right)}{\sqrt{\frac{2}{3}+\frac{\sqrt{2}}{3\upi}\sigma}}, \label{TE}
\end{eqnarray}
which is maximized at 
\begin{eqnarray}
\sigma_{\mathrm{opt}} \equiv \frac{\varGamma_{\mathrm{opt}}}{\sqrt{\mu}}   = 2 \sqrt{2} \upi \approx 8.885766. \label{sigma}
\end{eqnarray}

This calculation thus gives the optimal aspect ratio $\varGamma_{\mathrm{opt}}$ that maximizes $\Nu_\m$ at a given $\mu$.
Using $\sigma=\sigma_{\mathrm{opt}}$ in (\ref{TE}) gives $\Nu_\M(\Pe)$:  
\begin{eqnarray}
\Nu_\M &=& 1 + 0.1875 \, \Pe. 
\label{NuPeUlt} 
\end{eqnarray}
Figure~\ref{fig:MapNum} shows that (\ref{NuPeUlt}) gives the maximum possible transport with remarkable accuracy. 
Finally, combining (\ref{PeUlt3}) and (\ref{sigma}) determines $\varGamma_{\mathrm{opt}}(\Pe)$, the optimal cell aspect ratio at a given $\Pe$:
\begin{eqnarray}
\varGamma_{\mathrm{opt}} = 3.8476 \; \Pe^{-1/2}. \label{Gammaopt}
\end{eqnarray}
Thinner cells produce the maximum transport as the P\'eclet number increases.

\subsection{Example: Application to Porous Media Convection}\label{sec:pm}
We now show how Rayleigh--B\'enard convection in a fluid-saturated porous layer is an example of transport with fixed energy in order to express $\Nu_\m$ and $\Nu_\M$ as functions of the Rayleigh number $\Ra$ and compare the optimal transport bounds with the results of previous analytical and numerical investigations.  

Rayleigh--B\'enard convection in a fluid-saturated porous layer heated from below and cooled from above is often modeled by the system
\begin{eqnarray}
\bnabla \bcdot \vel &=& 0, \label{DivPM}\\
\frac{1}{\Pran} \, (\dot{\vel} + \vel \bcdot \bnabla \vel) + \vel  &=& - \bnabla p + \Ra \, T \zhat, \label{Bouss}\\
\dot{T} + \vel \bcdot \bnabla T &=& \Delta T, \label{AdPM}
\end{eqnarray}
with the same boundary conditions as have been used in the preceding sections: $T|_{z=0} = 1$, $T|_{z=1} = 0$, 
and $\zhat \cdot \vel |_{z=0} = \zhat \cdot \vel |_{z=1} = 0$.
In this system $\Pran$ is the Prandtl--Darcy number and $\Ra$ is the Rayleigh number (see, e.g., \cite{Doering98} and references therein for more details). 

Taking the inner product of (\ref{Bouss}) with $\vel$ and averaging over long times and over the spatial domain with impermeable walls yields
\begin{eqnarray}
\left< |\vel |^2 \right> = \Ra \, \left< w T \right>.
\end{eqnarray}
The transient term vanishes owing to the long-time averaging (when the kinetic energy is $o(t)$ as $t \rightarrow \infty$) and the nonlinear and pressure terms vanish because of the spatial integration. 
Then using the definition of $\Pe$ for fixed energy problems (\ref{ke}) and with $\Nu$ given by (\ref{Nu}) from \S\ref{sec:math} we identify
\begin{eqnarray}
\Pe^2 = \Ra\, (\Nu-1). \label{RaNuPe}
\end{eqnarray}

The Nusselt number $\Nu$, when calculated by long-time averaging, is solely a function of $\Ra$ and the domain aspect-ratio $\varGamma$ (and possibly initial data).
Consequently, in steady and statistically-steady flows equation~(\ref{RaNuPe}) shows that the time-averaged P\'eclet number is fixed for given values
of $\varGamma$ and $\Ra$ (and possibly initial data)---note that $\Ra$ depends on the fluid properties and the imposed temperature difference between the walls, not on the flow.
Therefore, steady convection in porous media occurs with `fixed' energy.

\begin{table}
\begin{center}
\def~{\hphantom{0}}
\begin{tabular}{lccc}
                     & {$\Nu(\Ra)$}  & {$\varGamma(\Ra)$}  & { $\Nu(\Ra,\varGamma_{\mathrm{fixed}})$} \\ [3pt]
{\underline{Boundary layer stability argument}} & &  &\\
\citet{Malkus,Howard}; and &  && \\ 
{\citet{Horne78}} & {$\sim C \, \Ra$}  && \\ 
&&&\\
{\underline{Background method (upper bounds)}} & & & \\
{\citet{Doering98}} & {$\le 0.035 \, \Ra$} & &  \\
{\citet{Otero}}     & {$\le 0.029 \, \Ra$} & &  \\
{\citet{Wen}} & {$\lesssim 0.017 \, \Ra$} & &  \\
&&&\\ 
{\underline{Unsteady simulations (DNS)}} &&& \\
{\citet{Otero}: $\Ra \le 10^4$} &  {$\sim C \, \Ra^{0.9}$}&  & \\
{\citet{Duncan}: $\Ra \le 4 \times 10^4$} &  {$\sim 0.007 \, \Ra$}& { $\sim C \, \Ra^{-0.4}$}& \\
&&&\\
{\underline{Steady unicellular analysis}} & && \\
{\citet{Fowler}} &  &  &  {$\sim C(\varGamma) \, \Ra^{1/3}$}  \\  
{\citet{Lind}} & {$\sim C \, \Ra^{2/3}$} & {$\sim C \, \Ra^{-1/2}$} &  {$\sim C(\varGamma) \, \Ra^{1/3}$}  \\  
&&&\\
{\underline{Current work}} &&& \\
{Numerical and asymptotic analyses} & {$\le 1 + 0.035 \, \Ra$} & {$\sim 8.89 \, \Ra^{-1/2}$}       &  {$\; \le 1 + \frac{0.68}{\varGamma} \, \Ra^{1/2}$}  \\  
\end{tabular}
\caption{Comparison of the results of the current work with the scalings for porous media convection obtained using various other methods.}
\label{tab:PM}
\end{center}
\end{table}

Employing (\ref{RaNuPe}), $\Pe$ can be replaced with $\Ra$ in (\ref{NuFixedGamma}) and (\ref{NuPeUlt})--(\ref{Gammaopt}) so that
\begin{eqnarray}
\Nu_\m(\Ra,\varGamma) &=& 1 + \frac{\sqrt{3}\upi}{8\varGamma} \, \Ra^{1/2}, \label{NumRa}\\
\Nu_\M(\Ra) &=& 1+ 0.0352 \, \Ra, \label{NuMRa}\\
\varGamma_{\mathrm{opt}} &=& 8.89 \, \Ra^{-1/2}. \label{GaRa}
\end{eqnarray}
Interestingly, $\varGamma \sim \Ra^{-1/2}$ is the scaling of the shortest-wavelength unstable mode about the conduction solution in porous media convection.

Table~\ref{tab:PM} compares the optimal (steady) transport derived here with pertinent results obtained using other methods.
The classical argument of \citet{Malkus} and \citet{Howard} based on the marginal stability of the boundary layer predicts $\Nu \sim \Ra$ for Rayleigh--B\'enard convection in a fluid saturated porous layer \citep{Horne78}.
The background method---which, we note, does {\it not} enforce the full advection-diffusion equation for the temperature---also implies an upper bound on $\Nu$ that scales linearly with $\Ra$.
The prefactors in the upper bound have been improved over the years \citep{Doering98,Otero,Wen}.

We emphasize that in this work (\ref{DivPM}) and the steady version of (\ref{AdPM}) were solved for one cell, but the momentum equation (\ref{Bouss}) was {\it not} imposed.
Related analysis of steady cellular flows of the full Rayleigh-B\'enard system was done recently by \citet{Lind} who used numerical continuation in a single convection cell to solve the steady version of (\ref{DivPM})--(\ref{AdPM}) in the limit of infinite Prandtl--Darcy number, finding $\Nu \sim \Ra^{2/3}$ and $\varGamma \sim \Ra^{-1/2}$.
Furthermore, \citet{Lind} showed that $\Nu$ scales as $\Ra^{1/3}$ for steady convection when the cell size $\varGamma$ is fixed.
Comparing these scalings with those obtained in the current work shows that steady convection in porous media does {\it not} transport as much heat as is possible by a steady flow with the same energy.

On the other hand the latest 2D direct numerical simulations (DNS) of infinite Prandtl--Darcy number Rayleigh--B\'enard convection in a porous layer, for $\Ra$ as high as $4 \times 10^4$, show that $\Nu \sim \Ra$ for ``turbulent'' flows and that the emergent cells that form in the bulk have an aspect ratio that scales approximately as $\Ra^{-0.4}$ \citep{Duncan}.
Comparing the steady \citep{Lind} and unsteady \citep{Duncan} solutions of (\ref{DivPM})--(\ref{AdPM}), it is evident that unsteadiness enhances the transport. The maximum possible steady transport scales linearly with $\Ra$ which, perhaps curiously, coincides with the unsteady results.
However, the unsteady transport is around $5$ times smaller than the maximum possible steady transport at a given $\Ra$ (current work), and the convection cells of the unsteady flow are wider than the optimal cells with aspect ratio $\varGamma_\mathrm{opt}$ (see Table~\ref{tab:PM}).
A recent investigation, for example, suggests that cells with aspect ratio smaller than $\varGamma$ where $\varGamma \sim Ra^{-5/14}$ are dynamically unstable at high $\Ra$ \citep{hewitt2013stability} which might explain the wider cells and (slightly) less-than-optimal transport observed in the direct numerical simulations of unsteady flows.  

\vspace{-0.15in}
\section[Fixed Enstrophy]{Optimal Transport with Fixed Enstrophy} \label{sec:problem2}
In the second version of the optimal transport problem we investigate steady flows with fixed enstrophy.
Then equations~(\ref{ad})--(\ref{bcv}) and (\ref{en}) become
\begin{eqnarray}
\vel \bcdot \bnabla \theta &=& \Delta \theta + w, \label{adS2} \\
\bnabla \bcdot \vel &=& 0, \label{divS2} \\
\Pe^2 &=& \left< |\bnabla \vel|^2\right>, \label{enS2} \\
\theta(x,0) &=& \theta(x,1) = 0, \label{bctS2} \\
w(x,0) &=& w(x,1) = 0. \label{bcvS2}
\end{eqnarray}   
Exactly as for the fixed-energy problem, simple analysis gives an {\it a priori} upper bound on $\Nu$.  Starting from (\ref{Nu}),
\begin{eqnarray}
\Nu &=& 1+\left< wT \right> = 1 + \left<  w (T-1/2) \right> \nonumber \\
&\leq& 1 + \left<  w^2 \right>^{1/2} \left< (T-1/2)^2 \right>^{1/2} \nonumber \\
&\leq& 1 + \frac{\left< |\nabla w|^2 \right>^{1/2}}{2\upi} \leq 1 + \frac{\left< |\bnabla \vel|^2\right>^{1/2}}{2\upi} \nonumber \\
&=&1 + \frac{\Pe}{2\upi}, \; \; \; \; \label{NuPe2}
\end{eqnarray}
where as before, incompressibility and the Cauchy--Schwarz inequality were used in the first line and the maximum principle for the temperature was invoked to deduce the first expression in the second line. 
Poincar\'e's inequality ($\left< (\partial w/\partial z)^2 \right> \ge \pi^ 2 \left<  w^2 \right>$) was then applied to derive the second term on the second line and (\ref{enS2}) used to obtain the final result.
As will be seen, this upper bound is much too high for this problem.

\subsection{Variational Formulation for Steady Flows} \label{sec:var2} 
The variational formulation of the fixed-enstrophy problem involves maximizing $\Nu$ with constraints (\ref{adS2})--(\ref{enS2}) and boundary conditions (\ref{bctS2})--(\ref{bcvS2}).
The relevant functional is
\begin{eqnarray}
\mathcal{F} = \left< w \theta  - \phi(x,z) \left(\vel \bcdot \bnabla \theta - \Delta \theta - w \right) + p(x,z) \left(\bnabla \bcdot \vel \right) + \frac{\mu}{2} \left(|\bnabla \vel|^2-\Pe^2 \right) \right>, 
\label{var2}
\end{eqnarray} 
where again, $\phi(x,z)$, $p(x,z)$, and $\mu$ are Lagrange multipliers.
The Euler--Lagrange equations are
\begin{eqnarray}
0 =  \frac{\delta \mathcal{F}}{\delta \vel} &=& \left( \theta + \phi \right) \zhat + \theta \, \bnabla \phi - \bnabla p + \mu \Delta \vel,  \label{duS2} \\
0 =  \frac{\delta \mathcal{F}}{\delta \theta} &=& \vel \bcdot \bnabla \phi  + \Delta \phi + w,\label{dtS2} \\
0 =  \frac{\delta \mathcal{F}}{\delta \phi} &=& -\vel \bcdot \bnabla \theta + \Delta \theta + w, \label{dphiS2}\\
0= \frac{\delta \mathcal{F}}{\delta p} &=& \bnabla \bcdot \vel, \label{div2S2} \\
0 = \frac{\partial \mathcal{F}}{\partial \mu} &=& \frac{1}{2}\left(\left<|\bnabla \vel|^2 \right>-\Pe ^2\right),  \label{dmuS2}
\end{eqnarray}
where $\phi$ vanishes at $z=0$ and $z=1$.
To eliminate the surface term coming from $\bnabla \bcdot (\vel \bnabla \vel)$ we can use {\it either} the stress-free ($\partial u/\partial z=0$) or no-slip ($u=0$) boundary conditions at $z=0$ and $z=1$.
Therefore, the full set of boundary conditions is
\begin{eqnarray}
w(x,0)=w(x,1)&=& 0, \label{bc1S2} \\
\theta(x,0)=\theta(x,1)&=& 0, \label{bc2S2} \\
\phi(x,0)=\phi(x,1)&=& 0, \label{bc3S2}
\end{eqnarray}
together with either
\begin{eqnarray}
u(x,0)=u(x,1) = 0 \label{noslip} \; \; \; \; \; \mbox{(no-slip)}
\end{eqnarray}
or
\begin{eqnarray}
\frac{\partial u}{\partial z}\bigg|_{z=0}=\frac{\partial u}{\partial z}\bigg|_{z=1} = 0 \label{freeslip} \; \; \; \; \; \mbox{(stress-free)}.
\end{eqnarray}
In the following, we restrict attention to the stress-free boundary condition (\ref{freeslip}), leaving the no-slip problem for future work. 

\subsection{The Limit of Small $\Pe$: Asymptotic Solution}
In the limit of small $\Pe$ when $|\vel| \ll 1$ we can linearize the Euler--Lagrange equations:
\begin{eqnarray}
- \mu \, \Delta \vel + \bnabla p &=& (\theta +\phi) \zhat, \label{l1S2} \\
\Delta \phi + w &=& 0, \label{l2S2} \\
\Delta \theta + w &=& 0, \label{l3S2}\\
\bnabla \bcdot \vel &=&  0. \label{l4S2}
\end{eqnarray}
Subtracting (\ref{l3S2}) from (\ref{l2S2}) and using (\ref{bc2S2})--(\ref{bc3S2}) establishes that $\theta=\phi$ in the small--$\Pe$ regime.
Then eliminating $p$ from (\ref{l1S2}) results in
\begin{equation}
- \mu \Delta \Delta w = \theta_{xx}+\phi_{xx} = 2 \, \theta_{xx}, \label{13S22}
\end{equation}     
which along with equations~\eqref{l3S2}--\eqref{l4S2} and boundary conditions (\ref{bc1S2})--(\ref{bc2S2}) and (\ref{freeslip}) can be solved analytically to find ($\vel$,$\theta$) in this small--$\Pe$ limit. Following the same procedure and notation as used in \S\ref{sec:smallPe} (see Appendix~\ref{app:ls}), we obtain    
\begin{eqnarray}
\Nu = 1 + \frac{k^2}{(m^2\upi^2+k^2)^{3}} \, \Pe^2, \label{smallPeBS2}
\end{eqnarray}
which, for given values of $\Pe$ and $\varGamma=\upi/k$, is maximized at $m=1$.
As a result, using the notation defined in \S\ref{sec:goal},
\begin{eqnarray}
\Nu_\m(\Pe,\varGamma) = 1 + \frac{\varGamma^4}{\upi^4(\varGamma^2+1)^{3}} \, \Pe^2. \label{smallPeS2}
\end{eqnarray}
The largest value of $\Nu_\m(\Pe,\varGamma)$ is achieved at $\varGamma_{\mathrm{opt}}=\sqrt{2}$: 
\begin{eqnarray}
\Nu_\M(\Pe)=1+\frac{\Pe^2}{(27\upi^4/4)}. \label{smallPemaxS2}
\end{eqnarray} 
Maximum transport in the limit of small $\Pe$ is achieved by an array of cells with aspect ratio $\varGamma_{\mathrm{opt}}=\sqrt{2}$ \citep[see][figure~3.1]{PedramMA}.
This flow resembles Rayleigh--B\'enard convection in a pure fluid layer (with stress-free boundary conditions) at the onset of linear instability \citep{Drazin}.
The factor $27\upi^4/4$ in (\ref{smallPemaxS2}) is, in fact, precisely the critical $\Ra$ of the instability.    

\subsection{Small to Large $\Pe$: Numerical Solution}\label{sec:numresS2}
Following the same steps as in the fixed-energy case, and using $\omega = \Delta \psi$, (\ref{duS2})--(\ref{div2S2}) simplify to
\begin{eqnarray}
\mathrm{J}(\theta,\phi) - \mu \, \Delta \omega + (\theta+\phi)_x &=& 0, \label{N1S2} \\
\Delta \psi - \omega &=& 0,  \label{psiomega} \\
-\mathrm{J}(\psi,\theta) - \Delta \theta + \psi_x &=& 0, \label{N3S2} \\ 
-\mathrm{J}(\psi,\phi) + \Delta \phi - \psi_x &=& 0, \label{N2S2} 
\end{eqnarray}  
and boundary conditions (\ref{bc1S2})--(\ref{bc3S2}) and (\ref{freeslip}) become
\begin{eqnarray}
\psi(x,0) = \psi(x,1) &=& 0, \\ 
\omega(x,0) = \omega(x,1) &=& 0, \\ 
\theta(x,0) = \theta(x,1) &=& 0, \\
\phi(x,0) = \phi(x,1) &=& 0, 
\end{eqnarray}
where $\omega$ has been introduced to avoid the occurrence of fourth order derivatives and to simplify the implementation of boundary conditions. 

Using the same continuation algorithm as given in \S\ref{sec:cont}, and following the steps described in \S\ref{sec:nummethod} and Appendix~\ref{app:NK}, equations~(\ref{N1S2})--(\ref{N2S2}) become 
\begin{eqnarray}\nonumber
 \begin{bmatrix}
  \mu \mathsfbi{\Delta}  & -\mathsfbi{I} & \mathsfbi{O} & \mathsfbi{O} \\
  \mathsfbi{O} &  \mu \mathsfbi{\Delta}  & -(\mathsfbi{I}+\boldsymbol{\phi}^N_z)\mathsfbi{D}_x + \boldsymbol{\phi}^N_x \mathsfbi{D}_z & -(\mathsfbi{I}-\boldsymbol{\theta}^N_z)\mathsfbi{D}_x - \boldsymbol{\theta}^N_x \mathsfbi{D}_z \\
 -(\mathsfbi{I}-\boldsymbol{\theta}^N_z)\mathsfbi{D}_x - \boldsymbol{\theta}^N_x \mathsfbi{D}_z & \mathsfbi{O} & \mathsfbi{\Delta} - \boldsymbol{\psi}^N_z \mathsfbi{D}_x + \boldsymbol{\psi}^N_x \mathsfbi{D}_z & \mathsfbi{O} \\
 -(\mathsfbi{I}+\boldsymbol{\phi}^N_z)\mathsfbi{D}_x + \boldsymbol{\phi}^N_x \mathsfbi{D}_z & \mathsfbi{O} & \mathsfbi{O} & \mathsfbi{\Delta} + \boldsymbol{\psi}^N_z \mathsfbi{D}_x - \boldsymbol{\psi}^N_x \mathsfbi{D}_z \\
 \end{bmatrix}
 \begin{bmatrix}
   \udelta \boldsymbol{\psi} \\
   \udelta \boldsymbol{\omega} \\
  \udelta \boldsymbol{\theta} \\
\udelta \boldsymbol{\phi}
 \end{bmatrix}
&& \\ =
 \begin{bmatrix}
  - \mathsfbi{\Delta} \boldsymbol{\psi}^N + \boldsymbol{\omega}^N  \\
   - \mu \mathsfbi{\Delta} \boldsymbol{\omega}^N +(\mathsfbi{I}+\boldsymbol{\phi}^N_z)\boldsymbol{\theta}^N_x +(\mathsfbi{I}-\boldsymbol{\theta}^N_z)\boldsymbol{\phi}^N_x \\
   -  \mathsfbi{\Delta} \boldsymbol{\theta}^N +(\mathsfbi{I}-\boldsymbol{\theta}^N_z)\boldsymbol{\psi}^N_x + \boldsymbol{\psi}^N_z \boldsymbol{\theta}^N_x\\
   -  \mathsfbi{\Delta} \boldsymbol{\phi}^N +(\mathsfbi{I}+\boldsymbol{\phi}^N_z)\boldsymbol{\psi}^N_x - \boldsymbol{\psi}^N_z \boldsymbol{\phi}^N_x\\
 \end{bmatrix} \hspace{1cm}
&& \label{bigmat2}
\end{eqnarray}
The details of the matrix algebra and boundary condition implementation are the same as before (see \S\ref{sec:nummethod}). The results presented here were obtained using $M=61$ or $81$.  As before, the iterative solution always converged in less than $6$ iterations.

\subsubsection{Numerical Results}

\begin{figure}
\centering
  \subfloat[]{\label{fig:res0psidS2}\includegraphics[width=.45\textwidth]{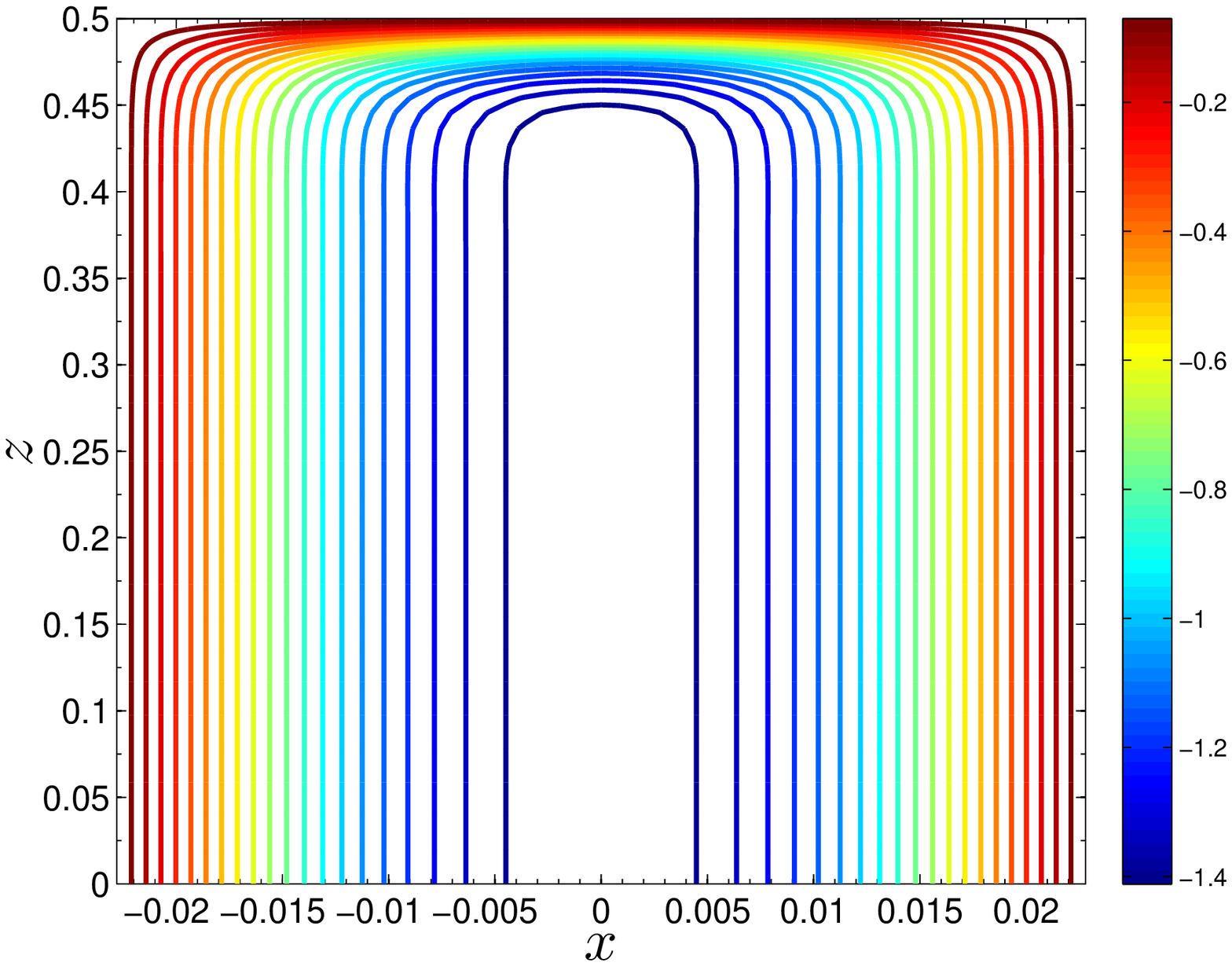}}  
  \subfloat[]{\label{fig:res0thtdS2}\includegraphics[width=.45\textwidth]{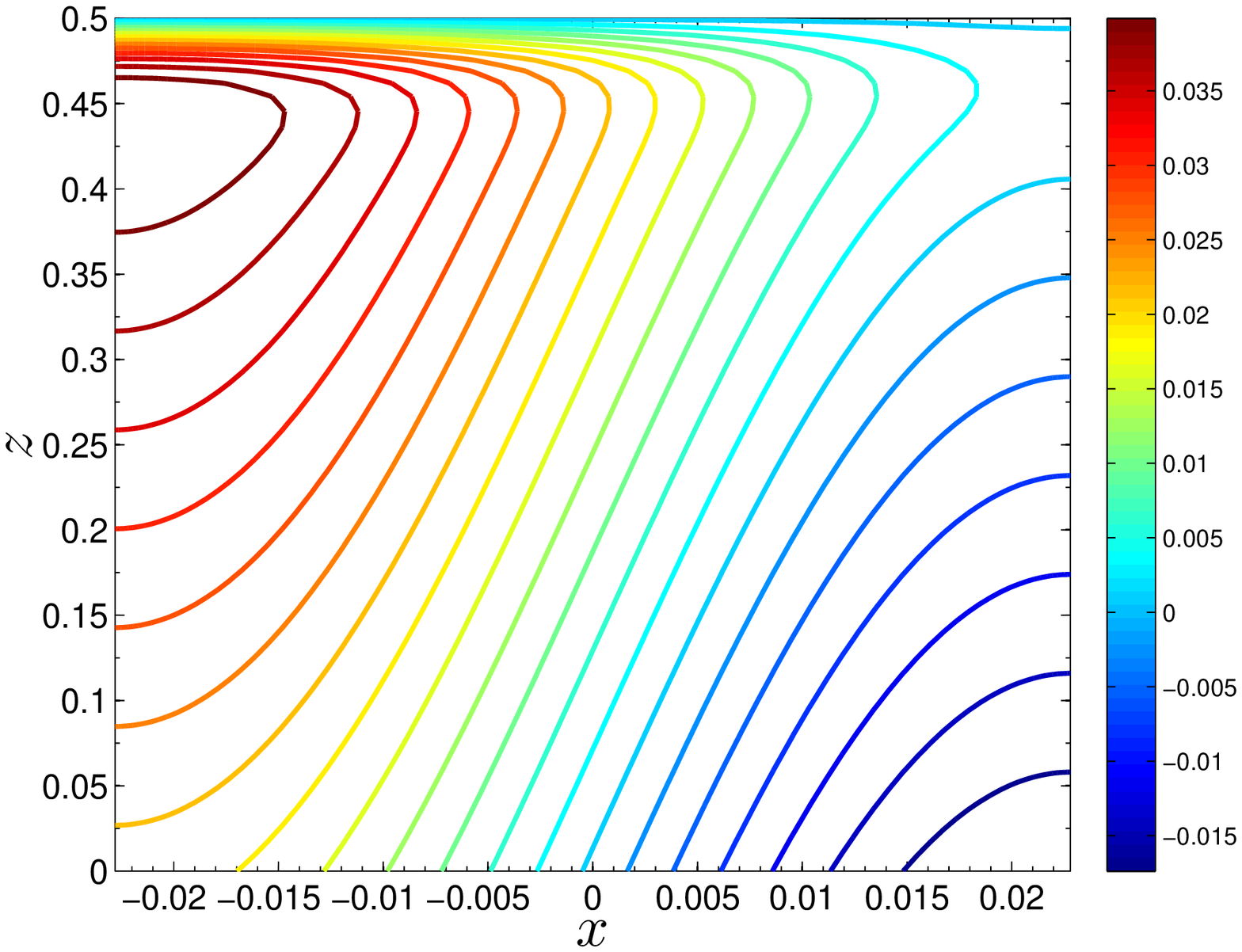}}  
\\           
  \subfloat[]{\label{fig:res0psieS2}\includegraphics[width=.45\textwidth]{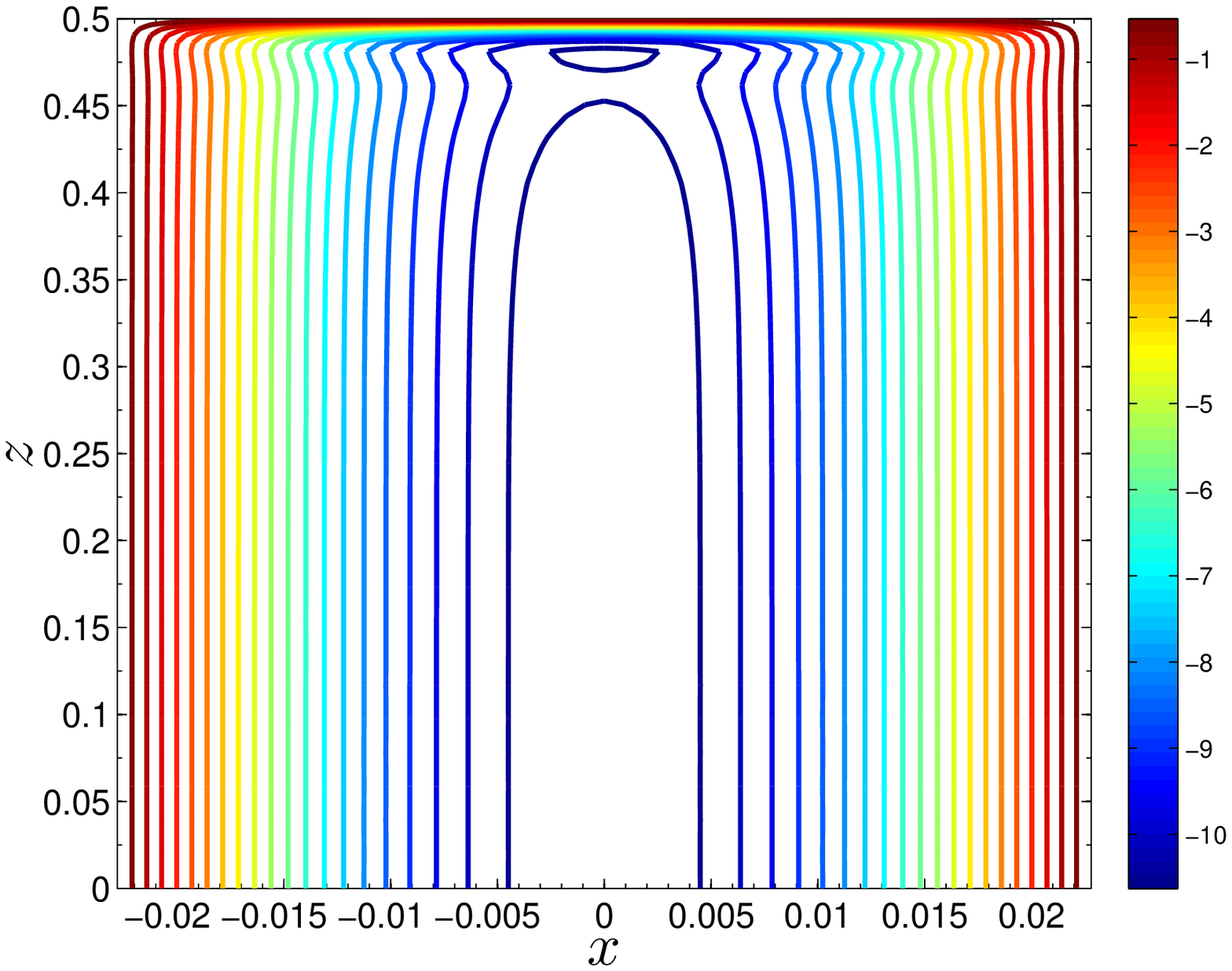}}  
  \subfloat[]{\label{fig:res0thteS2}\includegraphics[width=.45\textwidth]{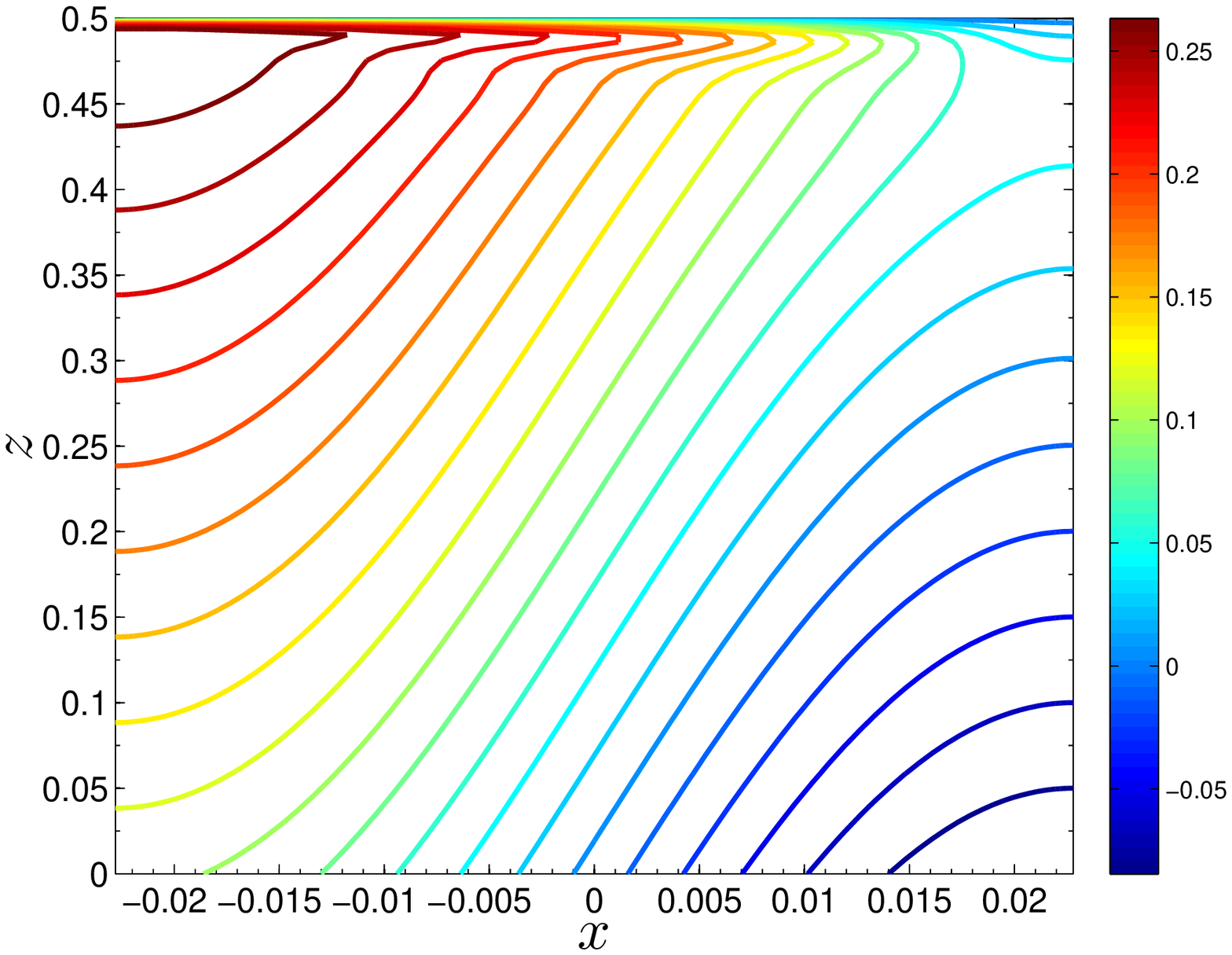}}  
\\           
  \subfloat[]{\label{fig:res0psibS2}\includegraphics[width=.45\textwidth]{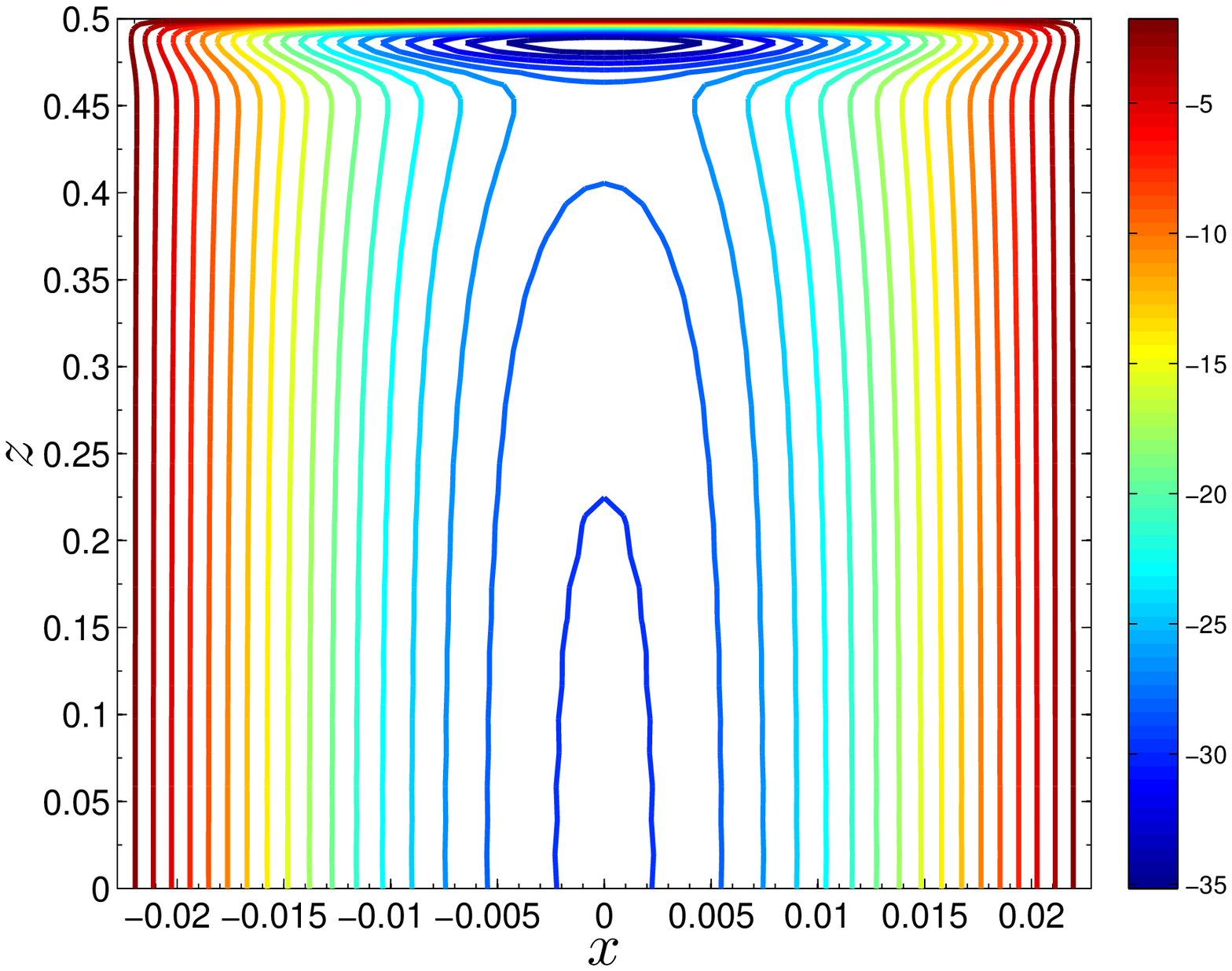}}  
  \subfloat[]{\label{fig:res0thtbS2}\includegraphics[width=.45\textwidth]{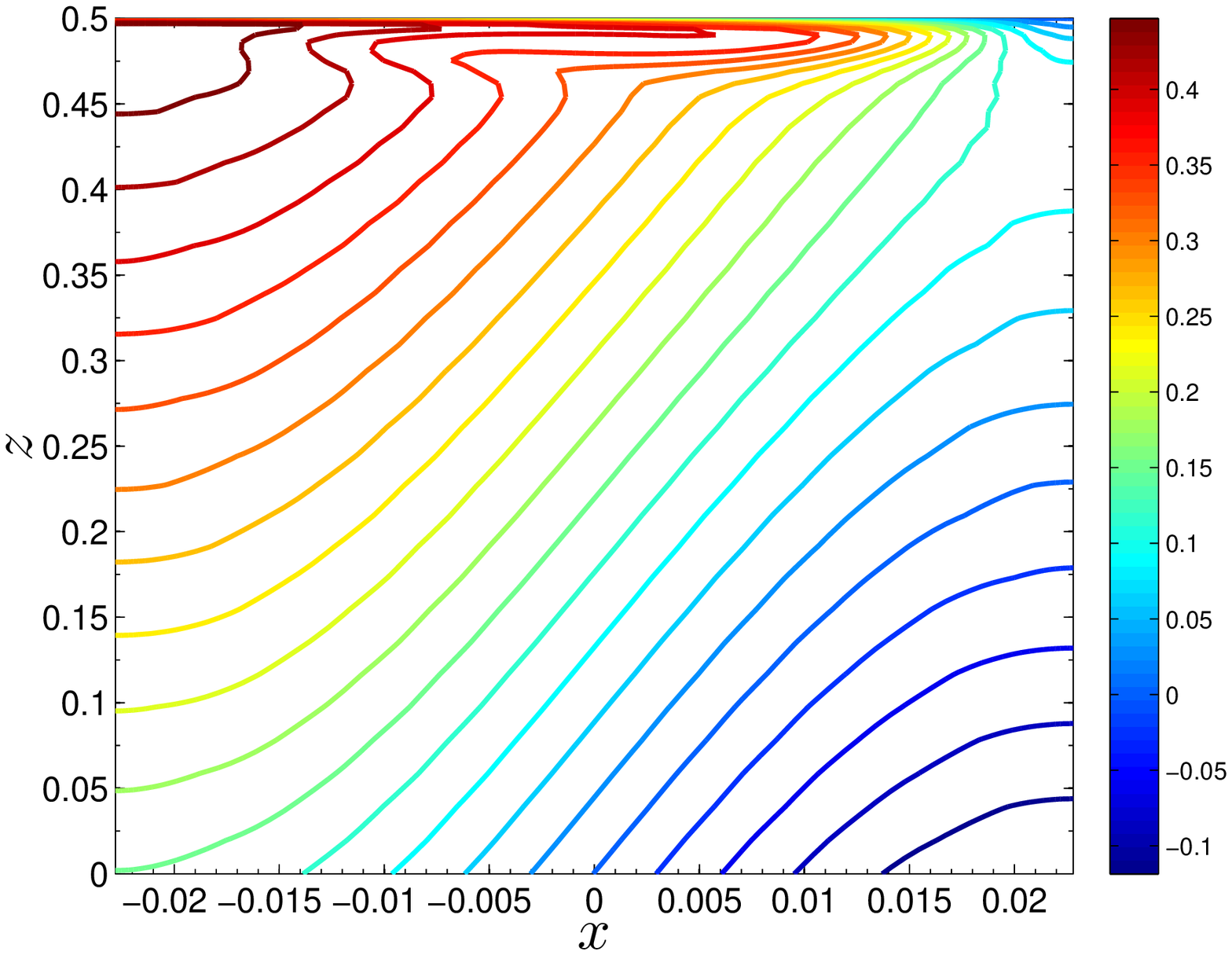}}              
  \caption{Evolution of the flow field with $\Pe$ for the case with $\varGamma=\sqrt{2}/\upi^3$. Panels on the left show $\psi$ and panels on the right show $\theta$ (only the upper half of the domain is shown for better illustration of the circulation zone). (a) and (b) $\Pe=4889.1$, $\Nu_\m=1.98$; (c) and (d) $\Pe=3.97 \times 10^4$, $\Nu_\m=40.1$; (e) and (f) $\Pe=1.43 \times 10^5$, $\Nu_\m=175.6$. The resolution is $81^2$.}
  \label{fig:res0S2}
\end{figure}

Figure \ref{fig:res0S2} shows  $\psi$ and $\theta$ for the case with $\varGamma=\sqrt{2}/\upi^3$ for low to high values of $\Pe$.
$\Nu_\m$ increases with $\Pe$, and the flow structure changes.
The enhancement of the heat transport is associated with the development of the boundary layers: the boundary layers thin as $\Pe$ increases and result in larger $\Nu_\m$.
However, in contrast to the fixed-energy problem a (re)circulation zone emerges between the boundary layers and the bulk at large values of $\Pe$.

\begin{figure}
\centering
  \subfloat[]{\label{fig:res1psiS2}\includegraphics[width=.45\textwidth]{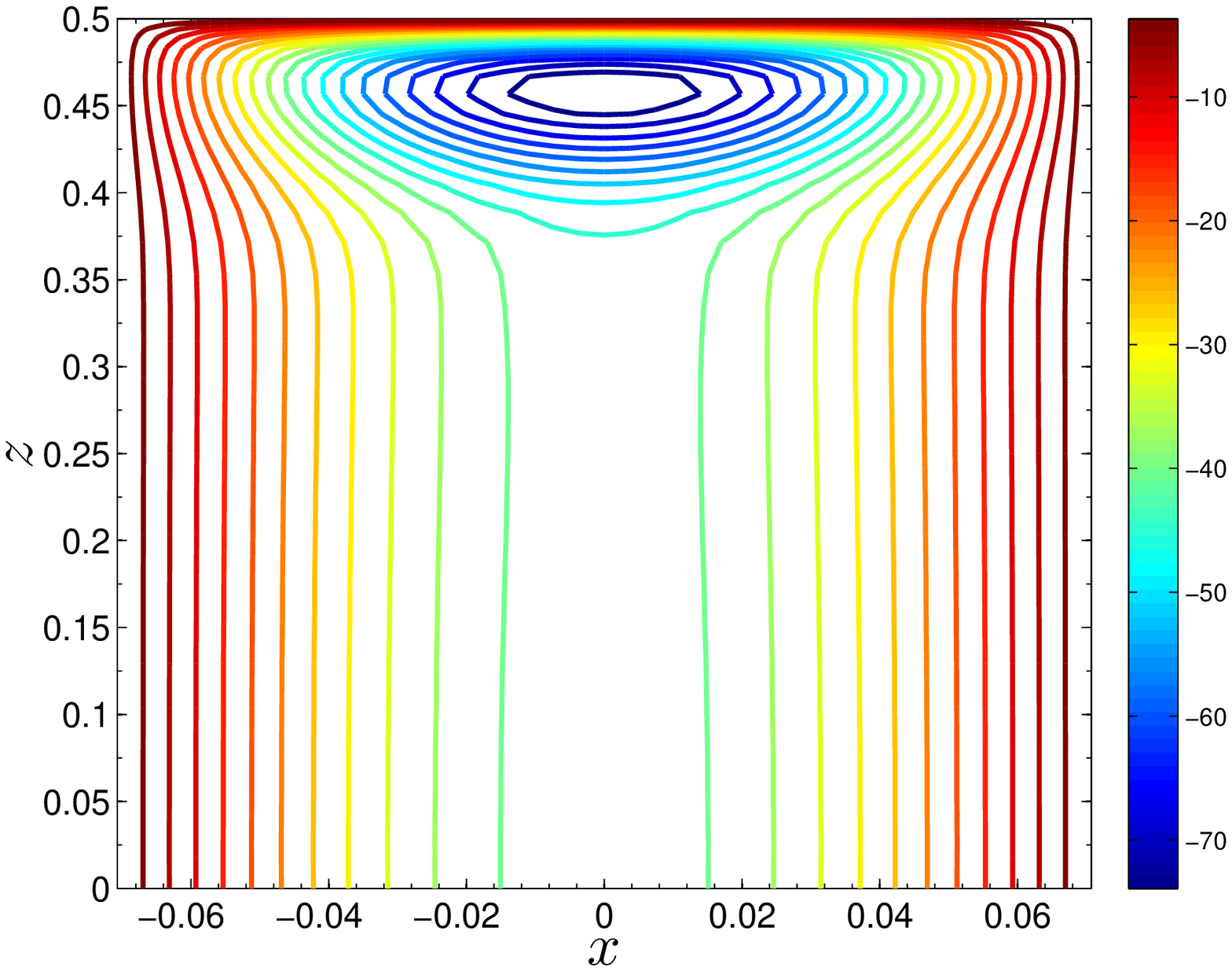}}  
  \subfloat[]{\label{fig:res1psi2S2}\includegraphics[width=.45\textwidth]{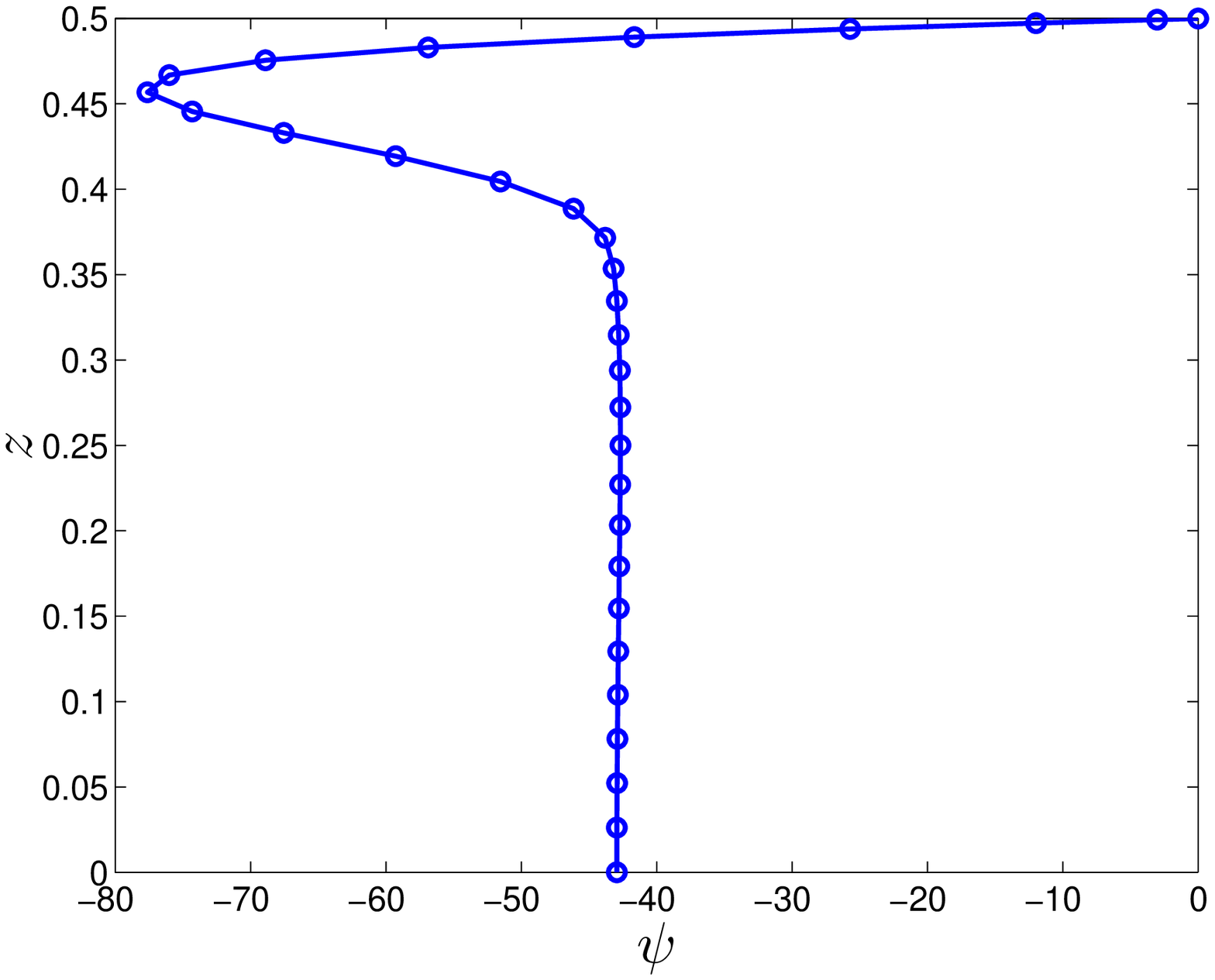}}  
\\           
  \subfloat[]{\label{fig:res1thtS2}\includegraphics[width=.45\textwidth]{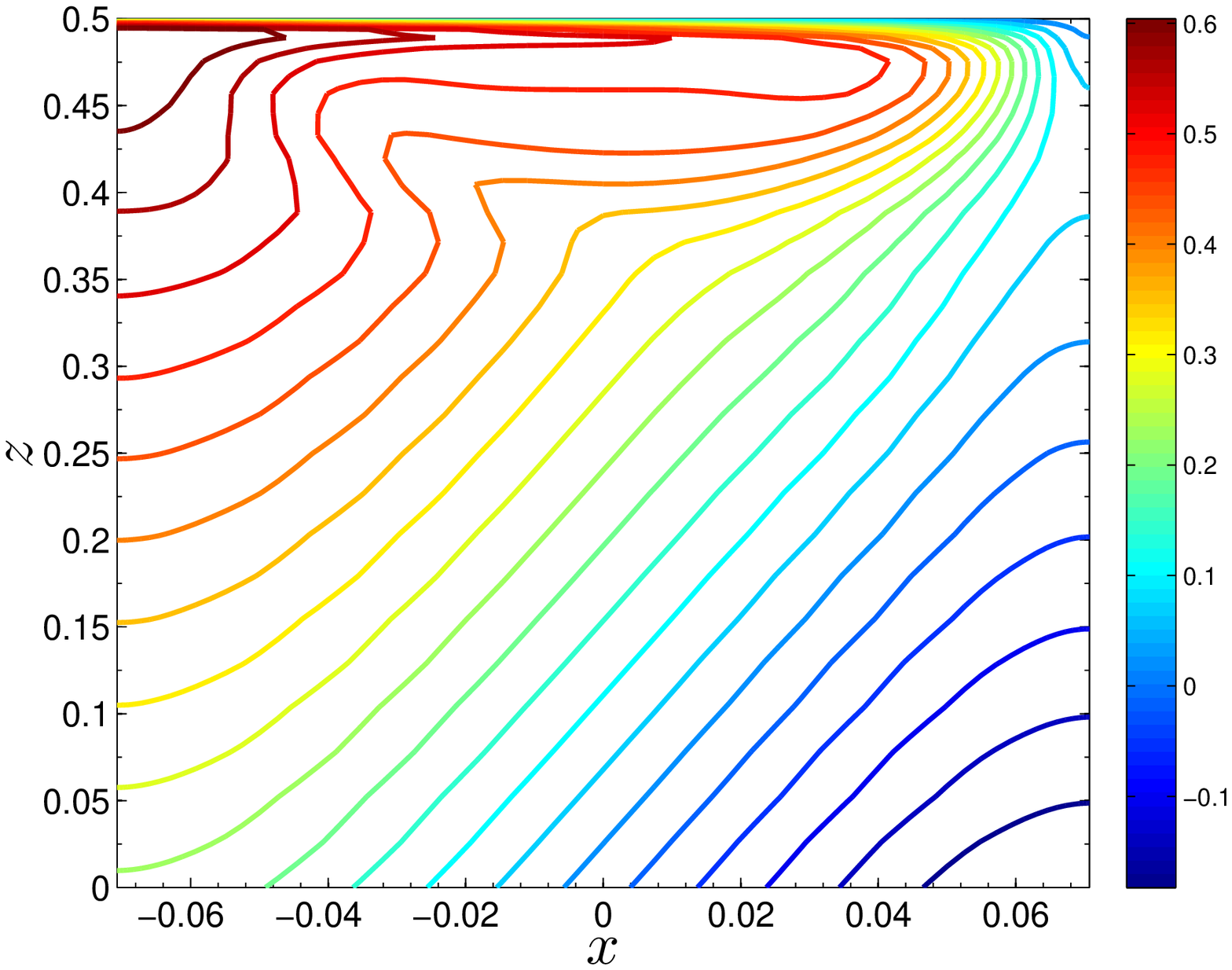}}   
  \subfloat[]{\label{fig:res1phiS2}\includegraphics[width=.45\textwidth]{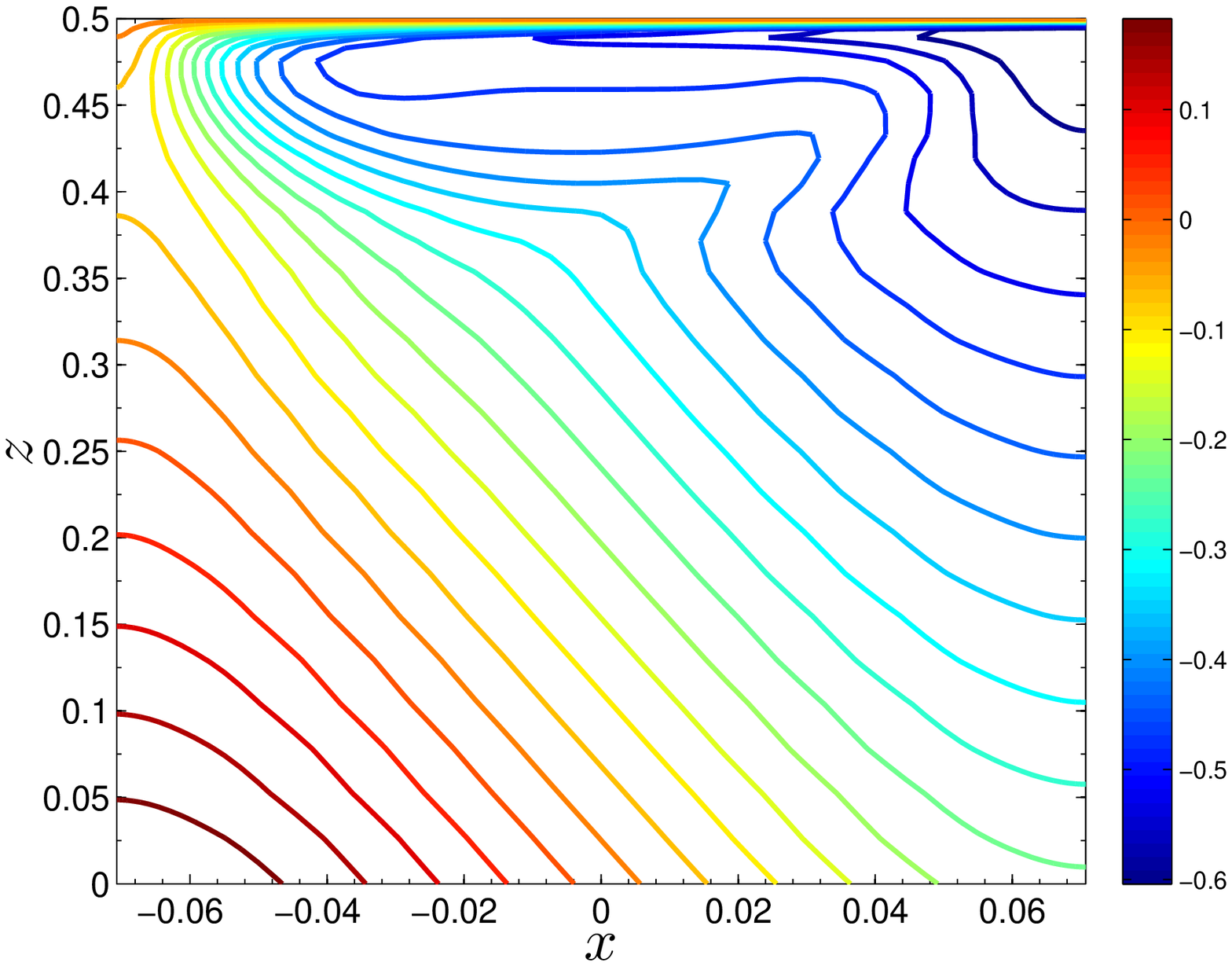}} 
\\           
  \subfloat[]{\label{fig:res1xiS2}\includegraphics[width=.45\textwidth]{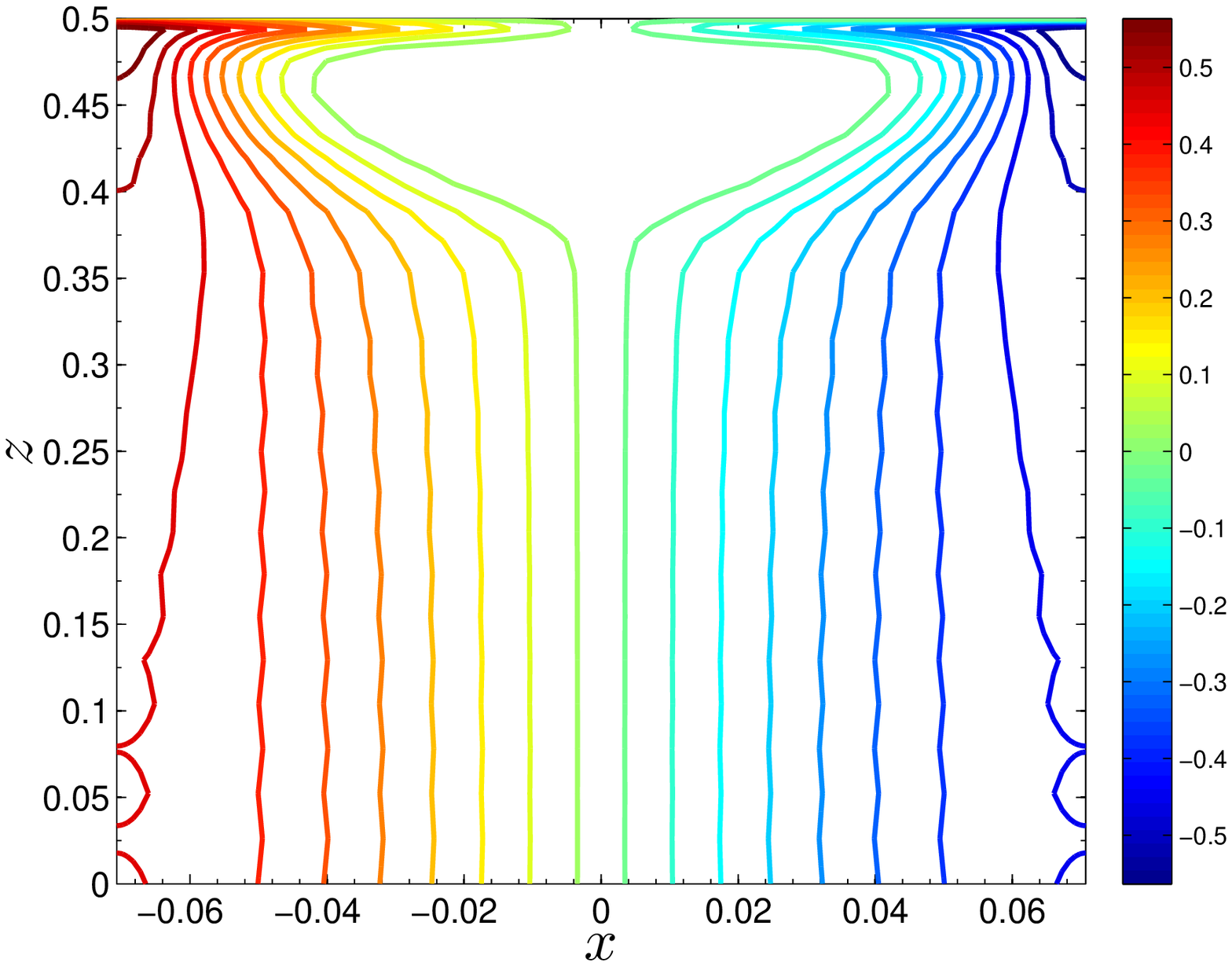}}   
  \subfloat[]{\label{fig:res1etaS2}\includegraphics[width=.45\textwidth]{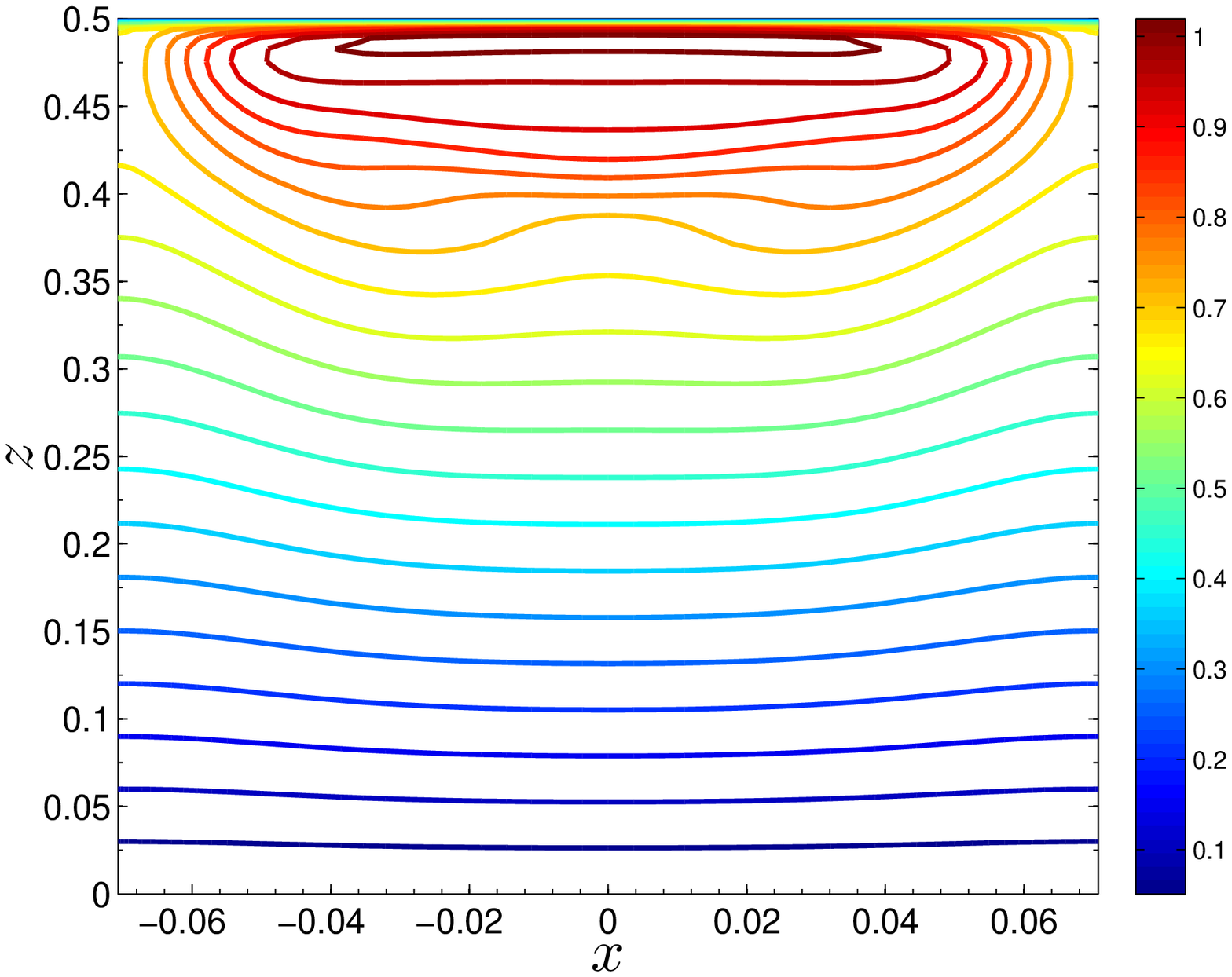}}             
  \caption{Flow field for $\varGamma=\sqrt{2}/\upi^2$, $\mu=4.83 \times 10^{-8}$, $\Pe=3.57 \times 10^4$, and $\Nu_\m=103.3.7$. Only the upper half of the domain is shown for better illustration of the circulation zone. (a) $\psi$, (b) $\psi$ along $x=0$, (c) $\theta$, (d) $\phi$, (e) $\xi \equiv \theta+\phi$, and (f) $\eta \equiv \theta - \phi$. The resolution is $61^2$.}
  \label{fig:res1S2}
\end{figure}

The high-$\Pe$ circulation zone complicates the flow structure for $\xi$ and $\eta$ as well.
Figure~\ref{fig:res1S2}
shows $(\psi,\theta,\phi,\xi,\eta)$ for $\varGamma =\sqrt{2}/\upi^2$ in the limit of large $\Pe$. (The small scale wiggles in the level sets are due in part to insufficient plotting resolution.)
The optimal flow field for the fixed enstrophy problem is thus more complicated than the optimal flow field for the fixed energy problem, mainly due to the presence of the circulation zone.  Nevertheless,
the bulk flows in the two problems are similar: $\psi$ and $\xi$ are nearly independent of $z$ and have a single mode dependence on $x$.
$\eta$ appears to be linear in $z$ as before and nearly $x$--independent. Appendix~\ref{app:isS2} presents the interior solution for this problem:
\begin{eqnarray}
\bar{\xi} &=& \pm \bar{\xi}_o \, \sin{(\upi \, x/\varGamma)},  \\
\bar{\eta}_o &=& 2-\left(\frac{\upi}{\varGamma}\right)^2 \, \sqrt{2\mu},   \\
\bar{\psi} &=& \frac{\pm\bar{\xi}_o}{(\upi/\varGamma)\sqrt{2\mu}} \, \cos{(\upi \, x/\varGamma)},  
\end{eqnarray}
which agrees with the numerical results.
This solution is determined up to an unknown constant $\bar{\xi}_o$ that should be determinable from the boundary layer solution, but owing to the complexity of this flow we have not yet succeeded in solving the boundary layer equations to complete the required matched asymptotic analysis.

\begin{figure}
\centering
  \includegraphics[width=1\textwidth]{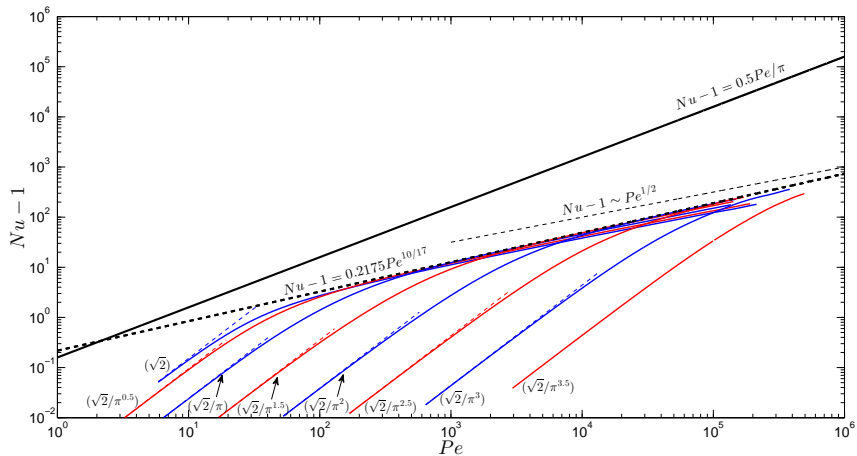}          
\caption{The numerically obtained $\Nu_\m$ as a function of $\Pe$ for various values of $\varGamma$ (blue and red lines). The labels show $(\varGamma)$. For each case, the short dashed line of the same color, visible for most cases, shows the analytical $\Nu_\m$ (\ref{smallPeS2}) in the small--$\Pe$ limit.
The thick black solid line shows the absolute upper bound (\ref{NuPe2}), and the thick black dashed line shows a fit to the envelope (i.e., $\Nu_\M$; see equation~(\ref{NuMPeS2})). The thin black dashed line indicates the $\Pe^{1/2}$ slope. All numerical results were obtained by continuing the linear solutions with $m=1$, and all results shown here have resolution $M=61$. Using a higher resolution $M=81$ results in negligible changes to the plot.}
\label{fig:MapNumS2}
\end{figure}

In the absence of an analytical solution we use the numerical results to find $\Nu_\m(\Pe,\varGamma)$ and $\Nu_\M(\Pe)$.
Figure \ref{fig:MapNumS2} is a plot of the numerically calculated $\Nu_\m(\Pe,\varGamma)$ for several values of $\varGamma$. Several conclusions drawn from the results 
presented in figure~\ref{fig:MapNum} also hold for the fixed-enstrophy problem: 
$\Nu_\m$ agrees with (\ref{smallPeS2}) in the limit of small $\Pe$ (benchmarking the code), 
and the absolute upper bound (\ref{NuPe2}) quantitatively overestimates transport albeit in the P\'eclet number scaling in this case.
For fixed $\varGamma$ and large $\Pe$, we observe that
\begin{eqnarray}
\Nu_\m(\Pe,\varGamma) = 1 + K(\varGamma) \, \Pe^{1/2}, \label{NumPeS2}
\end{eqnarray}
where $K(\varGamma)$ is a prefactor that can be determined from the numerical results.
A fit to the envelope made by the largest values of $\Nu_\m$ gives 
\begin{eqnarray}
\Nu_\M(\Pe) = 1+ 0.2175 \, \Pe^{0.58}.  \label{NuMPeS2}   
\end{eqnarray}
For reasons that will become clear in the next subsection immediately below, we conjecture that the measured exponent $0.58$ in \eqref{NuMPeS2} is really $10/17 = 0.5882 \dots$ for large values of $\Pe$. 

More data points in $\varGamma$, especially for smaller $\varGamma$, are needed to determine $\varGamma_{\mathrm{opt}}(\Pe)$ accurately from the numerical results.
Using just three points in the wide range of $\Pe = 1701-4.1 \times 10^4$, however, we obtain estimates $-0.361$ and $-0.358$ for the exponent of $\Pe$ in the scaling of $\varGamma_{\mathrm{opt}}(\Pe)$. 
For the same reasons alluded to above (that will becomes clear below) we conjecture that this exponent is really $6/17 = .3529 \dots$ for large $\Pe$.

\subsection{Example: Application to Rayleigh--B\'enard Convection}\label{sec:RB}
Classical Rayleigh--B\'enard convection in a layer of pure fluid heated from below and cooled from above is, at least in the steady case, an example of transport with fixed enstrophy.
\cite{Rayleigh1916} modeled this problem with the Boussinesq equations     
\begin{eqnarray}
\bnabla \bcdot \vel &=& 0, \label{DivRB}\\
\frac{1}{\Pran} \, (\dot{\vel} + \vel \bcdot \bnabla \vel) &=& - \bnabla p + \Delta \vel + \Ra\, T \zhat, \label{BoussRB} \\
\dot{T} + \vel \bcdot \bnabla T &=& \Delta T, \label{adRB}
\end{eqnarray}
where $\Pran$ is the Prandtl number. 
Taking the inner product of (\ref{BoussRB}) with $\vel$ and integrating over long time and over a domain with impermeable walls yields
\begin{eqnarray}
0 = -\left< |\bnabla \vel|^2 \right> + \Ra \, \left< w T \right>.
\end{eqnarray}
Using the definition (\ref{en}) of $\Pe$ for fixed enstrophy problems and the definition (\ref{Nu}) of $\Nu$ from \S\ref{sec:math},
\begin{eqnarray}
\Pe^2 = \Ra \, (\Nu-1). \label{RaNuPeS2}
\end{eqnarray}
The Nusselt number $\Nu$ is a function of $\Ra$ and aspect-ratio $\varGamma$ and as a result $\Pe$ is fixed for given values of these parameters.
(As before the control parameter $\Ra$ depends on the fluid properties and the imposed temperature difference between the walls, but not on the flow.)
Hence, steady Rayleigh--B\'{e}nard convection in a pure fluid layer occurs with fixed enstrophy.  

Employing (\ref{RaNuPeS2}) to replace $\Pe$ with $\Ra$ in (\ref{NumPeS2}) and (\ref{NuMPeS2}), the bounds are
\begin{eqnarray}
\Nu_\m(\Ra,\varGamma) &=& 1 + (K(\varGamma))^{4/3} \, \Ra^{1/3}, \label{NumRaS2}\\
\Nu_\M(\Ra) &=& 1+ 0.1152 \, \Ra^{5/12}. \label{NuMRaS2}
\end{eqnarray}
Furthermore, using the rather crude estimates for the exponent of $\Pe$ in $\varGamma_\mathrm{opt}(\Pe)$ gives 
\begin{eqnarray}
\varGamma_\mathrm{opt}(\Ra) \sim \Ra^{-0.2546}.
\end{eqnarray}
Interestingly, $\varGamma \sim k^{-1} \sim \Ra^{-0.25}$ is the scaling of the shortest-wavelength unstable mode for Rayleigh--B\'enard convection in pure fluids with stress-free boundaries, which corresponds to the conjecture that $\varGamma_\mathrm{opt}(\Pe) \sim \Pe^{-6/17}$.

Table~\ref{tab:RB} compares the bounds discovered here with the results of other analyses of Rayleigh--B\'enard convection.
The classical marginally stable boundary layer argument of \citet{Malkus} and \citet{Howard} predicts $\Nu \sim \Ra^{1/3}$ uniformly in the Prandtl number.
On the other hand, the argument by \citet{Ed} based on the assumption that transport is limited by the ballistic motions across the bulk (see also \citet{Kra})
suggests that $\Nu \sim (\Pran \, \Ra)^{1/2}$.
While both of these arguments are independent of the spatial dimension and velocity boundary conditions, \citet{Whitehead11,Whitehead12} recently used the background method to prove that $\Nu \lesssim \Ra^{5/12}$ uniformly in $\Pran$ for stress-free boundaries for steady or unsteady two-dimensional Rayleigh--B\'enard  convection (see also \cite{Otero02}) and for three-dimensional Rayleigh--B\'enard convection between stress-free boundaries at infinite $\Pran$ (see also \cite{Ier}).

As for the fixed-energy problem, $\Nu_\M(\Ra)$ obtained using the approach adopted in this paper has the same scaling in $\Ra$ as the upper bounds obtained using the background method.
For fixed $\varGamma$, \citet{Greg} analyzed steady unicellular Rayleigh--B\'enard convection and found that $\Nu \sim \Ra^{1/3}$ in agreement with the scaling of $\Nu_\m(\Ra,\varGamma)$ with $\Ra$ observed here.
This agreement suggests that steady Rayleigh--B\'enard convection in pure fluids in a cell of fixed aspect ratio (and with stress-free boundaries) transports as much, e.g., heat as any steady flow with a given amount of enstrophy, modulo prefactor.

\begin{table}
\begin{center}
\def~{\hphantom{0}}
\begin{tabular}{lcc}
                     & {$\Nu(\Ra,\Pran)$} & {$\Nu(\Ra,\varGamma_{\mathrm{fixed}})$} \\ [3pt]
{\underline{Classical theories}} & & \\ 
\citet{Malkus,Howard} & {$\sim C \, \Ra^{1/3}$} & \\ 
\citet{Ed,Kra} & {$\sim C \, (\Pran \, \Ra)^{1/2}$} & \\ 
&& \\
{\underline{Background method (upper bounds)}} & & \\
{\citet{Ier}: numerical, infinite $\Pran$} & {$\le C \, \Ra^{5/12}$} &   \\
{\citet{Otero02}: 2D numerical, finite $\Pran$} & {$\le 0.142 \, \Ra^{5/12}$} & \\
{\citet{Whitehead11}:} & {$\le 0.289 \, \Ra^{5/12}$}  &  \\
analytical, 2D finite $\Pran$ and 3D infinite $\Pran$ && \\
&& \\
{\underline{Steady unicellular analysis}} & & \\
{\citet{Greg}:  asymptotic, finite $\Pran$} & & {$\sim C(\varGamma) \, \Ra^{1/3}$}  \\  
&& \\
{\underline{Current work}} & &  \\  
{Numerical} & {$\le 1 + 0.115 \, \Ra^{5/12}$} &  {$\le 1 + ({K}(\varGamma))^{4/3} \, \Ra^{1/3}$}  \\ 
\end{tabular}
\caption{Comparison of the results of the current work with the scalings for Rayleigh--B\'enard convection in pure fluids with stress-free boundary conditions obtained using various other methods.}
\label{tab:RB}
\end{center}
\end{table}

\section{Summary and Discussions}\label{sec:conc}
In this work we have addressed some fundamental problems in fluid mechanics, namely how much heat can be transported between impermeable fixed-temperature walls by steady incompressible flows with a given amount of kinetic energy or enstrophy, and what the optimal flows look like.
We employed the calculus of variations to find flows that maximize the heat transport between the walls, answering the above questions for steady 2D flows with, in the fixed-enstrophy case, stress-free walls. For small energy or enstrophy, the resulting nonlinear Euler--Lagrange equations were linearized and solved analytically, revealing that the optimal flows are arrays of convection cells.
For larger energy or enstrophy budgets we solved the Euler--Lagrange equations using numerical continuation in a cell of a given aspect ratio $\varGamma$.
For the problem with fixed kinetic energy we were able to exploit the symmetries in the optimal flow to analytically solve the fully nonlinear equations using matched asymptotic analysis.
The analytical and numerical results agree remarkably well. Our results are presented in terms of the Nusselt number $\Nu$ and P\'eclect number $\Pe$.
We found that $\Nu_\M \sim \Pe$ for the fixed energy problem and $\Nu_\M \sim \Pe^{10/17}$ when the enstrophy is specified.

For each of the two primary flow-intensity constraints we have observed that there is a classical buoyancy-driven flow that satisfies the given constraint.
We have thus interpreted our results in terms of Rayleigh number $\Ra$ to see how the optimal transport compares with available upper bounds for Rayleigh--B\'enard convection.
In porous medium convection, which occurs at (mean) energy fixed by $\Nu$ and $\Ra$, we found $\Nu_\M \sim \Ra$ and $\varGamma_\mathrm{opt} \sim \Ra^{-1/2}$.
For Rayleigh--B\'enard convection in pure fluids, an example of fixed (mean) enstrophy transport, we found $\Nu_\M \sim \Ra^{5/12}$ and $\varGamma_\mathrm{opt} \sim \Ra^{-1/4}$.
The Rayleigh number scaling of the bounds on $\Nu$ agree with bounds derived by other methods for both problems, and also agree with the scaling found in the direct numerical simulations of porous media convection in the high--$\Ra$ regime.
Interestingly, for both problems the scalings of $\varGamma_\mathrm{opt}(\Ra)$ found here agree with the scalings of the shortest-wavelength unstable mode about the linear conduction profile (see also \cite{Wen}).



The results presented here offer new insights into steady optimal transport in two spatial dimensions, but several lines of research remain to be pursued to complete our understanding of the limits on fluid dynamical transport between impermeable walls. For example, the fixed enstrophy problem with no-slip boundary conditions is more challenging but of greater applicability for many applications and for laboratory experiments.
Indeed, the best known rigorous bounds for arbitrary Prandtl number Rayleigh--B\'enard convection between no-slip boundaries is $\Nu \lesssim \Ra^{1/2}$ (\cite{Howard63,DC1996,PK2003}) corresponding to $\Nu \sim \Pe^{2/3}$, but we do not know if this estimate is sharp.
Note that $2/3 > 10/17$ which leaves open the possibility that flow between no-slip boundaries may transport \emph{more} than flow between stress-free walls at the same bulk enstrophy level.
It is also natural to wonder if optimal transport in three dimensions takes place by three-dimensional flows.
Finally, it is necessary to determine the true optimal transport is realized by time-dependent, rather than steady, flows.
Unsteady optimal transport can be framed in the language of optimal control theory producing new challenges for both analysis and computational investigations. 

\section*{Acknowledgments}
Much of this work was completed at the 2012 Geophysical Fluid Dynamics (GFD) Program at Woods Hole Oceanographic Institution.
The GFD Program is supported by the US National Science Foundation (NSF) Award OCE-0824636 and the Office of Naval Research.
The Computational Fluid Dynamics Laboratory and the Department of Mechanical Engineering of UC Berkeley provided computational resources. 
This work was also supported by NASA Awards NXX10AB93G and NXX13AG56G (PH), and NSF Awards AST-1010046 and AST-1009907 (PH), PHY-0855335, DMS-0927587, and PHY-1205219 (CRD), and DMS-0928098 (GPC).
We gratefully acknowledge helpful discussions with C\'edric Beaume, Lindsey Corson, Duncan Hewitt, Genta Kawahara, and Felix Otto.

\appendix
\section{Equations for Newton-Kantorovich Iteration Scheme} \label{app:NK}
First we rewrite equations~(\ref{N1})--(\ref{N3}) as
\begin{eqnarray}
\Delta \psi &=& \mathrm{F}(\theta_x,\theta_z,\phi_x,\phi_z), \\ 
\Delta \theta &=& \mathrm{G}(\psi_x,\psi_z,\theta_x,\theta_z), \\ 
\Delta \phi&=& \mathrm{Q}(\psi_x,\psi_z,\phi_x,\phi_z), 
\end{eqnarray}   
where
\begin{eqnarray}
\mathrm{F} &=& \frac{1}{\mu} \, \left(-(1+\phi_z)\theta_x+(\theta_z-1)\phi_x \right), \\
\mathrm{G} &=& (1-\theta_z)\psi_x + \psi_z\theta_x, \\
\mathrm{Q} &=& (1+\phi_z)\psi_x - \psi_z \phi_x. 
\end{eqnarray}
Taylor expanding the nonlinear terms $\mathrm{F}$, $\mathrm{G}$, and $\mathrm{Q}$ about the solution of the $N$th iteration gives
\begin{eqnarray}
\Delta \psi^{N+1} &=& \mathrm{F}^N + \udelta \theta_x \, \mathrm{F}^N_{\theta_x} + \udelta \theta_z \, \mathrm{F}^N_{\theta_z} + \udelta \phi_x \, \mathrm{F}^N_{\phi_x} + \udelta \phi_z \, \mathrm{F}^N_{\phi_z} + \mathrm{h.o.t}, \label{dpsi} \\ 
\Delta \theta^{N+1} &=& \mathrm{G}^N + \udelta \psi_x \, \mathrm{G}^N_{\psi_x} + \udelta \psi_z \, \mathrm{G}^N_{\psi_z} + \udelta \theta_x \, \mathrm{G}^N_{\theta_x} + \udelta \theta_z \, \mathrm{G}^N_{\theta_z} + \mathrm{h.o.t}, \label{dtheta} \\
\Delta \phi^{N+1} &=& \mathrm{Q}^N + \udelta \psi_x \, \mathrm{Q}^N_{\psi_x} + \udelta \psi_z \, \mathrm{Q}^N_{\psi_z} + \udelta \phi_x \, \mathrm{Q}^N_{\phi_x} + \udelta \phi_z \, \mathrm{Q}^N_{\phi_z} + \mathrm{h.o.t},  \label{dphii}   
\end{eqnarray}
where the subscripts in $\mathrm{F}$, $\mathrm{G}$, and $\mathrm{Q}$ denote the Frechet derivatives (e.g., $F_{\psi_x} \equiv \partial F/\partial \psi_x$) and the superscript $N$ means evaluated at iteration $N$.
The ``$\udelta$'' of any quantity is the difference between its value at iterations $N+1$ and $N$ (e.g., $\udelta \psi \equiv \psi^{N+1}-\psi^{N}$).
The higher-order terms (h.o.t) are $O((\udelta (\cdot))^2)$ or smaller.

Calculating the Frechet derivatives and ignoring the higher-order terms, equations~(\ref{dpsi})--(\ref{dphii}) become
\begin{eqnarray}
\nonumber
\mu \Delta \udelta \psi + (1+\phi^N_z) \, \udelta \theta_x- \phi^N_x \, \udelta \theta_z + (1-\theta^N_z) \, \udelta \phi_x + \theta^N_x \, \udelta \phi_z &=& \\
 -\mu \Delta \psi^N - \left[(1+\phi^N_z)\theta^N_x + (1-\theta^N_z)\phi^N_x  \right], \label{e1} && \\
\nonumber
\Delta \udelta \theta - \psi^N_z \, \udelta \theta_x + \psi^N_x \, \udelta \theta_z - (1-\theta^N_z) \, \udelta \psi_x - \theta^N_x \, \udelta \psi_z &=& \\
 -\Delta \theta^N +(1-\theta^N_z)\psi^N_x + \psi^N_z \theta^N_x, \label{e2} &&\\
\nonumber
\Delta \udelta \phi - (1+\phi^N_z) \, \udelta \psi_x + \phi^N_x \, \udelta \psi_z + \psi^N_z \, \udelta \phi_x - \psi^N_x \, \udelta \phi_z &=& \\
 -\Delta \phi^N +(1+\phi^N_z)\psi^N_x - \psi^N_z \phi^N_x. \label{e3} &&
\end{eqnarray}

\section{Small--$\Pe$ Solution for the Fixed Enstrophy Problem} \label{app:ls}
A Fourier transform of equations~\eqref{l3S2} and \eqref{13S22} in the $x$ direction yields
\begin{eqnarray}
(\mathrm{D}_z^2-k^2) \, \hat{\theta}_k(z) + \hat{w}_k(z) &=& 0, \label{eq1S2} \\
-\mu (\mathrm{D}_z^2-k^2)^2 \, \hat{w}_k(z) + 2 k^2 \, \hat{\theta}_k(z) &=& 0, \label{eq2S2} 
\end{eqnarray}
with solutions of the form
\begin{eqnarray}
\hat{w}_k(z) &=& A_k \, \sin{(m \upi z)}, \label{wkS2} \\
\hat{\theta}_k(z) &=& B_k \, \sin{(m \upi z)}. \label{tkS2}
\end{eqnarray}     
Substitution into (\ref{eq1S2}) and (\ref{eq2S2}) gives
\begin{eqnarray}
\mu &=& (2 \, k^2)/(m^2\upi^2+k^2)^3, \label{muS2} \\
A_k &=& (m^2 \upi^2+k^2) \, B_k. \label{ABS2}
\end{eqnarray}
Then equation~(\ref{l4S2}) yields
\begin{eqnarray}
\hat{u}_k(z) &=& \mathrm{i} \, \frac{m \upi}{k} \, A_k \, \cos{(m \upi z)}. \label{ukS2}
\end{eqnarray} 

Substituting (\ref{ukS2}) and (\ref{wkS2}) into (\ref{dmuS2}) gives
\begin{equation}
\left< |\bnabla \vel|^2 \right> = \frac{1}{k^2} \left(m^2\upi^2 + k^2 \right)^2 A^2_k = \Pe^2 \Rightarrow A_k = \frac{k}{(m^2\upi^2+k^2)} \, \Pe, \label{AkS2}
\end{equation} 
which, combined with (\ref{ABS2}), yields
\begin{eqnarray}
B_k = \frac{k}{(m^2\upi^2+k^2)^{2}} \, \Pe. 
\label{BkS2}
\end{eqnarray}
The Nusselt number follows from (\ref{Nu}):
\begin{eqnarray}
\Nu = 1+A_k B_k = 1 + \frac{k^2}{(m^2\upi^2+k^2)^{3}} \, \Pe^2. \label{smallPeBS2A}
\end{eqnarray}

\section[Interior Solution]{Interior Solution for the Fixed Enstrophy Problem}\label{app:isS2}
Equations (\ref{N1S2})--(\ref{N2S2}) can be written in terms of $(\psi,\xi,\eta)$
\begin{eqnarray}
-\mathrm{J}(\xi,\eta) - 2 \mu \Delta^2 \psi + 2 \xi_x &=& 0, \label{F1S2} \\
\mathrm{J}(\psi,\xi) + \Delta \eta &=& 0,  \label{F2S2} \\
\mathrm{J}(\psi,\eta) + \Delta \xi - 2 \psi_x &=& 0,  \label{F3S2} 
\end{eqnarray} 
which except for $-2\mu \Delta^2$ instead of $+2\mu \Delta$ are the same as (\ref{F1})--(\ref{F3}). However, the higher derivative is expected to result in significant differences between the two problems. 

The numerical results suggest
\begin{eqnarray}
\psi &=& \bar{\psi}(x) \; A(x,z),\label{psibarS2}\\
\xi  &=& \bar{\xi}(x)  \; B(x,z), \label{xibarS2}\\
\eta &=& \bar{\eta}(z) \; C(x,z), \label{etabarS2}
\end{eqnarray}
where $(\bar{\psi},\bar{\xi},\bar{\eta})$ constitute the outer solution. 

Using ($\bar{\psi}(x)$,$\bar{\xi}(x)$,$\bar{\eta}(z)$) in (\ref{F1S2})--(\ref{F3S2}) gives
\begin{eqnarray}
2 \mu \, \bar{\psi}_{xxxx} - (2-\bar{\eta}_z) \, \bar{\xi}_x &=& 0, \label{A1S2} \\
\bar{\eta}_{zz} &=& 0, \label{A2S2} \\
\bar{\xi}_{xx} - (2-\bar{\eta}_{z}) \bar{\psi}_{x} &=& 0, \label{A3S2} 
\end{eqnarray}
which again implies that 
\begin{equation}
\bar{\eta}(z) = \bar{\eta}_o \, z, \label{etasolS2}
\end{equation}
where $\bar{\eta}_o$ is an unknown constant. Eliminating $\bar{\psi}$ between (\ref{A1S2}) and (\ref{A3S2}) yields
\begin{eqnarray}
\bar{\xi}_{xxxxx} - \left( \frac{\bar{\eta}_o-2}{\sqrt{2 \mu}} \right)^2 \, \bar{\xi}_{x} &=& 0. \label{xixS2} 
\end{eqnarray}  
Given the periodicity of $2 \varGamma$ in $x$, and $\xi_{x}(\pm \varGamma/2,z)=0$, this implies
\begin{eqnarray}
\bar{\xi} &=& \pm \bar{\xi}_o \, \sin{(\upi \, x/\varGamma)},  \label{xisolS2} \\
\bar{\eta}_o &=& 2-\left(\frac{\upi}{\varGamma}\right)^2 \, \sqrt{2\mu},  \label{eta_oS2}
\end{eqnarray}
where $\bar{\xi}_o>0$ is an unknown constant. Notice the difference between (\ref{eta_oS2}) and (\ref{eta_o}).

Equation~(\ref{A3S2}) yields
\begin{eqnarray}
\bar{\psi} &=& \frac{\pm\bar{\xi}_o}{(\upi/\varGamma)\sqrt{2\mu}} \, \cos{(\upi \, x/\varGamma)}.  \label{psisolS2}
\end{eqnarray}
As before, the interior flow field (i.e., the outer solution) is known up to an unknown constant $\bar{\xi}_o$, which is to be determined using the inner solution.

\bibliographystyle{jfm}

\bibliography{OptTr_v9}

\begin{thebibliography}{59}
\expandafter\ifx\csname natexlab\endcsname\relax\def\natexlab#1{#1}\fi

\bibitem[Boyd(2001)]{Boyd}
{\sc Boyd, J.P.} 2001 {\em Chebyshev and Fourier Spectral Methods\/}, 2nd edn.
  Dover.

\bibitem[Busse(1969)]{Busse69}
{\sc Busse, F.H.} 1969 On {H}oward's bound for heat transport by turbulent
  convection. {\em Journal of Fluid Mechanics\/} {\bf 37}, 457--477.

\bibitem[Busse(1970)]{Busse70}
{\sc Busse, F.H.} 1970 Bounds for turbulent shear flow. {\em Journal of Fluid
  Mechanics\/} {\bf 41}, 219--240.

\bibitem[Busse \& Joseph(1972)]{BJ1972}
{\sc Busse, F.H. \& Joseph, D.D.} 1972 Bounds for heat transport in a porous
  layer. {\em Journal of Fluid Mechanics\/} {\bf 54}, 521--543.

\bibitem[Caulfield \& Kerswell(2001)]{CP2001}
{\sc Caulfield, C.P. \& Kerswell, R.R.} 2001 Maximal mixing rate in turbulent
  stably stratified {C}ouette flow. {\em Physics of Fluids\/} {\bf 13},
  894--900.

\bibitem[Cheskidov {\em et~al.\/}(2007)Cheskidov, Petrov \& Doering]{CDP2007}
{\sc Cheskidov, A., Petrov, N.P. \& Doering, C.R.} 2007 Energy dissipation in
  fractal-forced flow. {\em Journal of Mathematical Physics\/} {\bf 48},
  065208.

\bibitem[Chini \& Cox(2009)]{Greg}
{\sc Chini, G.P. \& Cox, S.M.} 2009 Large {R}ayleigh number thermal convection:
  {H}eat flux predictions and strongly nonlinear solutions. {\em Physics of
  Fluids\/} {\bf 21}~(8).

\bibitem[Corson(2011)]{Lind}
{\sc Corson, L.T.} 2011 Maximizing the heat flux in steady unicellular porous
  media convection. Geophysical {F}luid {D}ynamics program report. Woods Hole
  Oceanographic Institution.

\bibitem[Cortelezzi {\em et~al.\/}(2008)Cortelezzi, Adrover \&
  Giona]{Optimal08}
{\sc Cortelezzi, L., Adrover, A. \& Giona, M.} 2008 Feasibility, efficiency and
  transportability of short horizon optimal mixing protocols. {\em Journal of
  Fluid Mechanics\/} {\bf 597}, 199--231.

\bibitem[D'Alessandro {\em et~al.\/}(1999)D'Alessandro, Dahleh \&
  Mezic]{d1999control}
{\sc D'Alessandro, D., Dahleh, M. \& Mezic, I.} 1999 Control of mixing in fluid
  flow: A maximum entropy approach. {\em Automatic Control, IEEE Transactions
  on\/} {\bf 44}~(10), 1852--1863.

\bibitem[Doering \& Constantin(1992)]{DC1992}
{\sc Doering, C.R. \& Constantin, P.} 1992 Energy dissipation in shear driven
  turbulence. {\em Physical Review Letters\/} {\bf 69}, 1648--1651.

\bibitem[Doering \& Constantin(1994)]{DC1994}
{\sc Doering, C.R. \& Constantin, P.} 1994 Variational bounds on energy
  dissipation in incompressible flows: {S}hear flow. {\em Physical Review E\/}
  {\bf 49}, 4087--4099.

\bibitem[Doering \& Constantin(1996)]{DC1996}
{\sc Doering, C.R. \& Constantin, P.} 1996 Variational bounds on energy
  dissipation in incompressible flows. {III}. {C}onvection. {\em Physical
  Review E\/} {\bf 53}, 5957--5981.

\bibitem[Doering \& Constantin(1998)]{Doering98}
{\sc Doering, C.R. \& Constantin, P.} 1998 Bounds for heat transport in a
  porous layer. {\em Journal of Fluid Mechanics\/} {\bf 376}, 263--296.

\bibitem[Doering {\em et~al.\/}(2003)Doering, Eckhardt \& Schumacher]{DES2003}
{\sc Doering, C.R., Eckhardt, B. \& Schumacher, J.} 2003 Energy dissipation in
  body-forced plane shear flow. {\em Journal of Fluid Mechanics\/} {\bf 494},
  275--284.

\bibitem[Doering \& Foias(2002)]{DF2002}
{\sc Doering, C.R. \& Foias, C.} 2002 Energy dissipation in body-forced
  turbulence. {\em Journal of Fluid Mechanics\/} {\bf 467}, 289--306.

\bibitem[Doering {\em et~al.\/}(2006)Doering, Otto \& Reznikoff]{DOR2006}
{\sc Doering, C.R, Otto, F. \& Reznikoff, M.G.} 2006 Bounds on vertical heat
  transport for infinite-{P}randtl-numbe~{R}ayleigh-{B}\'enard~convection. {\em
  Journal of Fluid Mechanics\/} {\bf 560}, 229--241.

\bibitem[Doering {\em et~al.\/}(2000)Doering, Spiegel \& Worthing]{DSW2000}
{\sc Doering, C.R., Spiegel, E.A. \& Worthing, R.A.} 2000 Energy dissipation in
  a shear layer with suction. {\em Physics of Fluids\/} {\bf 12}, 1955--1968.

\bibitem[Drazin \& Reid(2004)]{Drazin}
{\sc Drazin, P.G. \& Reid, W.H.} 2004 {\em Hydrodynamic Stability\/}, 2nd edn.
  Cambridge University Press.

\bibitem[Fowler(1997)]{Fowler}
{\sc Fowler, A.C.} 1997 {\em Mathematical Models in the Applied Sciences\/},
  1st edn. Cambridge University Press.

\bibitem[Gubanov \& Cortelezzi(2010)]{gubanov2010towards}
{\sc Gubanov, O. \& Cortelezzi, L.} 2010 Towards the design of an optimal
  mixer. {\em Journal of Fluid Mechanics\/} {\bf 651}, 27--53.

\bibitem[Gubanov \& Cortelezzi(2012)]{Cost12}
{\sc Gubanov, O. \& Cortelezzi, L.} 2012 On the cost efficiency of mixing
  optimization. {\em Journal of Fluid Mechanics\/} {\bf 692}, 112--136.

\bibitem[Gupta \& Joseph(1973)]{GJ1973}
{\sc Gupta, V.P. \& Joseph, D.D.} 1973 Bounds for heat transport in a porous
  layer. {\em Journal of Fluid Mechanics\/} {\bf 57}, 491--514.

\bibitem[Hagstrom \& Doering(2010)]{HD2010}
{\sc Hagstrom, G. \& Doering, C.R.} 2010 Bounds on heat transport in
  {B}\'enard-{M}arangoni convection. {\em Physical Review E\/} {\bf 81},
  047301.

\bibitem[Hagstrom \& Doering(2014)]{HD2013}
{\sc Hagstrom, G.I. \& Doering, C.R.} 2014 Bounds on surface stress-driven
  shear flow. {\em Journal of Nonlinear Science\/} {\bf 24}, 185–--199.

\bibitem[Hassanzadeh(2012)]{PedramMA}
{\sc Hassanzadeh, P.} 2012 Optimal transport from wall to wall. Master's
  thesis, University of California, Berkeley.

\bibitem[Hewitt {\em et~al.\/}(2012)Hewitt, Neufeld \& Lister]{Duncan}
{\sc Hewitt, D.R., Neufeld, J.A. \& Lister, J.R.} 2012 Ultimate regime of high
  {R}ayleigh number convection in a porous medium. {\em Physical Review
  Letters\/} {\bf 108}~(22).

\bibitem[Hewitt {\em et~al.\/}(2013)Hewitt, Neufeld \&
  Lister]{hewitt2013stability}
{\sc Hewitt, D.R., Neufeld, J.A. \& Lister, J.R.} 2013 Stability of columnar
  convection in a porous medium. {\em Journal of Fluid Mechanics\/} {\bf 737},
  205--231.

\bibitem[Horne \& O'Sullivan(1978)]{Horne78}
{\sc Horne, R.N. \& O'Sullivan, P.} 1978 Origin of oscillatory convection in a
  porous medium heated from below. {\em Physics of Fluids\/} {\bf 21},
  1260--1264.

\bibitem[Howard(1963)]{Howard63}
{\sc Howard, L.N.} 1963 Heat transport by turbulent convection. {\em Journal of
  Fluid Mechanics\/} {\bf 17}, 405--432.

\bibitem[Howard(1964)]{Howard}
{\sc Howard, L.} 1964 Convection at high {R}ayleigh numbers. In {\em
  Proceedings of the 11th International Congress of Applied Mechanics\/} (ed.
  H.~G\"ortler), pp. 1109--1115.

\bibitem[Ierley {\em et~al.\/}(2006)Ierley, Kerswell \& Plasting]{Ier}
{\sc Ierley, G.R., Kerswell, R.R. \& Plasting, S.C.} 2006
  Infinite--{P}randtl--number convection. {P}art 2. {A} singular limit of upper
  bound theory. {\em Journal of Fluid Mechanics\/} {\bf 560}, 159--228.

\bibitem[Kerswell(1996)]{Kerswell96}
{\sc Kerswell, R.R.} 1996 Upper bounds on the energy dissipation in turbulent
  precession. {\em Journal of Fluid Mechanics\/} {\bf 321}, 335--370.

\bibitem[Kerswell(1998)]{Kerswell98}
{\sc Kerswell, R.R.} 1998 Unification of variational principles for turbulent
  shear flows: The background method of {D}oering-{C}onstantin and the
  mean-fluctuation formulation of {H}oward-{B}usse. {\em Physica D\/} {\bf
  121}, 175--192.

\bibitem[Kraichnan(1962)]{Kra}
{\sc Kraichnan, R.H.} 1962 Turbulent thermal convection at arbitrary {P}randtl
  number. {\em Physics of Fluids\/} {\bf 5}, 1374--1389.

\bibitem[Krommes \& Smith(1987)]{KS1987}
{\sc Krommes, J.A. \& Smith, R.A.} 1987 Rigorous upper bounds for transport due
  to passive advection by inhomogeneous turbulence. {\em Annals of Physics\/}
  {\bf 177}, 246--329.

\bibitem[Lin {\em et~al.\/}(2011)Lin, Thiffeault \& Doering]{Optimal11}
{\sc Lin, Z., Thiffeault, J-L. \& Doering, C.R.} 2011 Optimal stirring
  strategies for passive scalar mixing. {\em Journal of Fluid Mechanics\/} {\bf
  675}, 465--476.

\bibitem[Malkus(1954)]{Malkus}
{\sc Malkus, W.V.R.} 1954 The heat transport and spectrum of thermal
  turbulence. {\em Proceedings of the Royal Society of London. Series A\/} {\bf
  225}, 196--212.

\bibitem[Mathew {\em et~al.\/}(2007)Mathew, Mezic, Grivopoulos, Vaidya \&
  Petzold]{Optimal07}
{\sc Mathew, G., Mezic, I., Grivopoulos, S., Vaidya, U. \& Petzold, L.} 2007
  Optimal control of mixing in {S}tokes fluid flows. {\em Journal of Fluid
  Mechanics\/} {\bf 580}, 261--281.

\bibitem[Nickerson(1969)]{Nickerson69}
{\sc Nickerson, E.C.} 1969 Upper bounds on torque in cylindrical~{C}ouette
  flow. {\em Journal of Fluid Mechanics\/} {\bf 38}, 807---815.

\bibitem[Okabe {\em et~al.\/}(2008)Okabe, Eckhardt, Thiffeault \&
  Doering]{Stirring08}
{\sc Okabe, T., Eckhardt, B., Thiffeault, J-L. \& Doering, C.R.} 2008 Mixing
  effectiveness depends on the source-sink structure: simulation results. {\em
  Journal of Statistical Mechanics - Theory and Experiment\/} {\bf 07}, P07018.

\bibitem[Otero(2002)]{Otero02}
{\sc Otero, J.} 2002 Bounds for the heat transport in turbulent convection. PhD
  thesis, University of Michigan, Ann Arbor.

\bibitem[Otero {\em et~al.\/}(2004)Otero, Dontcheva, Johnston, Worthing,
  Kurganov, Petrova \& Doering]{Otero}
{\sc Otero, J., Dontcheva, L.A., Johnston, H., Worthing, R.A., Kurganov, A.,
  Petrova, G. \& Doering, C.R.} 2004 High--{R}ayleigh--number convection in a
  fluid--saturated porous layer. {\em Journal of Fluid Mechanics\/} {\bf 500},
  263--281.

\bibitem[Petrov {\em et~al.\/}(2005)Petrov, Lu \& Doering]{PLD2005}
{\sc Petrov, N.P., Lu, L. \& Doering, C.R.} 2005 Variational bounds on the
  energy dissipation rate in body-forced shear flow. {\em Journal of
  Turbulence\/} {\bf 6}.

\bibitem[Plasting \& Kerswell(2003)]{PK2003}
{\sc Plasting, S.C. \& Kerswell, R.R.} 2003 Improved upper bound on the energy
  dissipation rate in plane~{C}ouette flow: the full solution to~{B}usse's
  problem and the~{C}onstantin-{D}oering-{H}opf problem with one-dimensional
  background field. {\em Journal of Fluid Mechanics\/} {\bf 477}, 363--379.

\bibitem[Plasting \& Young(2006)]{PY2006}
{\sc Plasting, S.C. \& Young, W.R.} 2006 A bound on scalar variance for the
  advection-diffusion equation. {\em Journal of Fluid Mechanics\/} {\bf 552},
  289--298.

\bibitem[Rayleigh(1916)]{Rayleigh1916}
{\sc Rayleigh, Lord.} 1916 On convection currents in a horizontal layer of
  fluid, when the higher temperature is on the under side. {\em Philosophical
  Magazine\/} {\bf 32}, 529--546.

\bibitem[Rollin {\em et~al.\/}(2011)Rollin, Dubief \& Doering]{RDD2011}
{\sc Rollin, B., Dubief, Y. \& Doering, C.R.} 2011 Variations on {K}olmogorov
  flow: turbulent energy dissipation and mean flow profiles. {\em Journal of
  Fluid Mechanics\/} {\bf 670}, 204--213.

\bibitem[Shaw {\em et~al.\/}(2007)Shaw, Thiffeault \& Doering]{Stirring07}
{\sc Shaw, T.A., Thiffeault, J.-L. \& Doering, C.R.} 2007 Stirring up trouble:
  Multi-scale mixing measures for steady scalar sources. {\em Physica D\/} {\bf
  231}, 143--164.

\bibitem[Siggers {\em et~al.\/}(2004)Siggers, Kerswell \& Balmforth]{SKB2004}
{\sc Siggers, J.H., Kerswell, R.R. \& Balmforth, N.J.} 2004 Bounds on
  horizontal convection. {\em Journal of Fluid Mechanics\/} {\bf 517}, 55--70.

\bibitem[Spiegel(1962)]{Ed}
{\sc Spiegel, E.~A.} 1962 Thermal turbulence at very small {P}randtl number.
  {\em Journal of Geophysical Research\/} {\bf 67}, 3063--3070.

\bibitem[Tang {\em et~al.\/}(2009)Tang, Caulfield \& Kerswell]{TCK2009}
{\sc Tang, W., Caulfield, C.P. \& Kerswell, R.R.} 2009 A prediction for the
  optimal stratification for turbulent mixing. {\em Journal of Fluid
  Mechanics\/} {\bf 634}, 487--497.

\bibitem[Tang {\em et~al.\/}(2004)Tang, Caulfield \& Young]{TCY2004}
{\sc Tang, W., Caulfield, C.P. \& Young, W.R.} 2004 Bounds on dissipation in
  stress driven flow. {\em Journal of Fluid Mechanics\/} {\bf 510}, 333--352.

\bibitem[Thiffeault {\em et~al.\/}(2004)Thiffeault, Doering \& Gibbon]{TDG2004}
{\sc Thiffeault, J.-L., Doering, C.R. \& Gibbon, J.D.} 2004 A bound on mixing
  efficiency for the advection-diffusion equation. {\em Journal of Fluid
  Mechanics\/} {\bf 521}, 105--114.

\bibitem[Trefethen(2001)]{Trefethen}
{\sc Trefethen, L.~N.} 2001 {\em Spectral Methods in MATLAB\/}, 1st edn. SIAM.

\bibitem[Wen {\em et~al.\/}(2013)Wen, Chini, Dianati \& Doering]{WCDD2013}
{\sc Wen, B., Chini, G.P., Dianati, N. \& Doering, C.R.} 2013 Computational
  approaches to aspect-ratio-dependent upper bounds and heat flux in porous
  medium convection. {\em Physics Letters A\/} {\bf 377}, 2931--2938.

\bibitem[Wen {\em et~al.\/}(2012)Wen, Dianati, Lunasin, Chini \& Doering]{Wen}
{\sc Wen, B., Dianati, N., Lunasin, E., Chini, G.P. \& Doering, C.R.} 2012 New
  upper bounds and reduced dynamical modeling for~{R}ayleigh-{B}\'enard
  convection in a fluid saturated porous layer. {\em Communications in
  Nonlinear Science and Numerical Simulation\/} {\bf 17}, 2191--2199.

\bibitem[Whitehead \& Doering(2011)]{Whitehead11}
{\sc Whitehead, J.P. \& Doering, C.R.} 2011 Ultimate state of two--dimensional
  {R}ayleigh--{B}\'enard convection between free--slip fixed--temperature
  boundaries. {\em Physical Review Letters\/} {\bf 106}, 244501.

\bibitem[Whitehead \& Doering(2012)]{Whitehead12}
{\sc Whitehead, J.P. \& Doering, C.R.} 2012 Rigid bounds on heat transport by a
  fluid between slippery boundaries. {\em Journal of Fluid Mechanics\/} {\bf
  707}, 241--259.

\end{thebibliography}

\end{document}